\documentclass[iop,apj,twocolumn]{emulateapj}
\usepackage{times}
\usepackage{epsfig}
\usepackage{amsmath, amsthm, amssymb}
\usepackage[plainpages=false, colorlinks=true, anchorcolor=blue,
  linkcolor=blue, citecolor=blue, bookmarks=false]
{hyperref}
\usepackage{color}
\usepackage{microtype}
\usepackage{float}
\usepackage[caption = false]{subfig}
\usepackage{graphicx}
\usepackage[export]{adjustbox}

\graphicspath{{figures/}}

\newcommand{\tycho}{{\tt TYCHO}}
\newcommand{\snsph}{{\tt SNSPH}}
\newcommand{\supernu}{{\tt SuperNu}}
\newcommand{\maverick}{{\tt Maverick}}

\begin{document}

\title{Light Curves and Spectra from a Unimodal Core-Collapse SuperNova}
\author{Ryan T. Wollaeger$^{1}$, Aimee L. Hungerford$^{1}$, Chris L. Fryer$^{1}$,
Allan B. Wollaber$^{1}$, Daniel R. van~Rossum$^{2}$, and Wesley Even$^{1}$}

\affil{$^{1}$Center for Theoretical Astrophysics, Los Alamos National Laboratory,
  P.O. Box 1663, Los Alamos,
  NM 87545; wollaeger@lanl.gov}
\affil{$^{2}$Flash Center for Computational Science, Department of Astronomy
  \& Astrophysics, University of Chicago, Chicago,
  IL, 60637}

\begin{abstract}
  To assess the effectiveness of optical emission as a probe of spatial
  asymmetry in core-collapse supernovae (CCSNe), we apply the radiative
  transfer software, \supernu, to a unimodal CCSN model.
  The \snsph\ radiation-hydrodynamics software was used to
  simulate an asymmetric explosion of a 16 M$_{\odot}$ ZAMS binary star.
  The ejecta has 3.36 M$_{\odot}$ with 0.024 M$_{\odot}$ of
  radioactive $^{56}$Ni, with unipolar asymmetry along the z-axis.
  For 96 discrete angular views, we find the ratio between maximum
  and minimum peak total luminosities is $\sim$1.36.
  The brightest light curves emerge from views orthogonal to the z-axis.
  Multigroup spectra from UV to IR are obtained.
  We find a shift in wavelength with viewing angle in a near-IR Ca II
  emission feature, consistent with Ca being mostly in the unimode.
  We compare emission from the grey gamma-ray transfer in
  \supernu\ and from the detailed gamma-ray transfer
  code \maverick.
  Relative to the optical light curves, the brightness of the
  gamma-ray emission is more monotonic with respect to viewing angle.
  UBVRI broad-band light curves are also calculated.
  Parallel with the unimode, the U and B bands have excess luminosity
  at $\gtrsim 10$ days post-explosion, due to $^{56}$Ni on the unimode.
  We compare our CCSN model with SN 2002ap, which is thought to
  have a similar ejecta morphology.
\end{abstract}

\section{Introduction}
\label{sec:intro}

%-- explosion, asymmetry
In the basic picture of core-collapse supernovae (CCSNe), silicon
burning in a $\gtrsim8$ M$_{\odot}$ star eventually produces a
$1.44$ M$_{\odot}$ iron core, the core's electron degeneracy pressure can no
longer counteract gravity, and the core collapses to a neutron star or
black hole (BH) in fractions of a second \citep{colgate1966,bethe1990,janka2007}.
The energy released by the collapse should sometimes result in an explosion.
Although this basic picture of CCSNe from 1966 remains intact,
the details have evolved, and our understanding of the engine has improved
over the past 5 decades.
Many of these advances arose from theorists trying to explain
observations of SN 1987A.
For instance, there is extensive mixing of the $^{56}$Ni in SN 1987A,
evident in observations from the early emission of $\gamma$-rays to broad widths
of the iron lines (for a review, see~\citet{hungerford2003}).
In addition, spectropolarimetry and speckle imaging from SN 1987A
indicate there are asymmetries in the ejecta
\citep{cropper1988,chevalier1989,wang2002}.
These asymmetries played a pivotal role in understanding the engine
behind these supernovae and, to explain them, scientists began to study
the convective instabilities above the newly formed neutron star in the
collapse of the stellar core~\citep{benz1994,herant1994,herant1995}.
These convective instabilities are important for the supernova engine, and
this early evidence for asymmetries played a crucial role in defining
the current CCSN engine (for a review, see~\citet{fryer2007}).

Since SN 1987A, evidence for strong supernova asymmetries has grown
(for a review, see~\citet{ellinger2012}).
Optical and x-ray imaging of Cassiopeia A (Cas A), a supernova remnant
originating from an SN IIb of a red supergiant star \citep{krause2008},
have uncovered details of asymmetry and structure in the high-velocity
regions of the ejecta~\citep{fesen2001,laming2003,fesen2006}.
Analysis of the spectropolarimetry in SN 2005bf~\citep{tanaka2009},
the double-peaked oxygen lines in SN 2008D
\citep{modjaz2009,maund2009,couch2011}, and the oxygen lines in the
nebular spectra of SN 2009jf~\citep{sahu2011,valenti2011} indicate
these events are asymmetric SNe Ib.
The strongest evidence to date has been the recent observations of the
$^{44}$Ti distribution in Cas A, a direct tracer of the inner
supernova engine, showing evidence for asymmetries in the explosion
\citep{grefenstette2014}.

Considerable effort has been invested in developing software
that is capable of accurately simulating multidimensional
neutrino radiation-hydrodynamics in the presence of gravity
(see, for instance,~\citet{livne1993,fryxell2000,fryer2006,
  burrows2006,ott2008,almgren2010,zhang2011,abdikamalov2012,
  zhang2013,dolence2015})
and performing computationally intensive studies
with this software~\citep{fryer2002,fryer2007,burrows2012,bruenn2013,
  couch2013a,couch2013b,dolence2015,couch2015}.
For recent reviews of the CCSN field, see
\citet{janka2012,burrows2013,janka2016,muller2017}.
The work in the past two decades has revealed the importance
of multidimensional effects and asymmetry in the explosion
mechanism for various progenitor models and
radiation-hydrodynamics methods.
With additional radiative transfer modeling, the results of
the explosion simulations can be compared to detections
of gamma-rays and UV/optical/IR emission from SNe.

%-- theoretical evidence for asymmetric explosions:
Many surveys, such as the Sloan Digital Sky Survey
\citep{frieman2008,sako2008} and Palomar Transient
Factory, regularly detect UV/optical/IR emission from
CCSNe~\citep{york2000,law2009}.
The abundance of CCSN data from these surveys motivate
theoretical efforts in modeling UV/optical/IR light curves
and spectra.
The properties of the light curves have been studied analytically
\citep{arnett1980,arnett1982,chatzopoulos2012},
and numerically~\citep{kasen2009,dessart2012,hillier2012,
  frey2013,kleiser2014,maeda2002,maeda2006,rapoport2012}.
For instance, to reproduce SN IIP standard candle relationships
\citep{hamuy2002}, \cite{kasen2009} simulate radiative transfer
in SN IIP models that vary in progenitor mass, explosion energy,
and metallicity.
\cite{frey2013} have applied a radiation-diffusion-hydrodynamics
solver with detailed radiative transfer post-processing
\citep{fryer2009,fryer2010,frey2013} to simulate several
SN Ic models.

%-- hypernovae with grbs
Multidimensional radiative transfer and hydrodynamics simulations
have shed light on observations of bright CCSNe (or hypernovae; HNe)
associated with gamma-ray bursts (GRBs), or GRB-HNe, as well.
These simulations were motivated by theoretical insight
into the HN explosion mechanism~\citep{rapoport2012}.
The GRB jet should cause unimode asymmetry in the
HN ejecta~\citep{khokhlov1999}, and accretion may shape disk
wind into lobes around the BH spin axis~\citep{proga2003}.
Observations of GRB-HNe, such as SN 1998bw, indicate
significant asymmetry in the ejecta~\citep{mazzali2001}.
In particular, for SN 1998bw, light curves from 1D models fit
well at early time but decline too rapidly in the tail
\citep{sollerman2000,nakamura2001,maeda2002}.
Models of bright CCSNe (or hypernovae) associated with gamma-ray
bursts have been simulated with 2D cylindrical hydrodynamics and
nucleosynthesis for the explosion, followed by 1D NLTE
\citep{maeda2002} or with 3D transport
(grey \citet{maeda2006}; multifrequency \citet{rapoport2012}).
These models introduce artificial asymmetry in the initial
conditions for the explosion, rescaling the equatorial and
polar velocity components~\citep{maeda2002,rapoport2012}.
\cite{rapoport2012} make the first application of the
multidimensional radiative transfer Monte Carlo code {\tt Artis}
\citep{kromer2009} to an SN Ic model of a GRB-HN, using
the models with parameterized asymmetry from~\cite{maeda2002}.
In two of these models, velocity is increased along the
polar axis, to emulate the unimode remains from the GRB
\citep{maeda2002,rapoport2012}.
\cite{rapoport2012} find that the asphericity in their model causes
a brighter, bluer pre-maximum light curve in views close in
alignment with the polar ($z$-)axis, but higher peak luminosities
for views towards the equatorial ($xy$-)plane.
This result is consistent with the equatorial views seeing
radiation from a larger cross-section of the ejecta
\citep{kromer2009}.

\begin{figure}
\begin{center}
  \subfloat[]{\includegraphics[height=70mm]{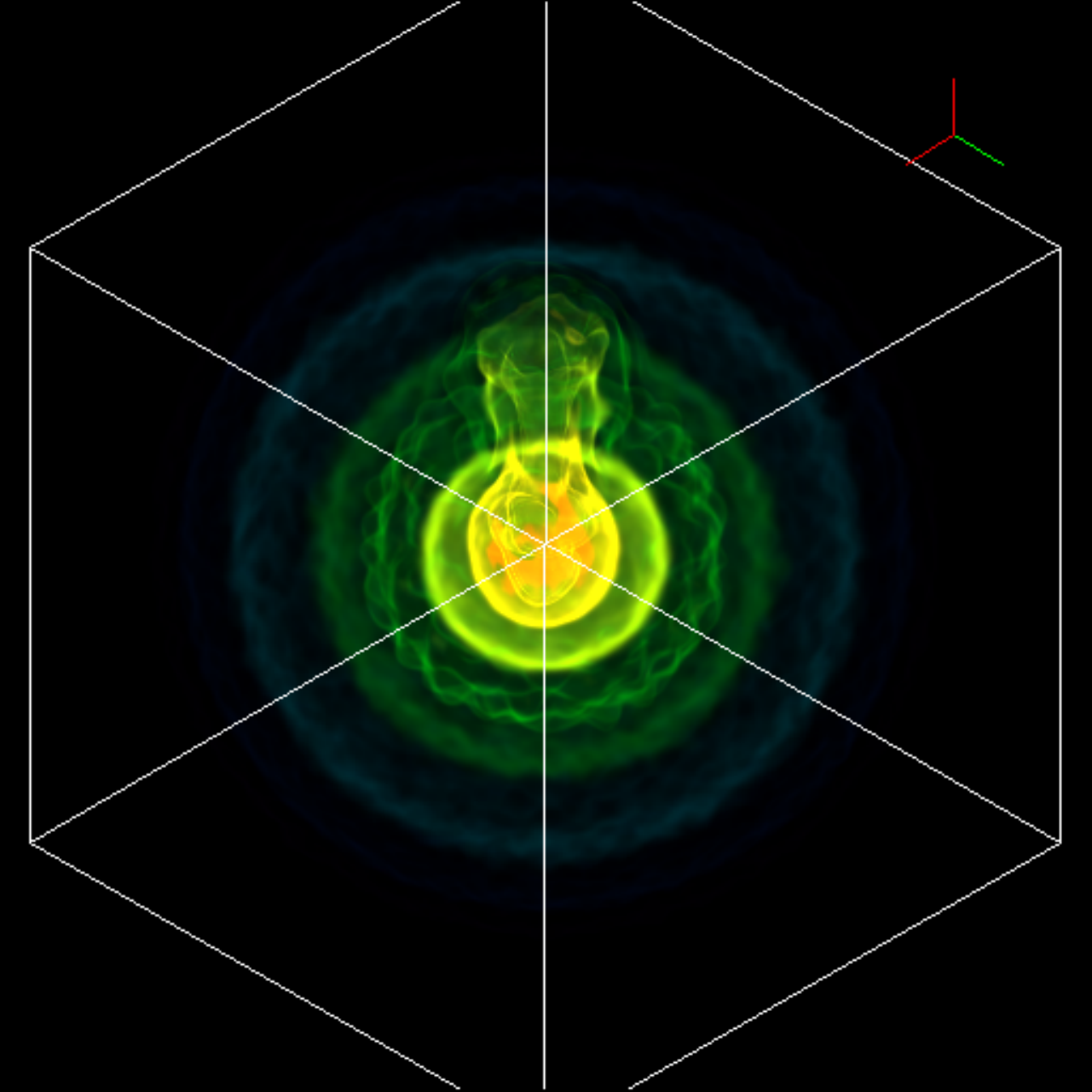}\label{fg1a}}\\
\end{center}
\subfloat[]{\includegraphics[height=42mm]{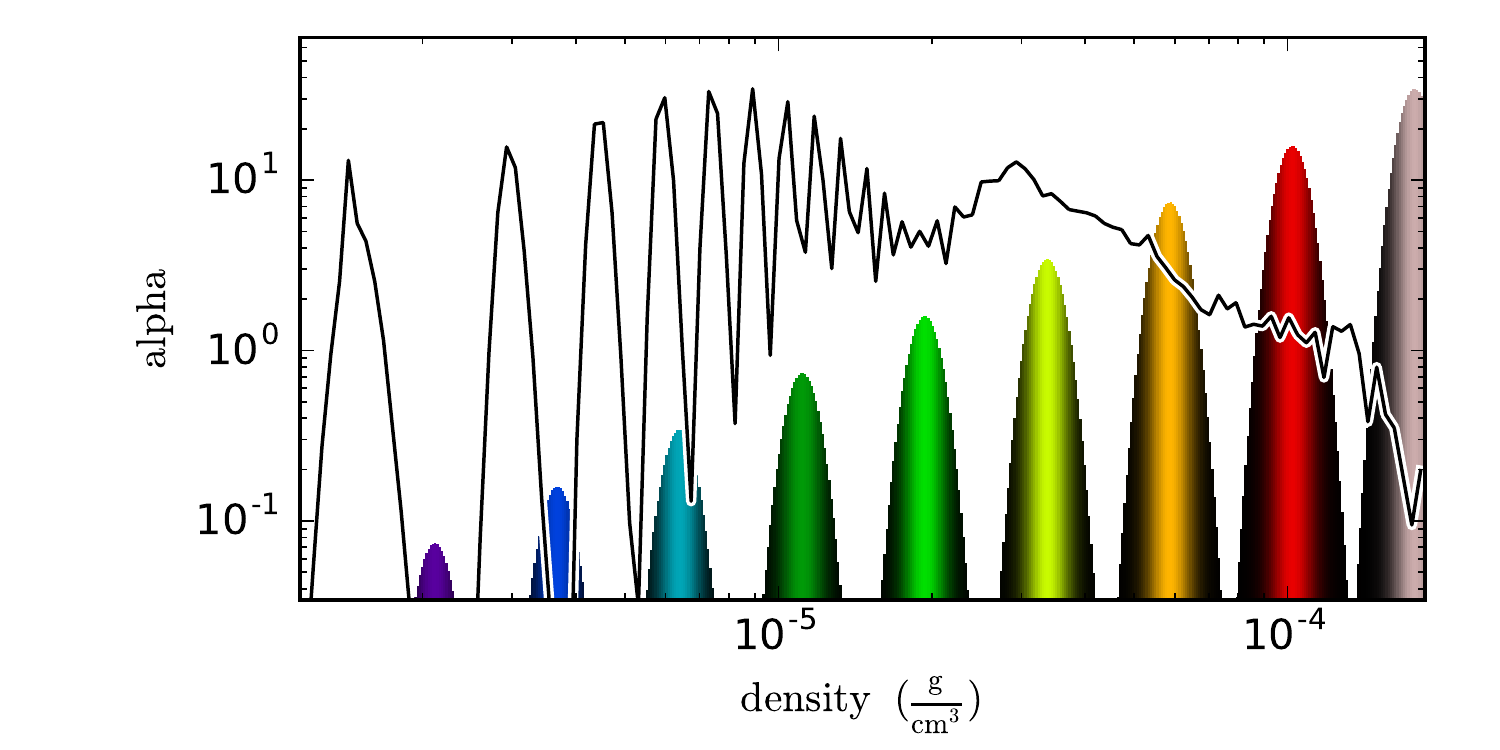}\label{fg1b}}
\caption{
  A yt volume rendering and transfer function
  of our CCSN ejecta.
  In Fig.~\ref{fg1a}, an isometric volume rendering of
  density when the ejecta becomes homologous.
  The domain is outlined with a box with edges parallel
  to the $x$, $y$, and $z$ axes.
  The unimode can be seen extending upward along the $z$-axis.
  In Fig.~\ref{fg1b}, a transfer function for the
  coloration, where higher alpha corresponds to
  more opaque coloring.
}
\label{fg1}
\end{figure}

%-- novel work here (see abstract)
We obtain synthetic light curves and spectra from the
outflow of a CCSN model, with a unimodal component
in the ejecta.
The model is more representative of a standard
CCSN than an HN.
In particular, the structure is motivated by the models
of~\cite{fryer2004} and~\cite{blondin2003}, where asymmetric
modes are shown to develop from convective and standing accretion
shock instabilities, respectively (see also~\cite{dolence2013}
for asymmetry from large-scale, neutrino-heating driven plumes).
The ejecta is derived from a 16 M$_{\odot}$ ZAMS mass non-rotating
star that is stripped to $\sim 5$ M$_{\odot}$ from
mass loss to a binary companion.
Similar to the GRB-HN work of~\cite{maeda2002} and
\cite{rapoport2012}, asymmetry is artificially introduced for a
portion of the explosion, resulting in a unimode structure.
Fig.~\ref{fg1} has a volume rendering of the density at the
beginning of the homologous expansion phase, $\sim 2660$ seconds.
The ejecta has 3.36 M$_{\odot}$ with 0.024 M$_{\odot}$ of
radioactive $^{56}$Ni.
This is a factor of $\sim3$ lower in mass and a factor of
$\sim10$ lower in $^{56}$Ni mass than the models considered by
\cite{rapoport2012}.
The ejecta is hydrodynamically evolved to a homologous
state with the 3D smoothed particle radiation-hydrodynamics
code \snsph~\citep{fryer2006}.
With the constraint of homologous outflow,
we have assumed the ejecta is expanding in a vacuum,
and not through a CSM.
With these constraints, our model represents an extreme case
of asymmetry.
To supplement the previous multidimensional light curve
studies, we apply the radiative transfer software, \supernu,
in 1D spherical and 3D Cartesian geometries to post-process
the ejecta obtained from \snsph.
For the 1D simulations, we spherically average the ejecta.

We perform 1D simulations to determine appropriate spatial cell,
time step, and wavelength group sizes for our ejecta.
We refer to the surface where the inward integral of the Planck
opacity with respect to radius is 1 as the ``Planck photosphere''.
The Planck photosphere helps to roughly determine where spatial
grid cells need to be well sampled by SPH particles.
In order to have accurate radiative transfer in a cell, the
cell's density and temperature must be accurate, which in turn
implies there should be at least one SPH particle in proximity
to the cell.
However, cells that are poorly sampled by SPH but are optically
thin should not greatly impact the radiative transfer.

\begin{figure}
\includegraphics[height=70mm]{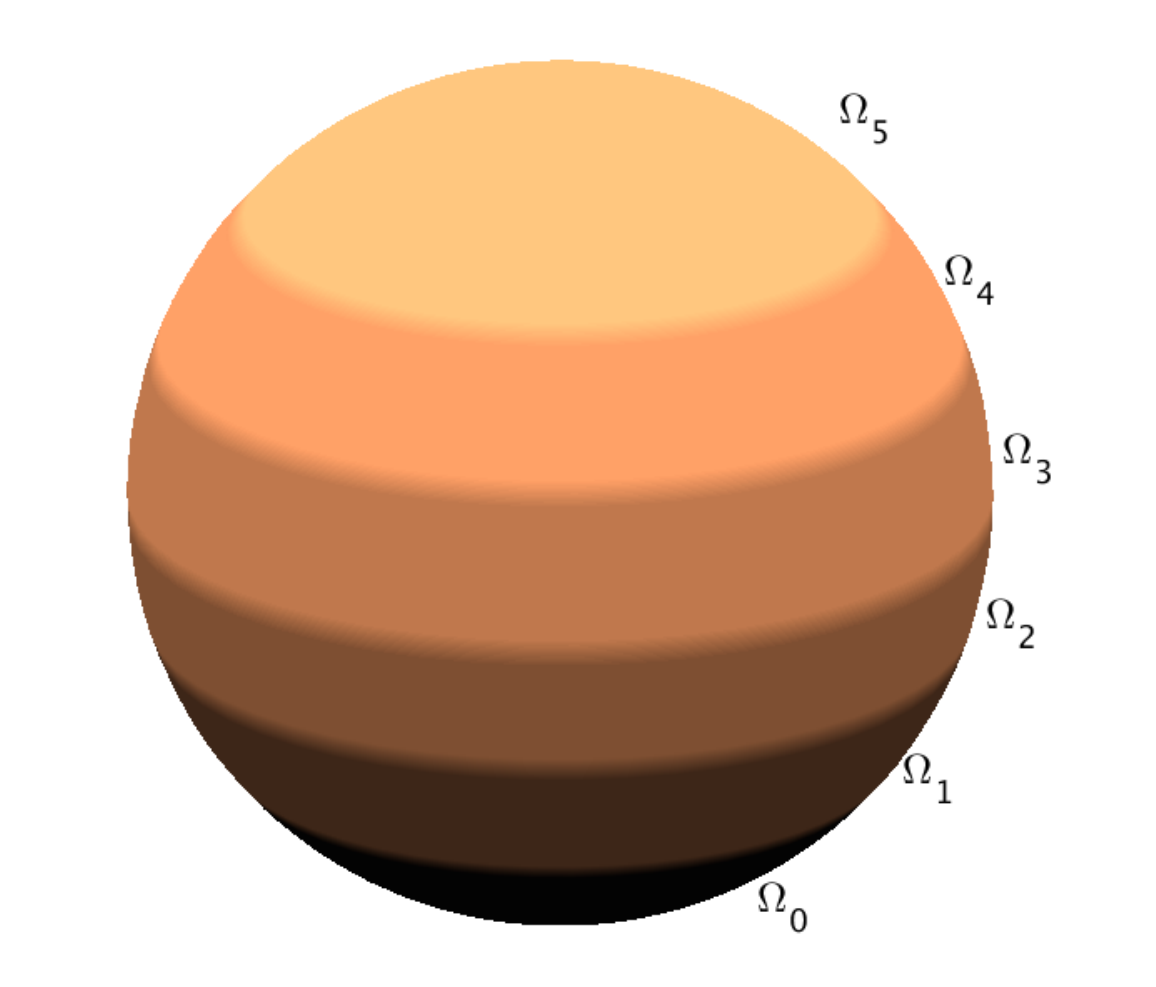}
\caption{
  A visualization of six uniform polar solid angle bins,
  with viewing angle tilted towards the northern
  hemisphere.
  Monte Carlo particles escaping the ejecta with a
  direction in the solid angle range of $\Omega_{i}$ are
  tallied in bin $\Omega_{i}$.
  This sphere of observational directions is independent
  of the geometry of the ejecta.
}
\label{fg2}
\end{figure}

Subsequently, we perform a $120^{3}$ cell 3D Cartesian radiative
transfer simulation.
For the 3D calculation, the viewing angles are partitioned into
96 polar bins, where the $z$-axis is taken to be aligned with the
unimode of $^{56}$Ni; the azimuthal variation of the light curve is
small relative to the polar variation.
Figure~\ref{fg2} has a depiction of 6 polar viewing bins; the
directions of escaping MC particles would determine where they
are tallied on this sphere.
The simulations reveal that the peak luminosity of the light curve
can vary significantly with viewing angle, but the shape
of the light curves and spectra do not change significantly.
Specifically, the locations of the peak luminosities only differ by
as much as $\sim$2 days.

%-- navigation paragraph
This article is organized as follows.
In Section~\ref{sec:methods}, we briefly summarize each
code in the software pipeline: \snsph, the mapping
of SPH data to the grid, and \supernu.
In Section~\ref{sec:ejecta}, we briefly present details of
the stellar evolution, explosion method, and the ejecta
structure and composition.
In Section~\ref{sec:1dconv}, we demonstrate in 1D that the
radiative transfer calculation is robust for several
numerical set-ups, and is in a converged regime.
In Section~\ref{sec:3dlcspec}, we show results from a
high-resolution, 3D Cartesian simulation and compare the
light curves to 1D.
In Section~\ref{sec:gamma}, for the 3D simulation, we
present gamma-ray light curves from the simple grey
gamma-ray treatment in \supernu,
compare to the optical light curves, and compare to light
curves from the detailed gamma-ray transfer code \maverick\
\citep{hungerford2003}.
In Section~\ref{sec:caii}, we discuss a Ca II emission feature
which has a viewing angle trend similar to the gamma-ray light
curves.
In Section~\ref{sec:ubvri}, we provide UBVRI broadband light
curves for views aligned parallel and perpendicular
with the unimode.
In Section~\ref{sec:sn2002ap}, we briefly compare the unimodal
CCSN model with the light curve and spectra of SN 2002ap, which
is thought to have similar properties.
In Section~\ref{sec:caveats}, we detail some of the
assumptions and caveats made in these calculations.
We conclude in Section~\ref{sec:conc} with some remarks
on the efficacy of probing the asymmetries in our unimodal
CCSN structure with optical/UV radiation relative to
gamma-ray radiation.

\section{Methods}
\label{sec:methods}

To obtain light curves and spectra, we perform the
calculations in steps.
First, we run an \snsph\ simulation of a CCSN event
until the explosion energy has driven the ejecta into a
homologous state.
The initial conditions for \snsph\ are from
the output of a model simulated in 1D through core-collapse,
with the stellar evolution code \tycho~\citep{young2005}.
From \snsph, the resulting structure is a set
of particles; each has a mass, density, velocity, temperature,
and chemical abundance.
The particles are then mapped to a spatial grid and an
average is obtained for each grid cell.
The gridded structure is used as an initial state for
the \supernu\ simulations.
Brief summaries of the codes are given in Sections
\ref{sec:snsph},~\ref{sec:sphgrid}, and~\ref{sec:supernu}
that follow.

\subsection{SNSPH}
\label{sec:snsph}

The \snsph\ code has an implementation of
smooth particle hydrodynamics (SPH) along with
physics features for SNe~\citep{fryer2006}.
SPH has seen considerable development and application
in astrophysics (see, for instance,
\citet{rosswog2009,owen2014,rosswog2015}).
SPH solves the hydrodynamics equations by interpolating
functions among particles; these particles are
a set of disordered points along with associated
hydrodynamical data
\citep{gingold1977,monaghan1992,monaghan2005}.
A tree structure is implemented to treat the gravitational
force between distant clusters of particles as one-body forces.
The tree structure is also used to determine particle neighbors
\citep{fryer2006}.

In \snsph, radiation transport for neutrinos and photons
is performed with flux-limited diffusion~\citep{herant1994,fryer2006}.
We do not apply the flux-limited diffusion for the \snsph
calculation presented here.
The initial conditions for \snsph\ are calculated from
a 1D code that simulates the collapse through the launch of
the explosion~\citep{herant1994,fryer1999}.
To achieve an explosion, energy is injected into the
convective region lying between the edge of the proto-neutron
star and the stalled shock position.
The convective region above the proto-neutron star was
assumed to be 0.1 M$_{\odot}$ in size.
\tycho\ and \snsph\ employed an 18
isotope (15 element) network for nucleosynthesis and
energy generation~\citep{ellinger2012}.

\subsection{SPH-Grid Mapping}
\label{sec:sphgrid}

We map the particles to a grid as though they are
point-particles.
The SPH smoothing kernels could be used reconstruct the
continuous eject properties, then these profiles could be
integrated over cells in the spatial grid.
For the unimodal CCSN model, resolution tests in 1D and 3D
indicate the point approximation is sufficient to resolve the
radioactively powered light curve.
Moreover, the point approximation generally leads to more
zero-mass cells in the outer regions of the ejecta, allowing
\supernu\ to employ a domain compression that lowers
the memory footprint of the simulation.
However, we must note that the point-approximation in 3D
Cartesian geometry performs poorly on the early thermal
shock-breakout light curve from 0 to 7 days post-explosion.
The shock-breakout region of the ejecta is sparsely populated
with particles, relative to the grid resolution required to
resolve the radioactively powered light curve.
We leave more sophisticated SPH-grid mapping and resolution of
the shock-breakout light curve as future work.

\begin{figure}
\subfloat[]{\includegraphics[height=70mm]{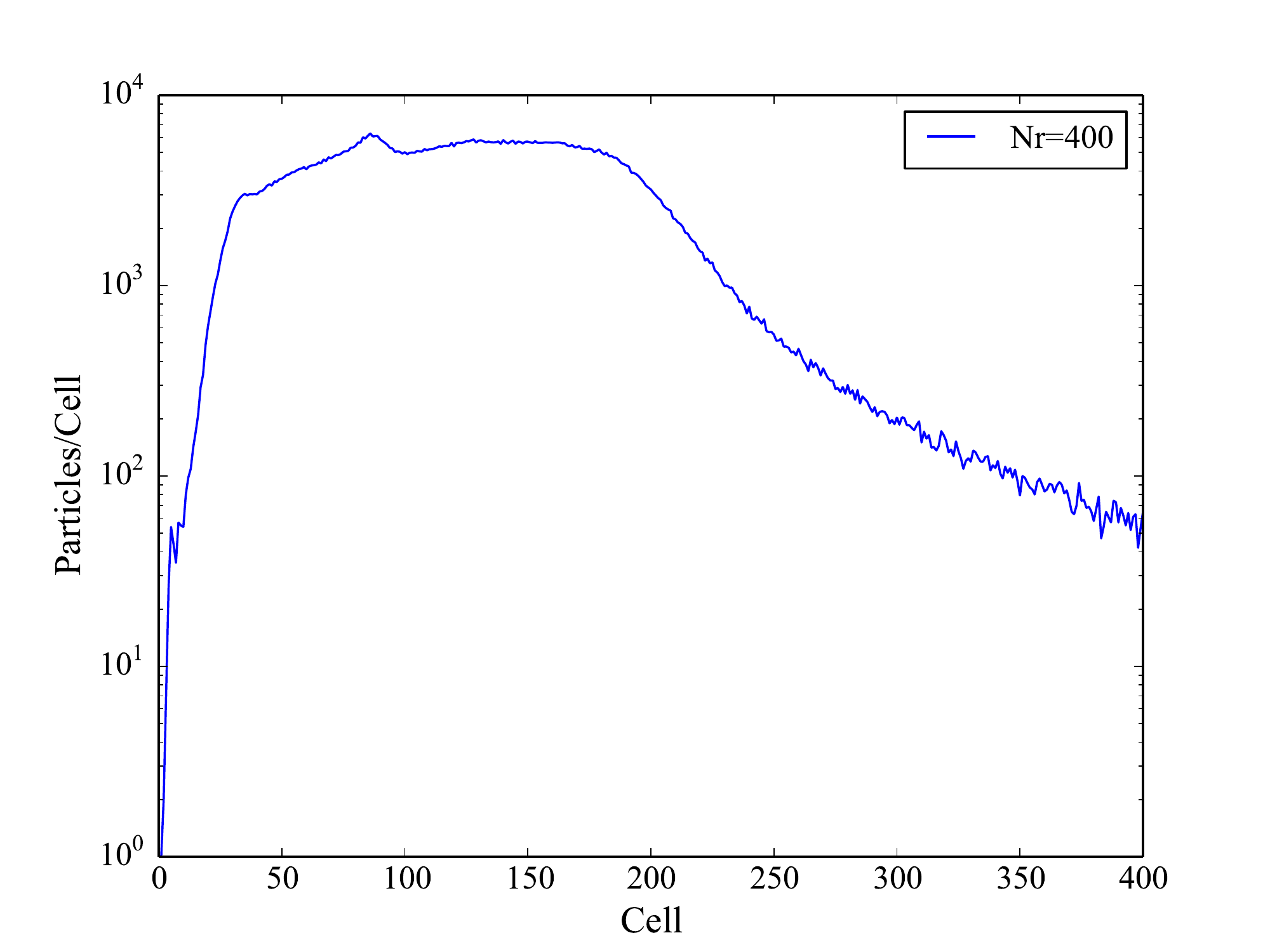}\label{fg3a}}\\
\subfloat[]{\includegraphics[height=70mm]{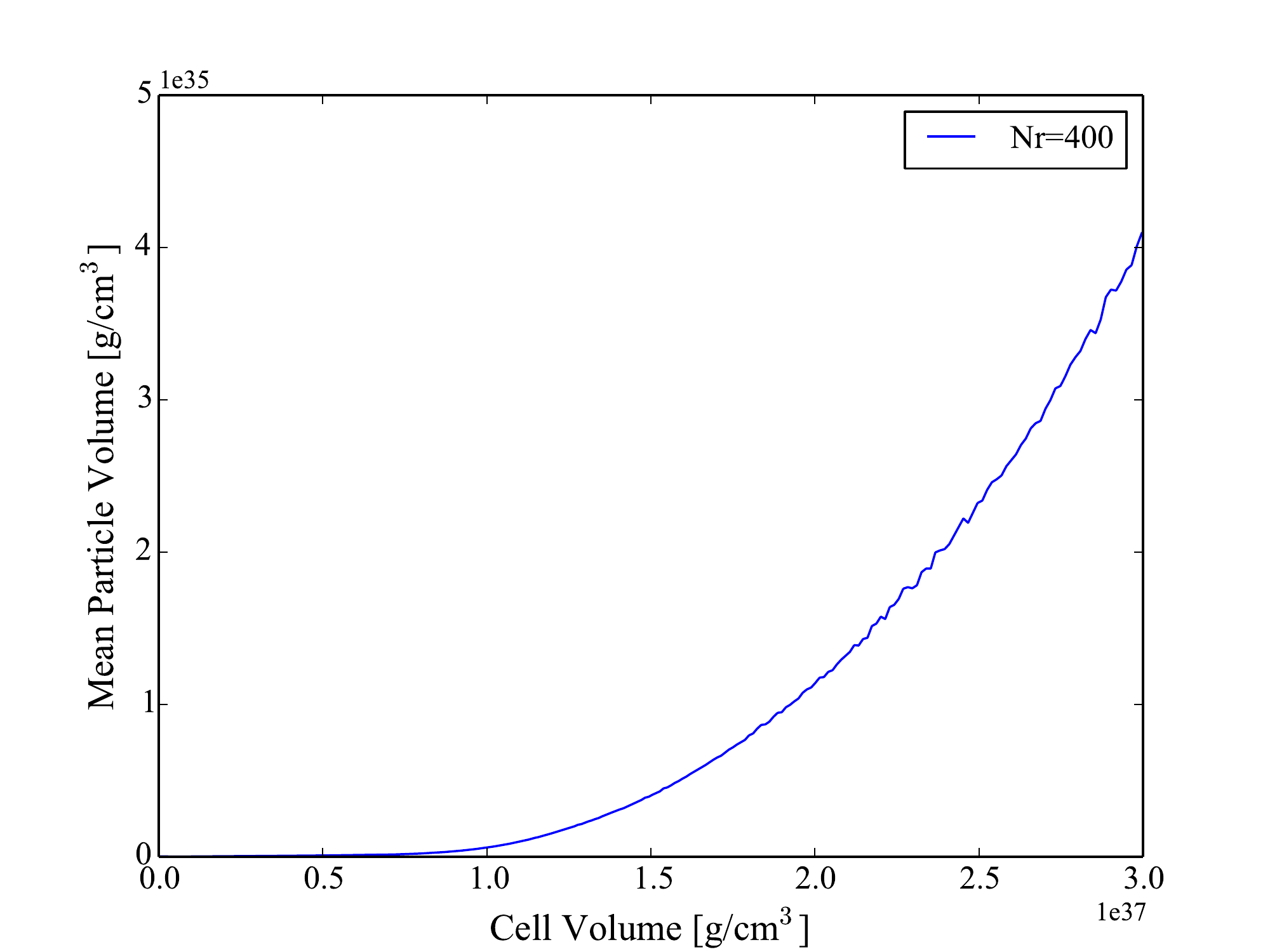}\label{fg3b}}
\caption{
  SPH particles mapped to a 400 cell 1D spherical grid.
  In Fig.~\ref{fg3a}, the number of SPH particles per cell vs cell number.
  In Fig.~\ref{fg3b}, the estimated SPH ``particle volume'',
  $m_{p}/\rho_{p}$, averaged per cell vs cell volume.
  Since the volume per particle is small and there are many particles
  per cell, the point approximation for SPH particles is reasonable.
}
\label{fg3}
\end{figure}

Plotted in Fig.~\ref{fg3} are results of the SPH-mapping for
the 400 spatial cell 1D spherical test; the 400 cell test is
the highest spatial resolution in 1D.
From Fig.~\ref{fg3a}, which has the number of particles per cell
versus cell index, we calculate that the Planck photosphere only
exists in radial cells with $>50$ SPH particles for the 80 day
span of the radiative transfer simulations (the Planck photosphere
moves from cell 400 to roughly cell 80 over the entire 80 day period).
Figure~\ref{fg3b} has the average of the estimated particle
volumes per cell versus cell volume, where a particle volume is
estimated as $m_{p}/\rho_{p}$.
From Fig.~\ref{fg3b}, it is discernible that the volume per particle
is much lower than the cell volumes for the spatial range simulated.
Considering Fig.~\ref{fg3}, the point-particle approximation appears
to be reasonable for spherical geometry.

In 3D Cartesian geometry with $120^{3}$ cells, the 1D radial average of
the estimated SPH particle volumes is about 100 times greater than the
size of the grid cells at the edge of the ejecta.
This discrepancy is indicative of the low sampling of particles per
cell in the outer region, where particle kernels would have larger
smoothing lengths.
However, by peak luminosity, the Planck photosphere has moved in
approximately halfway in radius through the ejecta, where the $120^{3}$
cell 3D Cartesian simulation has an average of about 3 particles
per cell.

\subsection{SuperNu}
\label{sec:supernu}

The \supernu\ code has an implementation of Implicit Monte Carlo
(IMC)~\citep{fleck1971,wollaber2016} for thermal radiative
transfer, and Discrete Diffusion Monte Carlo (DDMC)
\citep{densmore2007,densmore2012,abdikamalov2012} to accelerate
IMC in optically thick regions of phase space.
\supernu\ has features specialized for homologous outflows
and structured opacity~\citep{wollaeger2013,wollaeger2014}.
DDMC is one of several techniques that use diffusion to accelerate
Monte Carlo transport (see, for instance,~\citet{fleck1984,gentile2001,
densmore2007,densmore2008,densmore2012,abdikamalov2012,cleveland2014}.
Semi-relativistic effects are incorporated following the prescription
of~\cite{abdikamalov2012}, where Doppler shift and advection effects
are operator split in DDMC.

%-- review of 3D rad-tran codes for SN LCs; applications
The approximation of homologous outflow in \supernu\
is consistent with other multidimensional radiative transfer
codes specialized for synthesizing light curves and spectra from SNe
(see, for instance,
\citet{lucy2005,kasen2006,kromer2009,tanaka2013,roth2015}), and has
a reasonable motivation.
The equation for homologous outflow is
\begin{equation}
  \label{eq1}
  \vec{U}=\frac{\vec{r}}{t} \;\;,
\end{equation}
where $\vec{U}$, $\vec{r}$, and $t$ are ejecta velocity, spatial
coordinate, and time, respectively.
For SNe Ia, Eq.~\eqref{eq1} becomes valid on the order of a minute
after the explosion~\citep{ropke2005}; for CCSNe, Eq.~\eqref{eq1}
may not become valid for hours to days after the explosion
\citep{kifonidis2006,guzman2009}.
Equation~\eqref{eq1} greatly simplifies the hydrodynamics.
The gas energy equation,
\begin{equation}
  \label{eq2}
  \rho\frac{De}{Dt}+P\nabla\cdot\vec{U} = -g^{(0)}
  + \varepsilon_{\gamma}^{(0)} \;\;,
\end{equation}
where $\rho$, $e$, $P$, $-g^{(0)}$ and $\varepsilon_{\gamma}^{(0)}$
are gas density, internal energy, pressure, thermal radiative coupling,
and heating due to gamma-rays, respectively, is often simplified
by assume the left side is negligible
\citep{kasen2006,kromer2009,vanrossum2012}.
For IMC, \supernu\ only removes the gas pressure term
\citep{wollaeger2013}.

The IMC-DDMC implementation in \supernu\ has been extended to
multiple spatial dimensions, allowing for 1D, 2D, and 3D calculations
(van Rossum et al, in preparation).
Consequently, \supernu\ is equipped to potentially furnish insight
into supernova problems where multidimensional effects are important
and the flow of the ejecta is homologous~\citep{vanrossum2016}.
The \supernu\ software has produced light curves in 1D spherical
geometry for a pair-instability supernova model~\citep{kozyreva2016}
and in 2D cylindrical geometry for a spiral instability white dwarf
merger~\citep{kashyap2015,vanrossum2016}.

\section{Numerical Results}
\label{sec:numres}

In order to demonstrate the importance of multidimensional
structure in obtaining accurate light curves, we perform
radiative transfer in both 1D spherical and 3D Cartesian geometries.
In Section~\ref{sec:ejecta}, we describe the structure and
composition of the ejecta.
In Section~\ref{sec:1dconv}, resolution tests are presented
for the 1D spherical ejecta, which inform the 3D numerical set up.
In Section~\ref{sec:3dlcspec}, we present light curves and spectra
from a 3D Cartesian simulation, discuss their dependence on viewing
angle, and compare to the dependence of the grey gamma-ray light
curves brightness on viewing angle.
In Sections~\ref{sec:gamma}-\ref{sec:caveats}, we discuss: gamma-ray
light curves and spectra, a Ca II emission feature that shifts
with viewing angle, UBVRI broadband light curves, similarities of
the model to SN 2002ap, and the underlying assumptions and
caveats for the radiative transfer calculations.
Except for the UBVRI broad band light curves, all luminosity
is in erg/s.
For plots of 3D Cartesian luminosity in different solid angle
(flux) bins, $L$ is multiplied by $4\pi/\Omega$,
where $\Omega$ is the amount of solid angle for a viewing-angle bin.

All \supernu\ simulations were performed on the LANL Institutional
Computing (IC) supercomputer platforms Wolf and Mustang, using
$\sim$15000 CPU-hours for the 3D Cartesian simulations.

\subsection{Ejecta Properties}
\label{sec:ejecta}

Here we describe some features of the initial ejecta, which
is produced by \snsph\ and serves as the input for \supernu.
For \snsph, the initial conditions were obtained from the stellar
evolution code \tycho~\citep{young2005}.
The initial conditions for the stellar evolution are taken from
\cite{young2006}.
The progenitor is a non-rotating star with a ZAMS mass of 16
M$_{\odot}$ and solar metallicity.
The 16 M$_{\odot}$ star is evolved to 5 M$_{\odot}$ from mass loss
to a binary companion~\citep{young2006}.
At 5 M$_{\odot}$, the collapse and explosion is modeled with a 1D
code that includes a nuclear network and flux-limited neutrino
diffusion~\citep{herant1994,fryer1999,young2006}.
The subsequent propagation of the shock through the star is
modeled in 3D with \snsph.
As in the work of~\cite{hungerford2003} and~\cite{young2006}, asymmetry
is introduced artificially by increasing the velocity
along one axis when mapping from 1D to 3D (the $z$-axis in this work).
The unimodal asymmetry is generated from the velocity
mapping formulae, Eqs.~1 and 2, of~\citep{hungerford2005}.
With the seeded asymmetry, \snsph\ then forms the unimode, or
unimodal lobe~\citep{hungerford2005}.
The resulting ejecta of our CCSN model has 3.36 M$_{\odot}$
with 0.024 M$_{\odot}$ of $^{56}$Ni.

Table~\ref{tb1} has the mass fractions post explosion,
integrated over space, for the final composition (before radioactive
decay of $^{56}$Ni).
The last two rows in Table~\ref{tb1} indicate all of the Nickel
is radioactive $^{56}$Ni, which decays following
\begin{equation}
  \label{eq3}
  ^{56}\text{Ni}\rightarrow^{56}\text{Co}\rightarrow^{56}\text{Fe} \;\;.
\end{equation}
In Eq.~\eqref{eq3}, both decays produce the gamma-rays that
power the optical light curve~\citep{colgate1969,nadyozhin1994}.
\begin{table}
\caption{Integrated mass fractions of the 3.36 M$_{\odot}$ ejecta
at 2660 seconds post-explosion.}
\label{tb1}
\centering
%\resizebox{70 mm}{!}{
\begin{tabular}{|l|c|}
\hline
Element & Mass Fraction \\
\hline
H & 0.013953 \\
\hline
He & 0.345539 \\
\hline
C & 0.099356 \\
\hline
O & 0.492936 \\
\hline
Ne & 0.00944116 \\
\hline
Mg & 0.00313361 \\
\hline
Si & 0.00847409 \\
\hline
S & 0.00713697 \\
\hline
Ar & 0.00194831 \\
\hline
Ca & 0.0013786 \\
\hline
Sc & 3.11855e-08 \\
\hline
Ti & 3.88255e-05 \\
\hline
Cr & 6.65922e-05 \\
\hline
Fe & 0.00941617 \\
\hline
Ni & 0.00718054 \\
\hline
Ni-56 & 0.00718054 \\
\hline
\end{tabular}
%}
\end{table}

\begin{figure}
  \includegraphics[height=70mm]{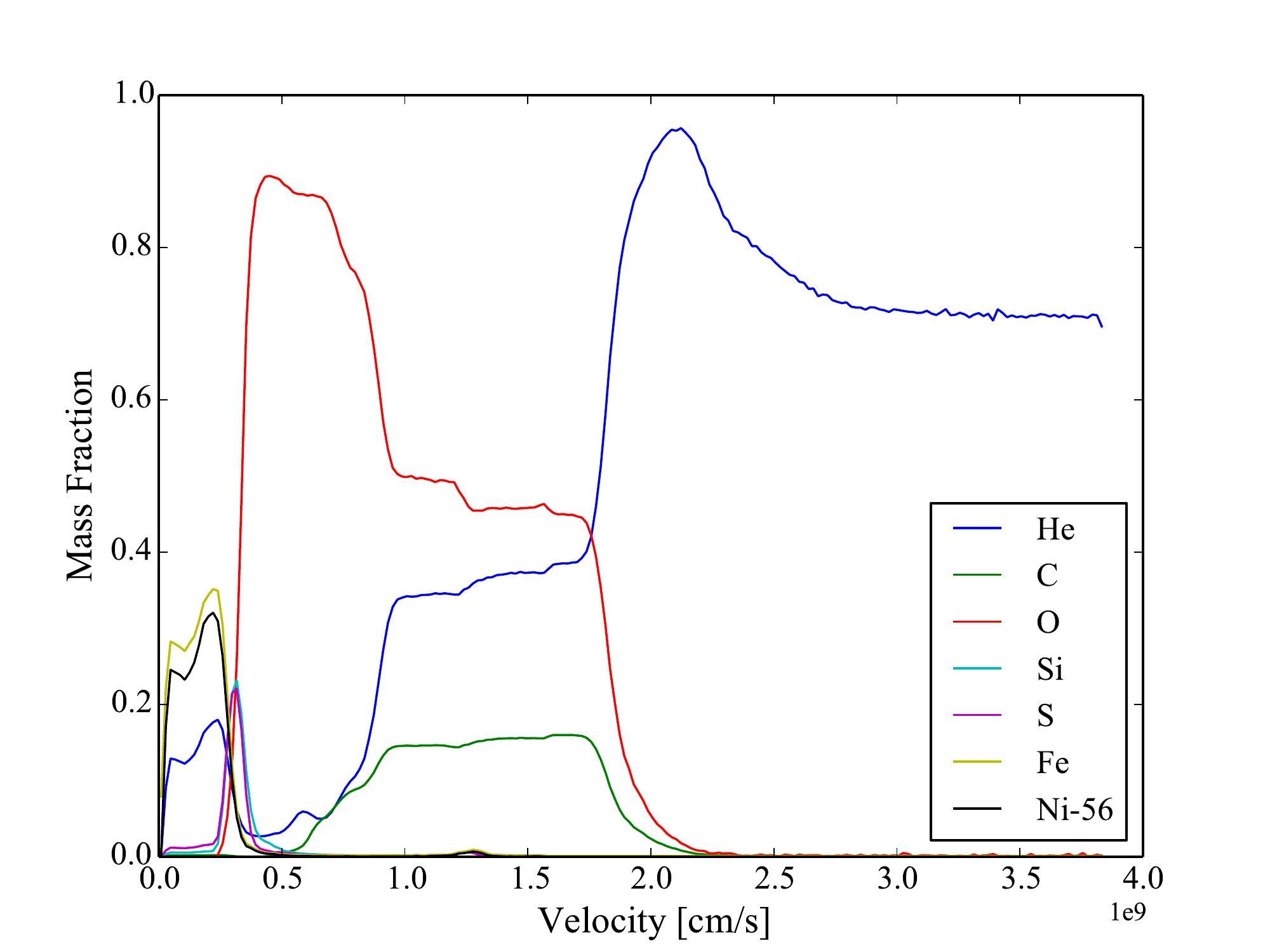}
  \caption{
    Angle averaged mass fractions versus radial velocity
    for some elements of the ejecta.
  }
  \label{fg4}
\end{figure}

Figure~\ref{fg4} has mass fractions versus radial velocity,
including $^{56}$Ni; it is evident that elements with higher
atomic mass tend to be closer to the center of the ejecta.

\begin{figure*}
\subfloat[]{\includegraphics[height=65mm]{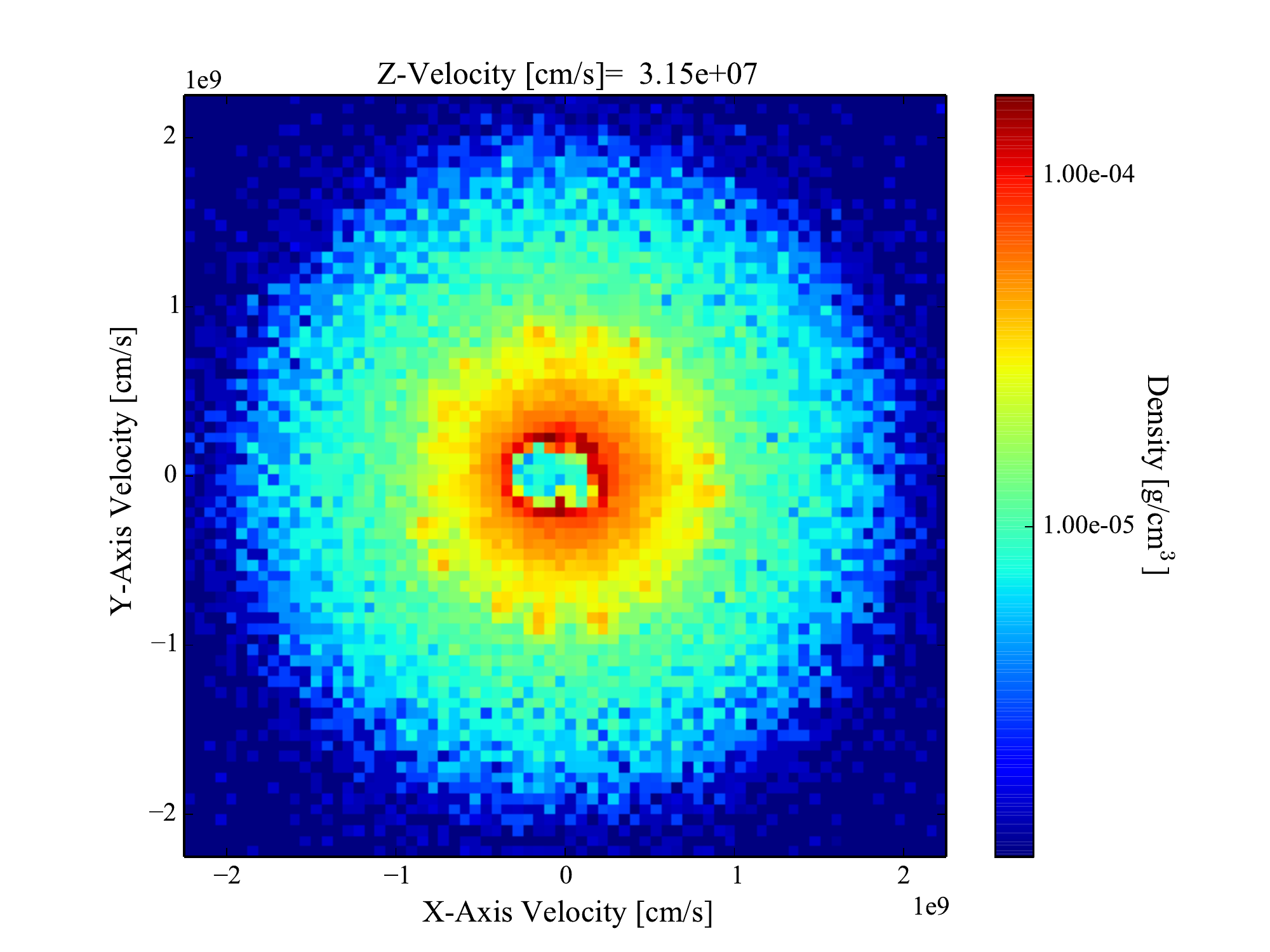}\label{fg5a}}
\subfloat[]{\includegraphics[height=65mm]{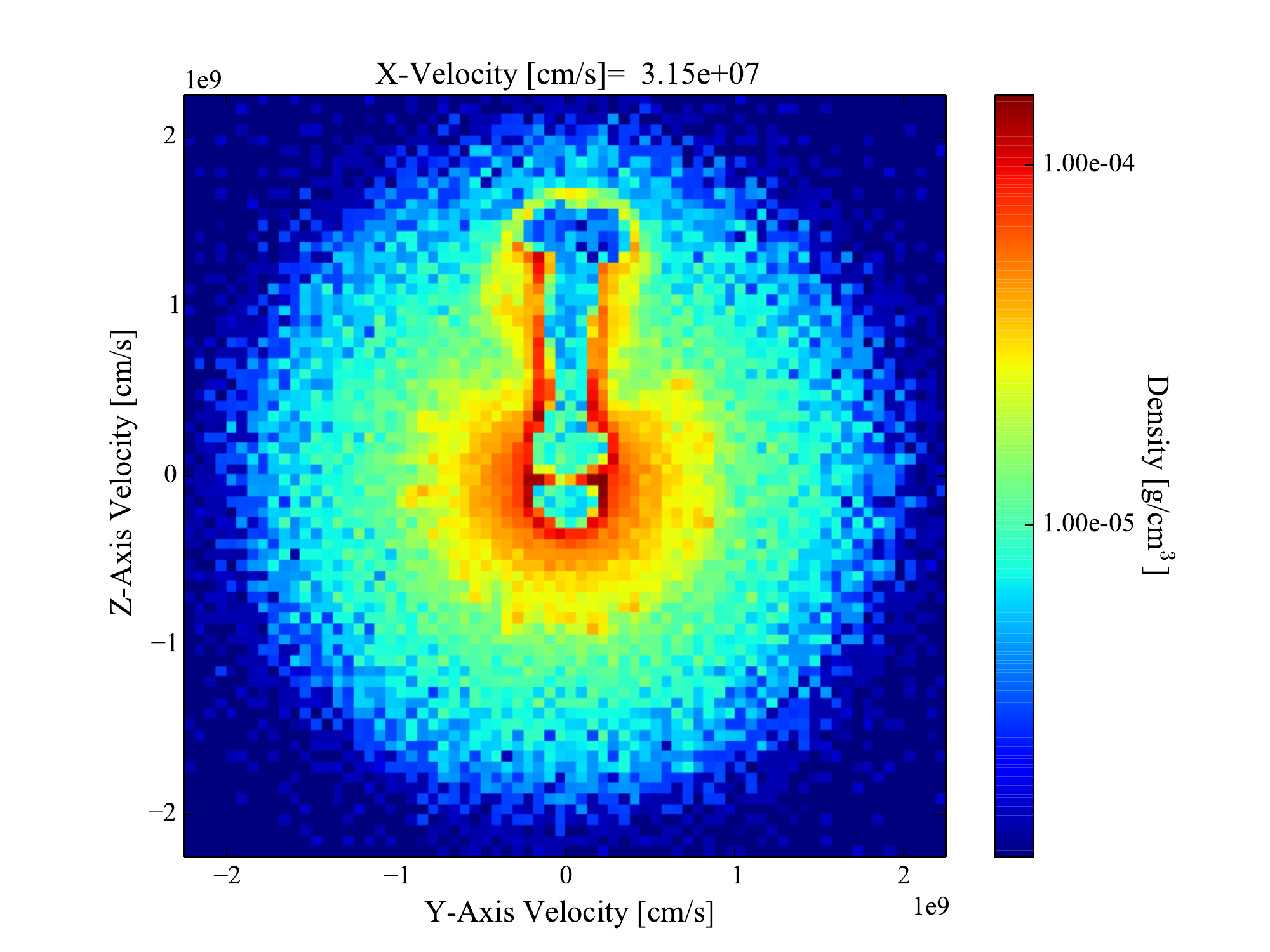}\label{fg5b}}\\
\subfloat[]{\includegraphics[height=65mm]{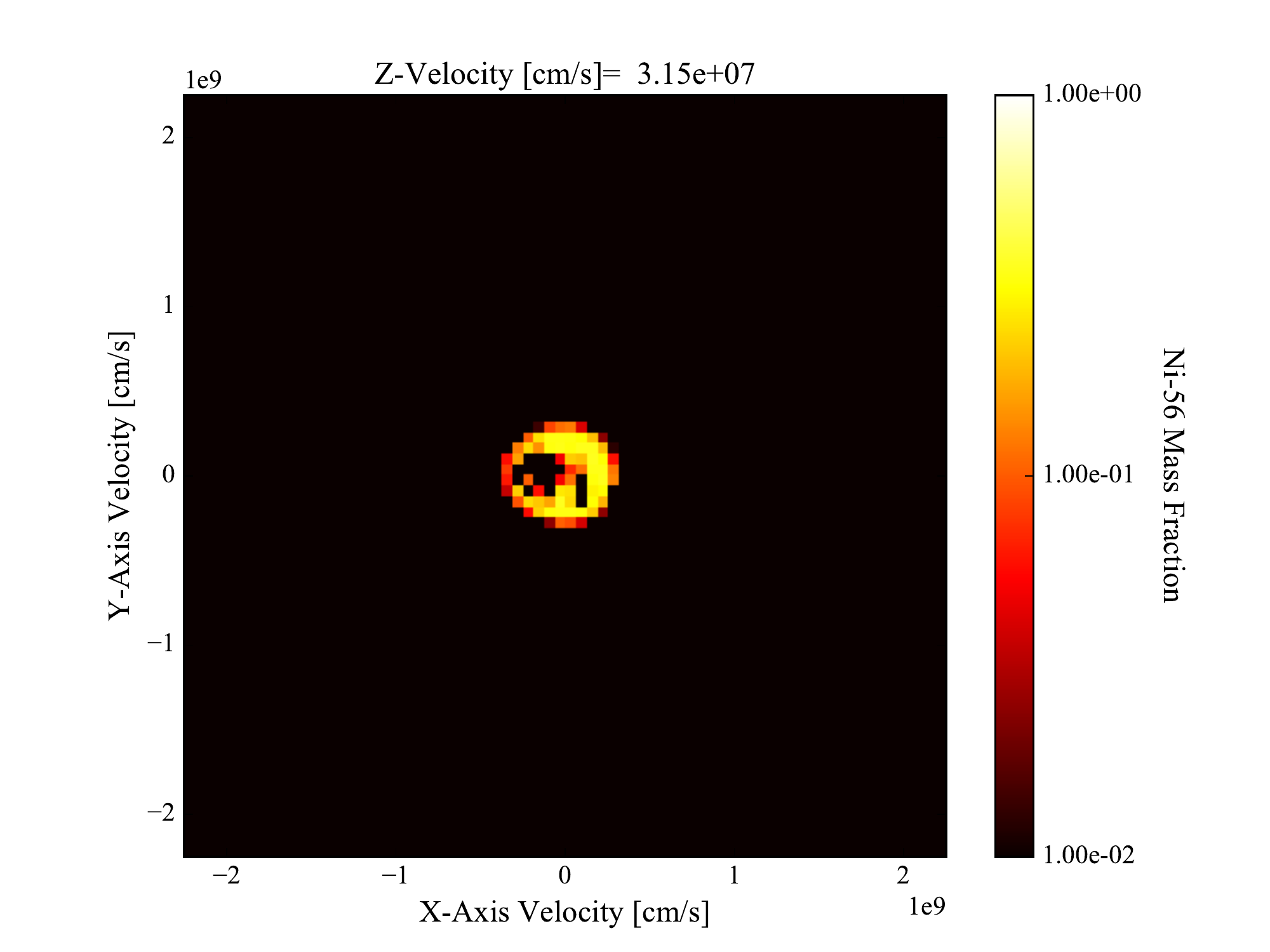}\label{fg5c}}
\subfloat[]{\includegraphics[height=65mm]{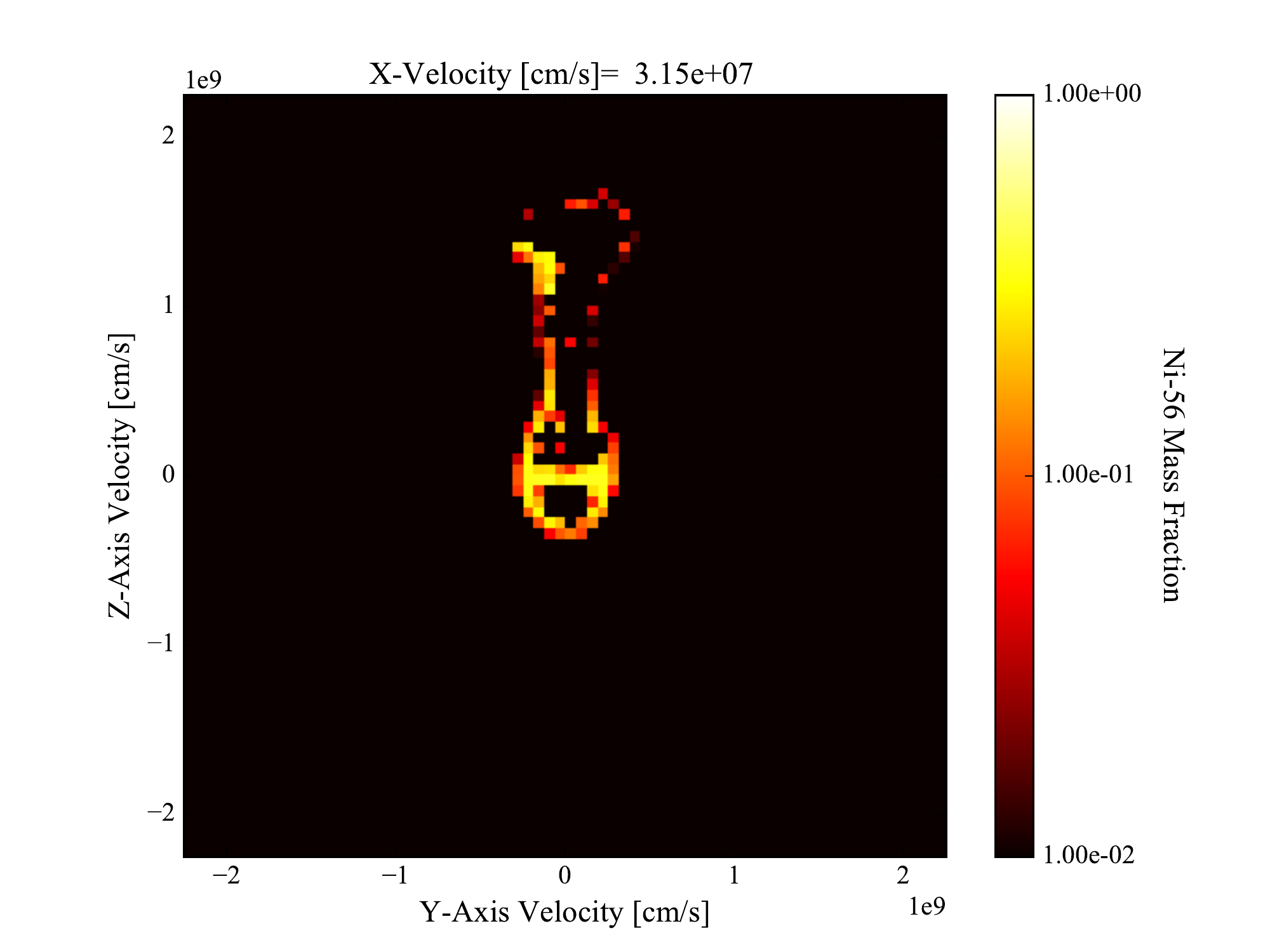}\label{fg5d}}\\
\subfloat[]{\includegraphics[height=65mm]{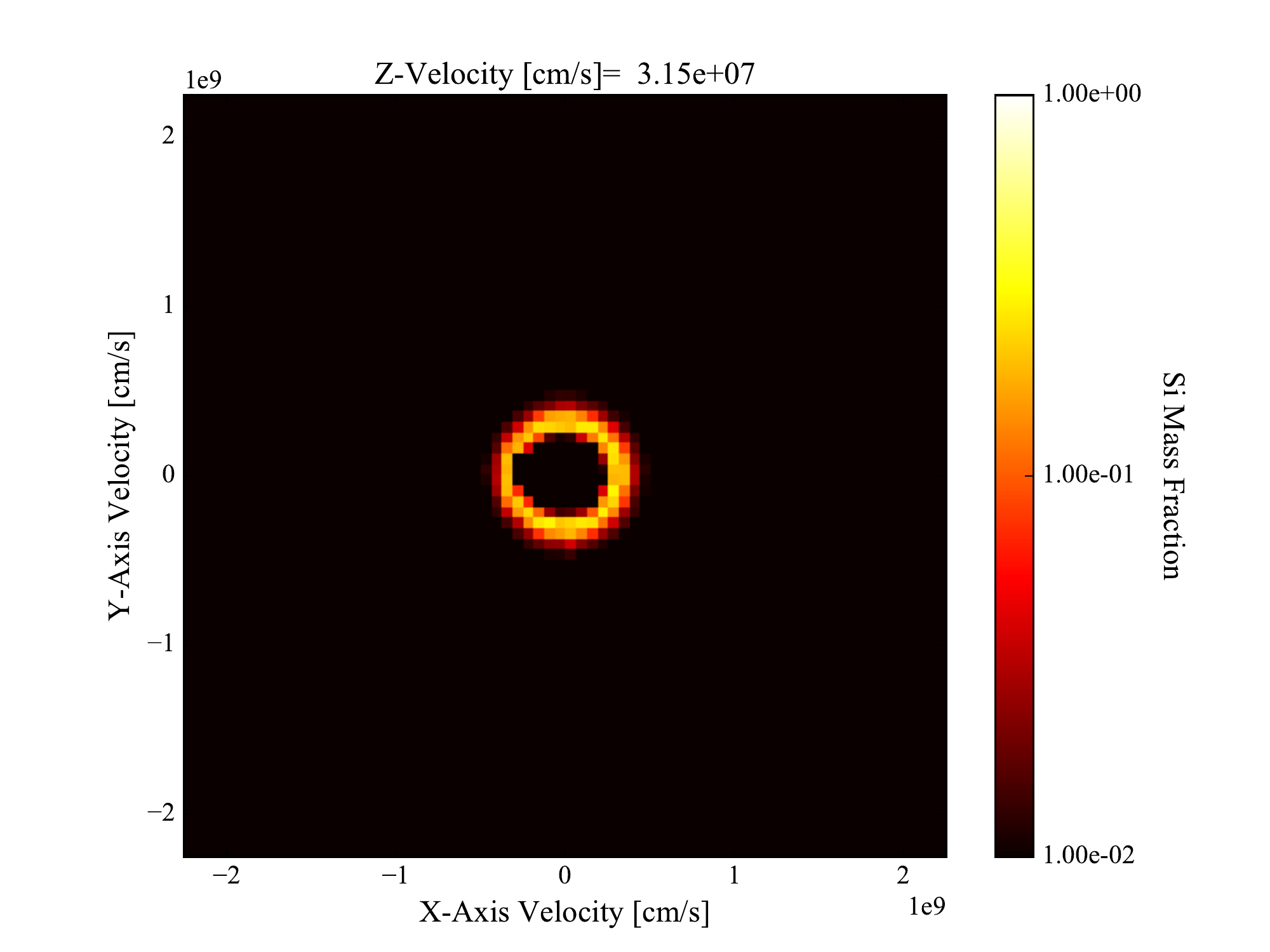}\label{fg5e}}
\subfloat[]{\includegraphics[height=65mm]{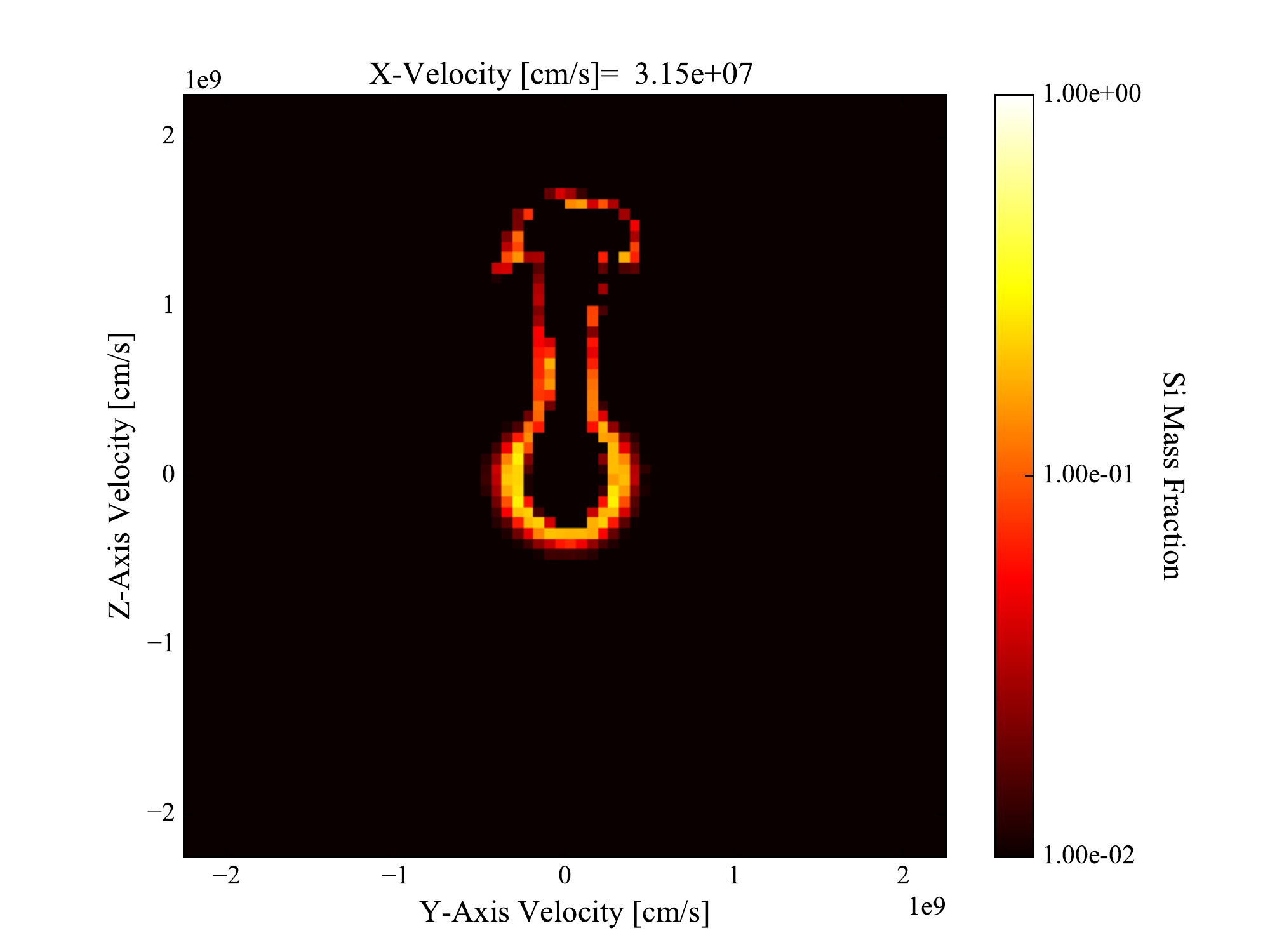}\label{fg5f}}
\caption{
  Close-up of color map of density for the unimodal CCSN structure
  at 2663 s mapped to a $120^{3}$ Cartesian grid.
  The explosion forms a unimode up the $z$-axis.
  The outflow is homologous at this time.
  In Fig.~\ref{fg5a}, density in $xy$-plane in a spatial cell nearest
  to the origin.
  In Fig.~\ref{fg5b}, density in $yz$-plane in a spatial cell nearest
  to the origin.
  In Fig.~\ref{fg5c} and~\ref{fg5d}, map of $^{56}$Ni mass fractions
  in the $xy$ and $yz$ planes, respectively.
  In Fig.~\ref{fg5e} and~\ref{fg5f}, map of Si mass fractions
  in the $xy$ and $yz$ planes, respectively.
  The high-density regions toward the center of the expansion and
  on the surface of the unimode contain radioactive $^{56}$Ni.
  The Si and S are almost entirely in the high-density surface of
  the unimode.
}
\label{fg5}
\end{figure*}

\begin{figure*}
\subfloat[]{\includegraphics[height=65mm]{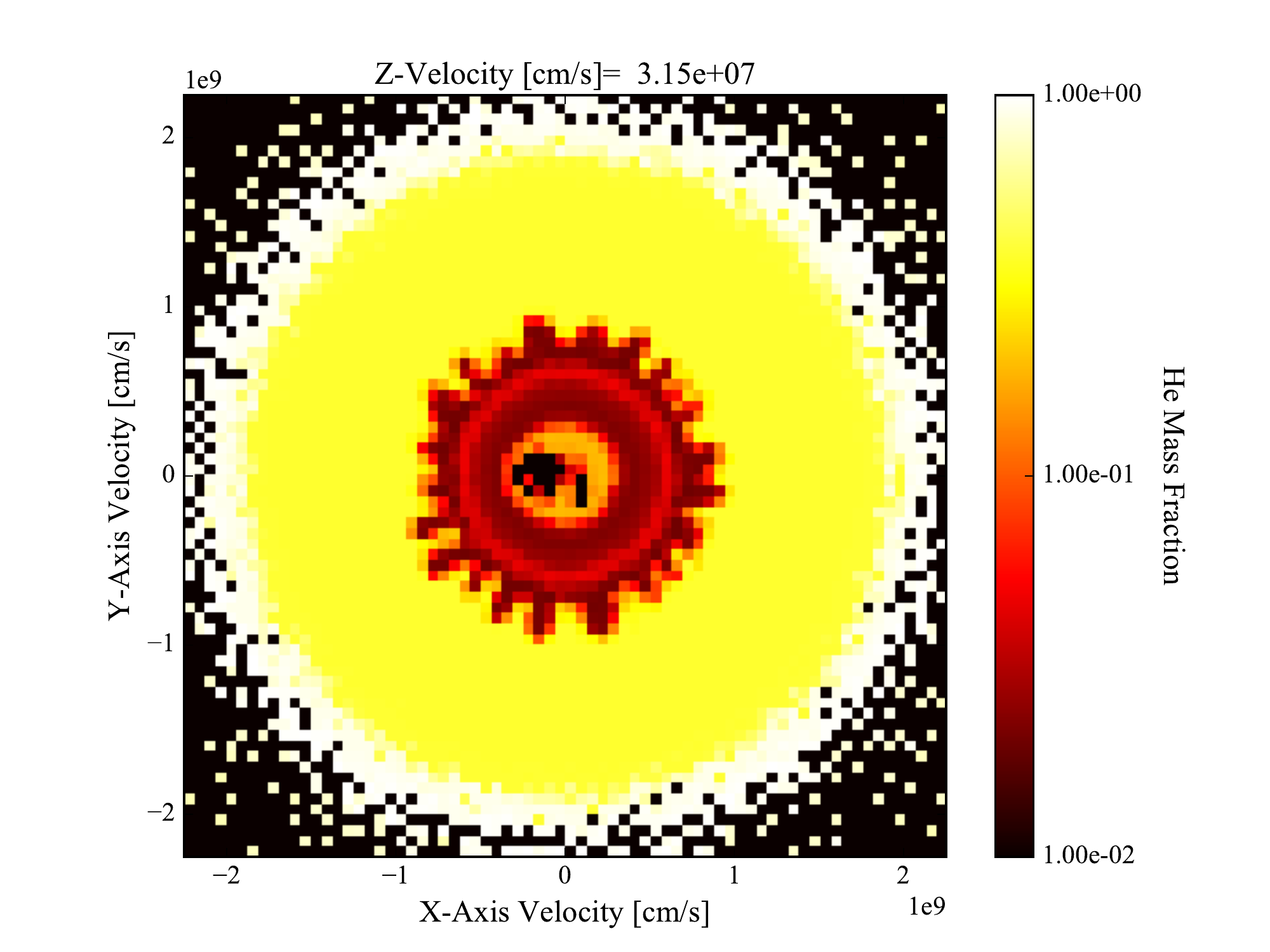}\label{fg6a}}
\subfloat[]{\includegraphics[height=65mm]{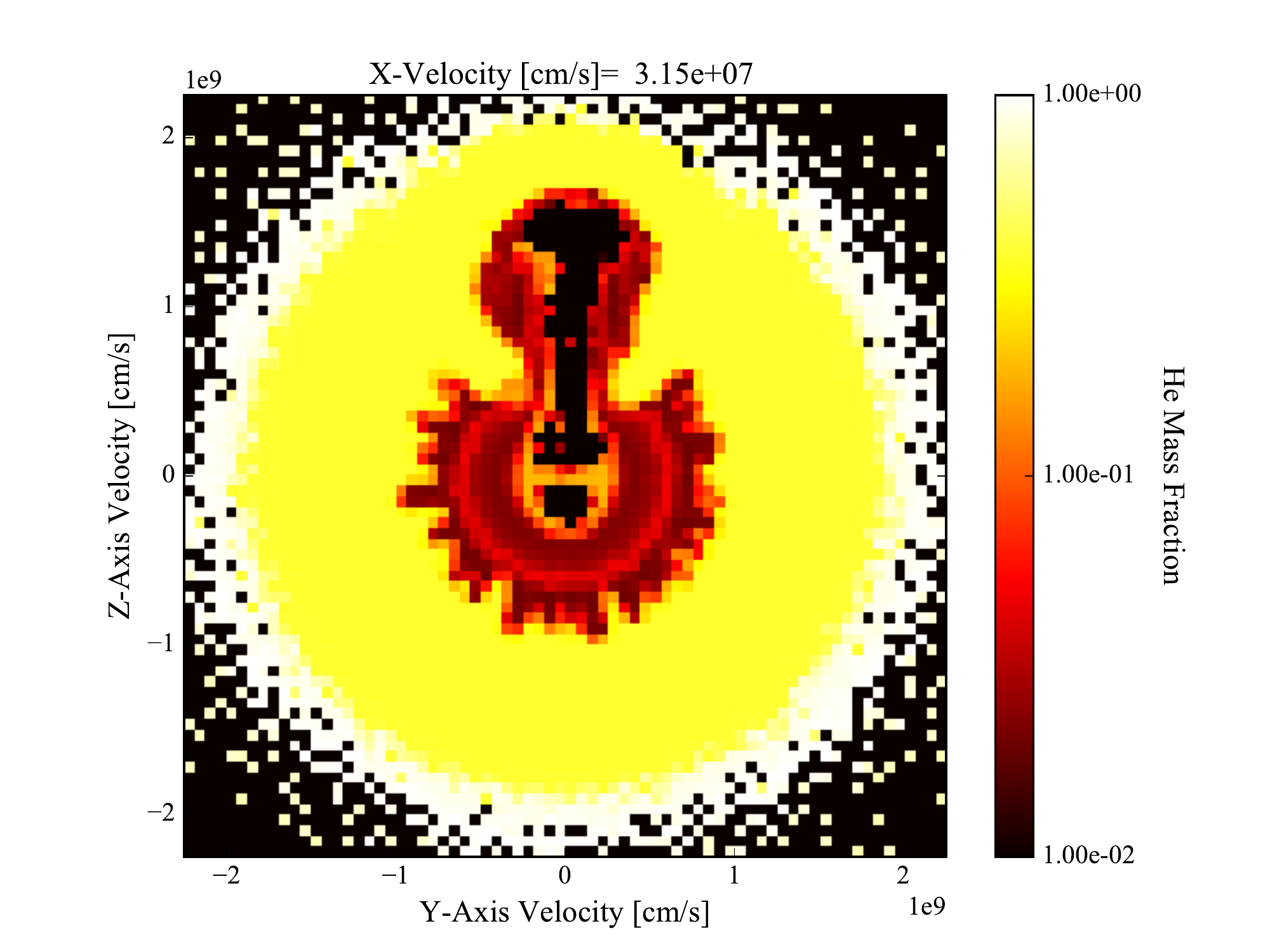}\label{fg6b}}\\
\subfloat[]{\includegraphics[height=65mm]{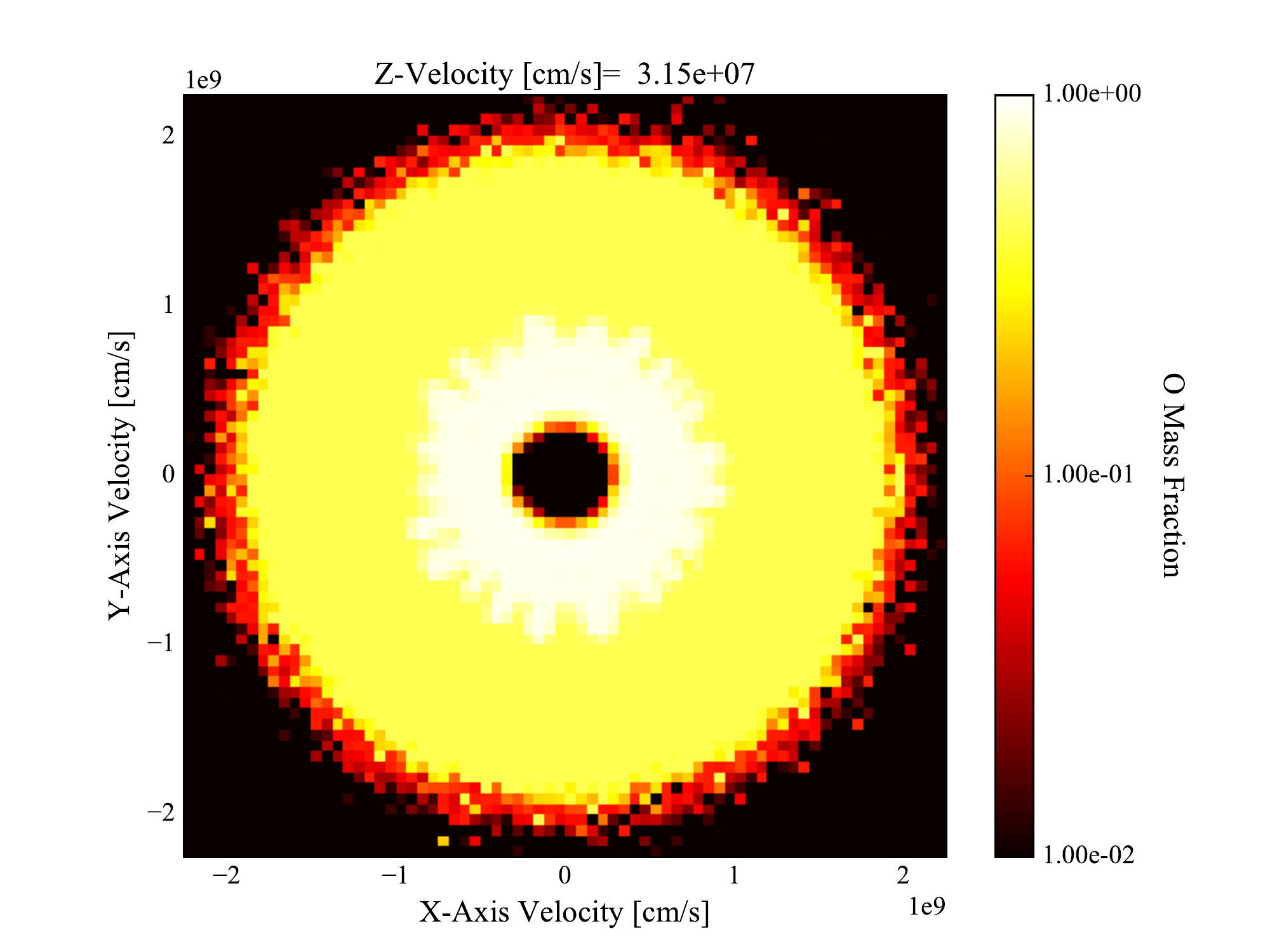}\label{fg6c}}
\subfloat[]{\includegraphics[height=65mm]{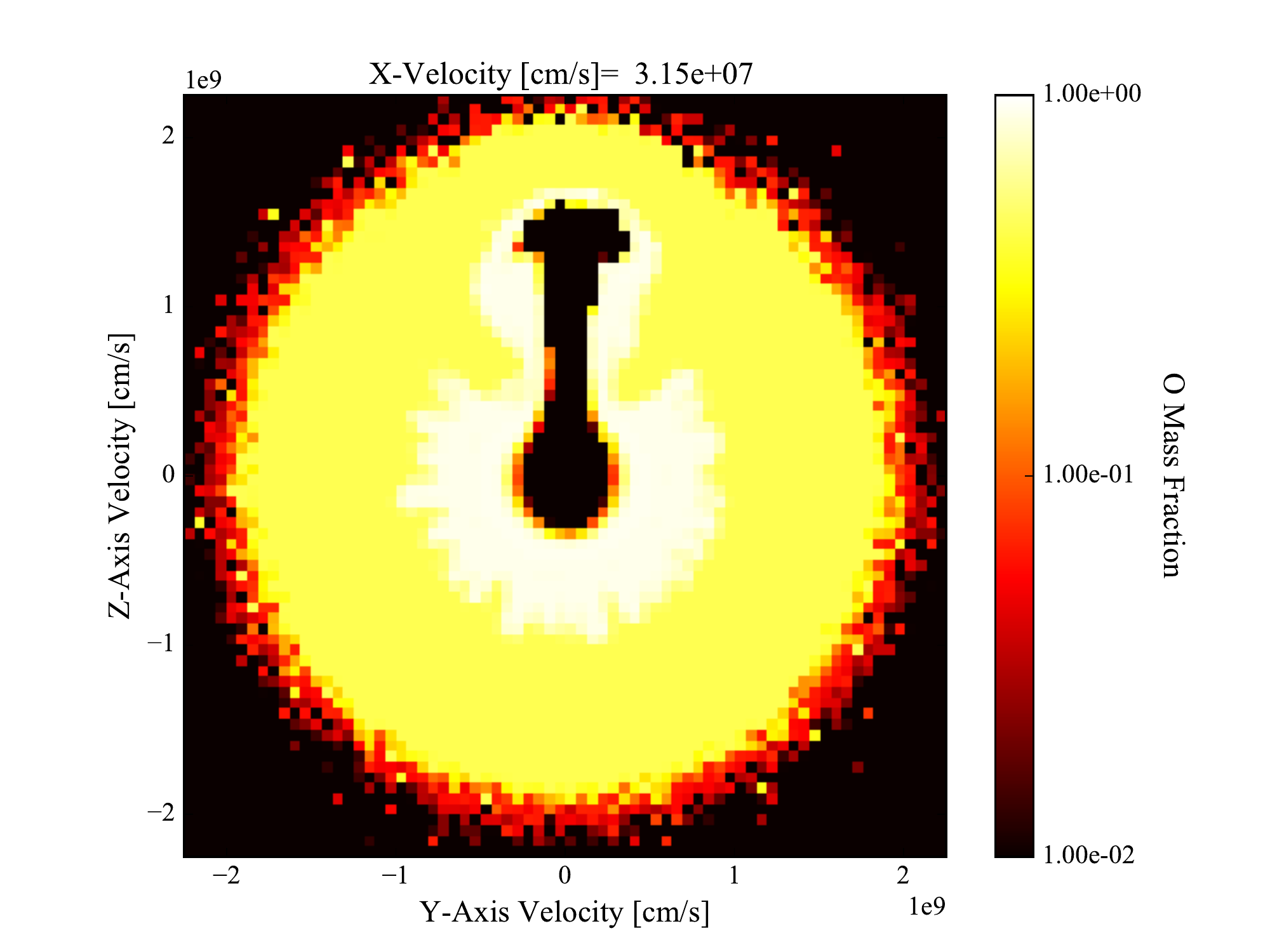}\label{fg6d}}\\
\subfloat[]{\includegraphics[height=65mm]{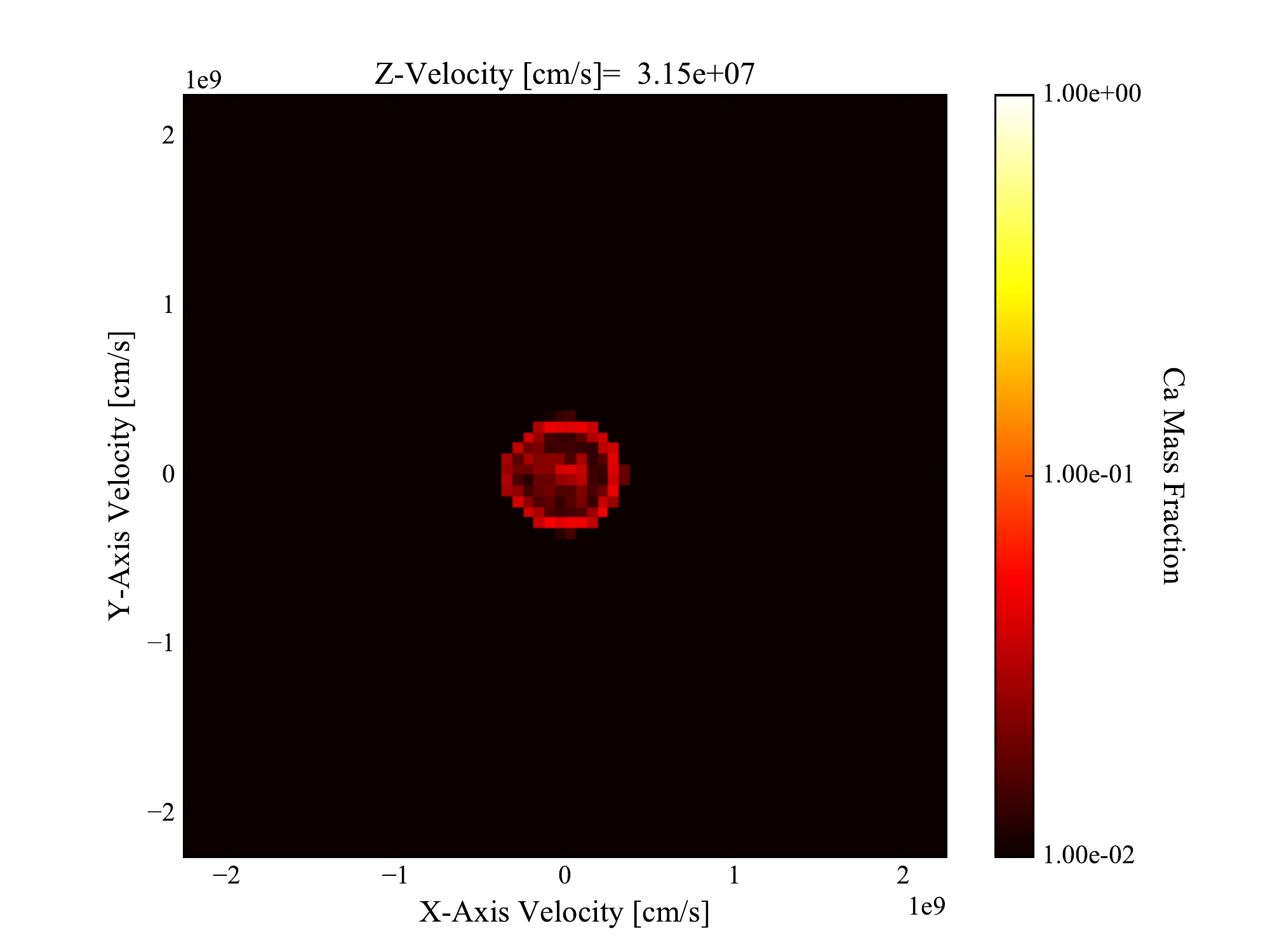}\label{fg6e}}
\subfloat[]{\includegraphics[height=65mm]{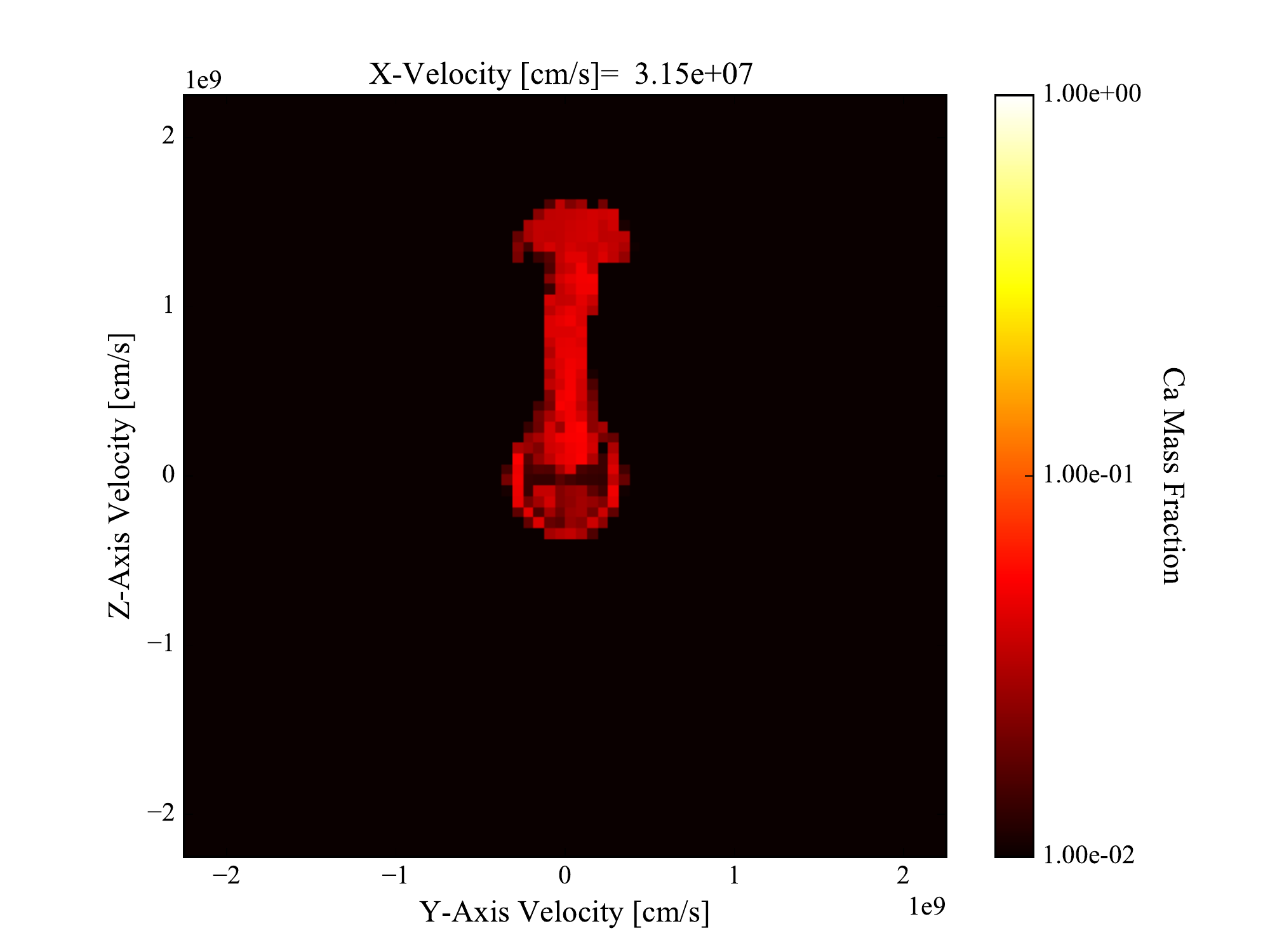}\label{fg6f}}
\caption{
  Close-up of color map of mass fractions for the unimodal CCSN structure
  at 2663 s mapped to a $120^{3}$ Cartesian grid.
  The explosion forms a unimode up the $z$-axis.
  The outflow is homologous at this time.
  In Figs.~\ref{fg6a} and~\ref{fg6b}, map of He mass fractions
  in the $xy$ and $yz$ planes, respectively.
  In Figs.~\ref{fg6c} and~\ref{fg6d}, map of O mass fractions
  in the $xy$ and $yz$ planes, respectively.
  In Figs.~\ref{fg6e} and~\ref{fg6f}, map of Ca mass fractions
  in the $xy$ and $yz$ planes, respectively.
}
\label{fg6}
\end{figure*}

The explosion was simulated to $\sim$2660 seconds by \snsph,
approximately when the ejecta becomes homologous (or when Eq.
\eqref{eq1} becomes applicable).
Figure~\ref{fg5} has maps of density, $^{56}$Ni abundance,
and Si abundance in the $xy$ and $yz$ velocity planes at this time.
A unimode has formed up the $z$-axis with a high density surface.
$^{56}$Ni is located near the origin and in the surface of the unimode.
Intermediate-mass elements Si and S are nearly exclusively located
in the surface of the unimode.
Figure~\ref{fg6} has additional maps of He, O, and Ca
abundances.
Ca is located mostly within the unimode, permitting the possibility
that Ca emission features could serve as a probe of the asymmetry.

\subsection{1D Spherical Simulations}
\label{sec:1dconv}

With the point approximation to SPH particles, we first map the SPH data
from \snsph\ to a 1D spherical spatial grid, and perform radiative
transfer on the resulting outflow.
We examine the variation in the 1D solutions with respect to
changes in spatial, temporal, and wavelength resolution.
Otherwise, the properties of the domain are the same for each simulation.
The ejecta is expanded for 80 days, the radial velocity spans 0 to
3.8$\times10^{9}$ cm/s, and the wavelength ($\lambda$) range for the
luminosity and spectra is 10$^{3}$ to 3.2$\times10^{4}$ $\AA$.
For all simulations, the spatial cells, $\Delta U$, and time step sizes,
$\Delta t$, are uniform, and the wavelength group intervals, $\Delta\lambda$,
are logarithmic.
We denote the number of spatial cells, time steps, and wavelength groups as
$N_{r}$, $N_{t}$, and $N_{g}$, respectively.
Table~\ref{tb2} has percent changes for luminosity averages around peak
luminosity for four resolutions for each of the three independent variables.
The $sk\in\{s1,s2,s3,s4\}$ values in Table~\ref{tb2} correspond to
$N_{r}\in\{50,100,200,400\}$ for radius, $N_{t}\in\{496,992,1984,3968\}$
for time, and $N_{g}\in\{125,250,500,1000\}$, where the test is
indicated in the left column by the resolution parameter.
We also test $N_{g}=2000$ and compare to $N_{g}=1000$ in the
$s4\rightarrow s5$ column of Table~\ref{tb2}.
For each resolution test, the other two resolution parameters are taken
from the base case: $N_{r}=100$, $N_{t}=1984$, $N_{g}=500$.
The averaged luminosity is the time-average of the computed values from
$t=16.3$ to $t=24.2$ days post-explosion.
For the spatial resolution test, the averaged luminosity appears to converge
in 100-200 cells.
At 400 spatial cells, the time step size is starting to become large
with respect to the gas crossing time for a cell near peak luminosity,
and the error slightly increases (see Section~\ref{sec:caveats}).
For the temporal resolution test, the percentage change in the solution
systematically decreases for the resolutions tested.
Unlike the other resolution tests, which are complicated by method
changes across groups and cells, lowering the time step sizes merely
improves the accuracy of the operator split
\citep{abdikamalov2012,wollaeger2013}.
For the group resolution test, the heights of the peaks in the spectra
vary with group resolution but do not shift in wavelength.
For \supernu, group resolutions on the order of 500 to 1000 yield
reasonable results for the SN regime of domain properties~\citep{wollaeger2014}.
Similarly, the multigroup transport code {\tt EDDINGTON} has applied
500 to 5000 groups for various SN problems
\citep{eastman1993,eastman1994,pinto2000}.
For this 1D test, for instance, the percent increase in the average around
peak luminosity from 1000 to 2000 groups is only 0.07$\%$.

\begin{table}
\caption{Percent change in peak luminosities versus resolution increase.}
\label{tb2}
\centering
%\resizebox{70 mm}{!}{
\begin{tabular}{|l|c|c|c|c|}
\hline
& $s1\rightarrow s2$ & $s2\rightarrow s3$ & $s3\rightarrow s4$
& $s4\rightarrow s5$ \\
\hline
N$_{r}$ & 3.62 & -0.04 & -0.41 & - \\
\hline
N$_{t}$ & 1.09 & 0.71 & 0.56 & - \\
\hline
N$_{g}$ & 3.21 & 3.34 & 1.35 & 0.07  \\
\hline
\end{tabular}
%}
\end{table}

\begin{figure}
\subfloat[]{\includegraphics[height=70mm]{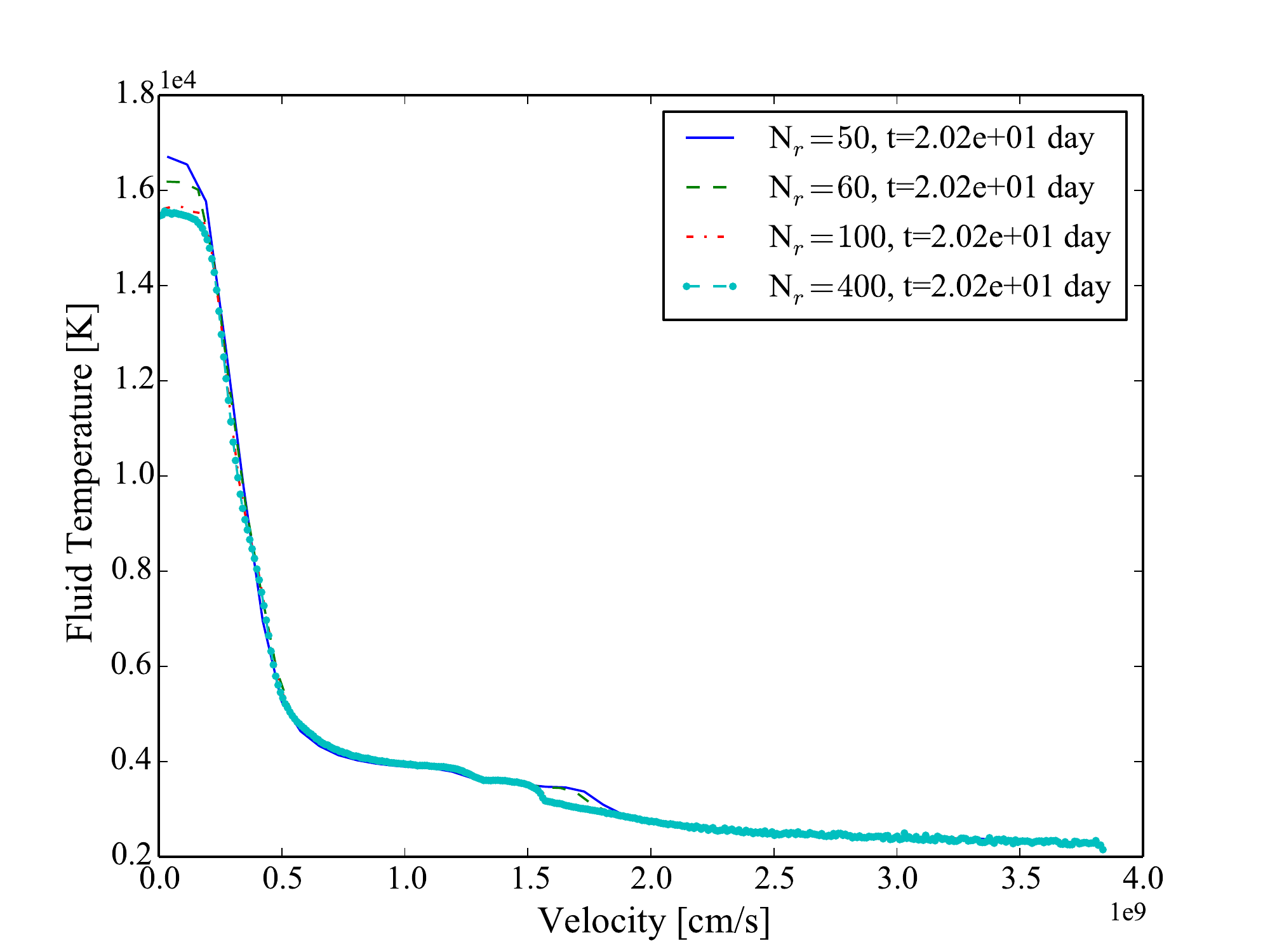}\label{fg7a}}\\
\subfloat[]{\includegraphics[height=70mm]{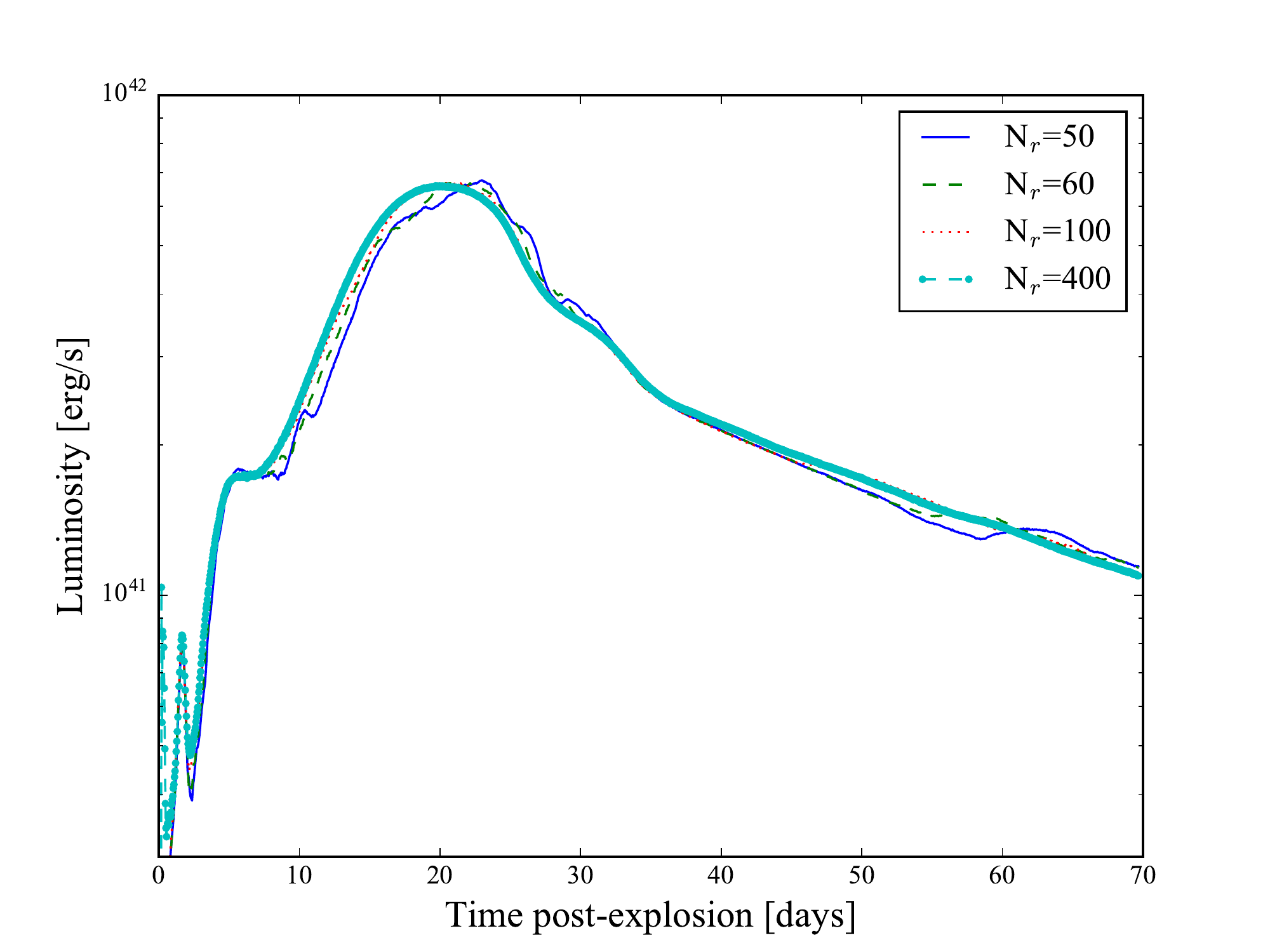}\label{fg7b}}
\caption{
  Four different spatial resolutions are depicted: $N_{r}\in\{50,60,100,400\}$,
  where $N_{r}$ is the number of cells in 1D spherical geometry.
  In Fig.~\ref{fg7a}, gas temperature near peak luminosity for the
  four spatial resolutions.
  In Fig.~\ref{fg7b}, bolometric luminosity for the four spatial
  resolutions.
  The solutions exhibit convergence, and the differences are
  small for the resolutions tested.
}
\label{fg7}
\end{figure}

\begin{figure}
\subfloat[]{\includegraphics[height=70mm]{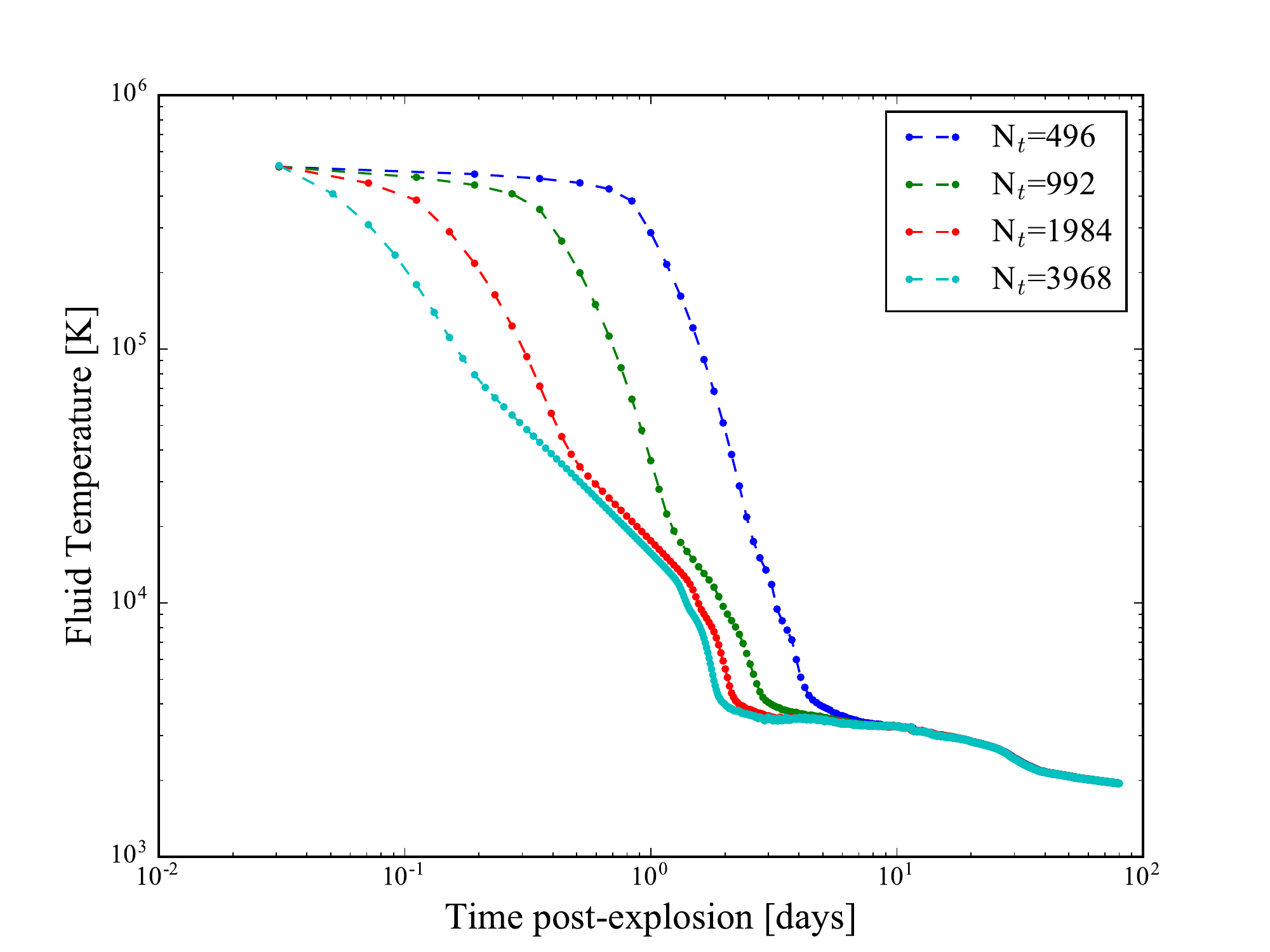}\label{fg8a}}\\
\subfloat[]{\includegraphics[height=70mm]{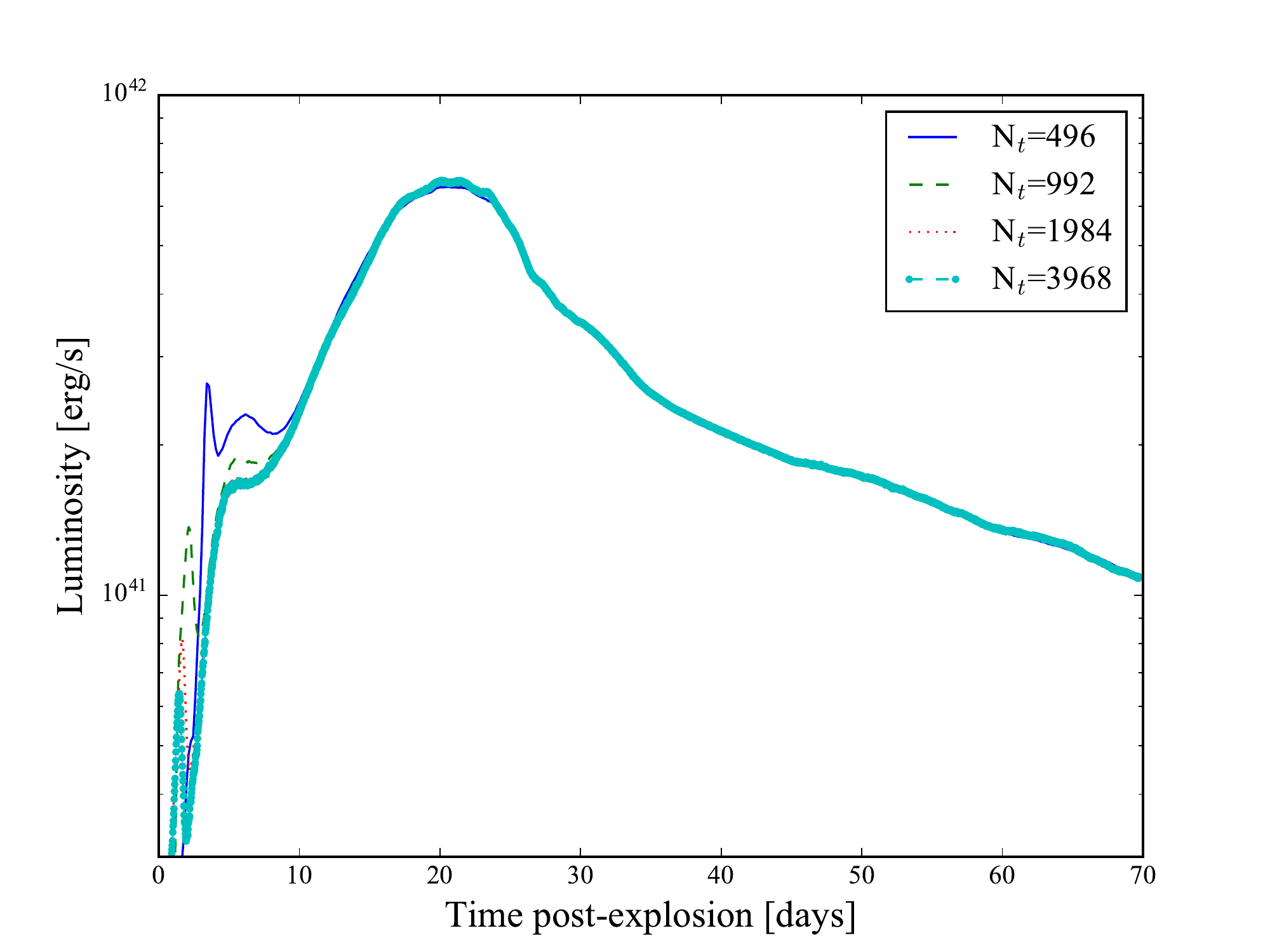}\label{fg8b}}
\caption{
  Four different temporal resolutions are depicted:
  $N_{t}\in\{496,992,1984,3968\}$,
  where $N_{t}$ is the number of uniform time steps over the 80 day
  time span (starting from 2663 s).
  In Fig.~\ref{fg8a}, gas temperature at the radial midpoint of
  the ejecta (near the Planck photosphere) vs time on a log-log
  scale.
  In Fig.~\ref{fg8b}, bolometric luminosity for the four temporal
  resolutions.
  The initial temperature relaxation time scale is not resolved for
  the two coarser time step resolutions; this affects the shock
  breakout light curve.
}
\label{fg8}
\end{figure}

\begin{figure}
\subfloat[]{\includegraphics[height=70mm]{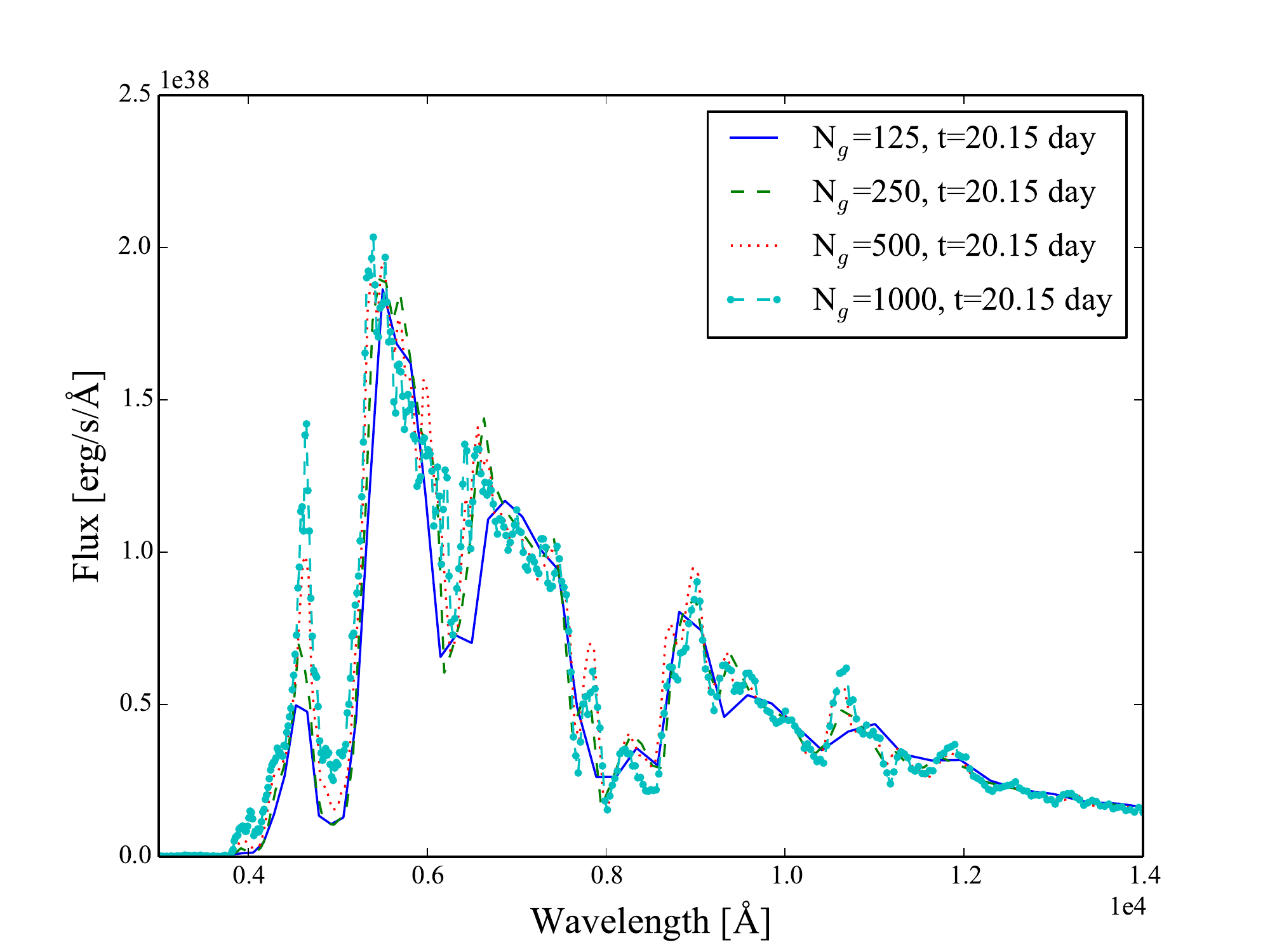}\label{fg9a}}\\
\subfloat[]{\includegraphics[height=70mm]{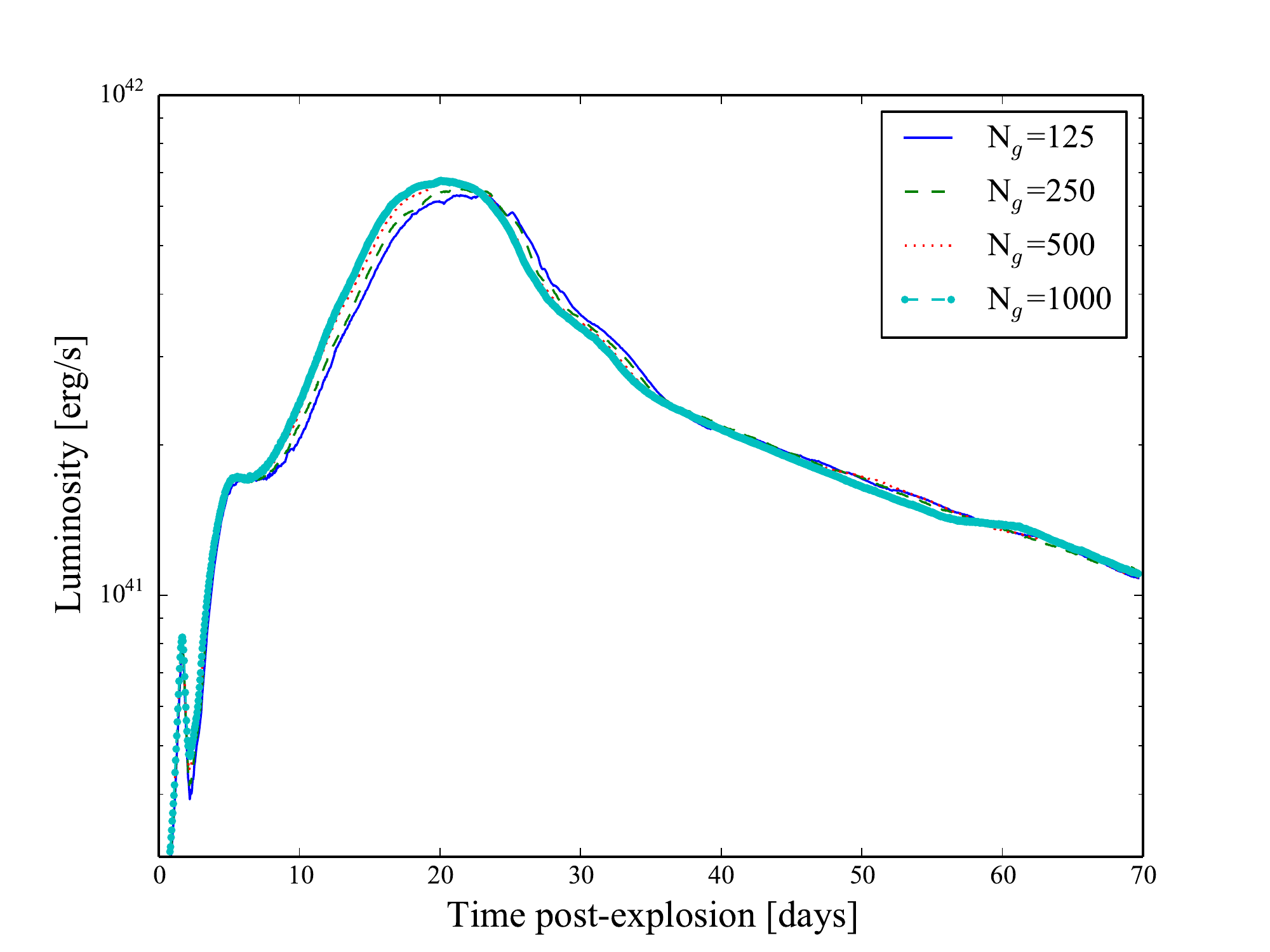}\label{fg9b}}
\caption{
  Four different spectral resolutions are depicted:
  $N_{g}\in\{125,250,500,1000\}$,
  where $N_{g}$ is the number of logarithmic groups over wavelength
  range, $\lambda\in[10^{3},3.2\times10^{4}]$ $\AA$.
  In Fig.~\ref{fg9a}, spectra of the different group resolutions at
  a time near peak luminosity.
  In Fig.~\ref{fg9b}, bolometric luminosity for the four spectral
  resolutions.
  At low group resolutions, artifacts appear in the shape of the light
  curves.
}
\label{fg9}
\end{figure}

In Figs.~\ref{fg7},~\ref{fg8}, and~\ref{fg9}, we have plotted light
curves for different parameter resolutions along with gas temperature
or spectra.
Figure~\ref{fg7} has gas temperatures near peak luminosity and light
curves for several spatial resolutions.
From Fig.~\ref{fg7a}, it is discernible that the radial temperature
is nearly converged at O(100) cells, in agreement with the light curve
results in Table~\ref{tb2}.
The gas temperature profile is also consistent with the ejecta composition
shown in Fig.~\ref{fg4}; high temperatures at $U\lesssim5\times10^{8}$ cm/s
correspond to the optically thick, radioactive core of $^{56}$Ni.

Figure~\ref{fg8a} has the time evolution of the gas temperature at
the radial midpoint of the ejecta.
For O(1 day) time scales, the different time step resolutions tested
do not quite resolve the thermal relaxation time, which is required
to obtain the early component of the light curve.
Consequently, the error from the artificial temperature evolution manifests
as the error in the O(1-10 day) early light curve transient shown
in Fig.~\ref{fg8b}.

From the group resolution test, spectra and light curves are shown
in Figs.~\ref{fg9a} and~\ref{fg9b}, respectively.
Following the error in Table~\ref{tb2}, the light curve changes considerably,
showing a systematic shift towards earlier times with increasing group resolution.
However, we also tested 2000 groups, and the percentage change from 1000 to 2000
is only 0.07$\%$.
In Fig.~\ref{fg9a}, the spectra at day 20.19 post-explosion (near peak luminosity)
show the variation in features with changing group resolution.

In general, it is difficult to ascribe convergence orders in resolution
tests over single variables in these simulations, given the interdependence
of the effect of the variables on the solution (see Section~\ref{sec:caveats}).
Moreover, the spatial resolution tests for 1D do not as directly
imply an appropriate resolution for 3D as the other tests, since radial
averaging to 1D changes the morphology.
In particular, small scale ejecta features are effectively spread out in
the 1D spherical angular spatial averaging, removing some of the resolution
problem.
Table~\ref{tb2} and the figures presented in this section are meant to
show that the radioactively powered light curves are not strongly sensitive
to changes in the resolutions around $\sim100$ radial points, $\sim1984$
time steps, and $\sim500$ groups.
Since the highest resolution 3D Cartesian test has $120^{3}$ cells, 1984 time
steps, and 500 groups, we expect that the error in the simulation will not be
too restrictive in assessing variation in peak luminosities and flux with
viewing angle.

\subsection{3D Cartesian Simulations}
\label{sec:3dlcspec}

We map the SPH data to a $120^{3}$ cell, 3D Cartesian spatial
grid and perform radiative transfer on the resulting structure.
In order to see how the asymmetry affects the observable light curve,
we partition the $4\pi$ steradian view of the structure into a
viewing-angle grid (the asymmetry is discernible from the density
profiles in Fig.~\ref{fg5}).
Each bin of the viewing-angle grid corresponds to a range of
viewing angles; light curves and spectra are the totals from Monte Carlo
tallies in these bins.
The viewing-angle grid has 96 polar bins of equal solid angle (a
viewing-angle grid with six uniform polar bins, tilted towards the
northern hemisphere, is depicted in Fig.~\ref{fg2}).
The polar viewing-angle bins are integrated in azimuthal viewing
angle, allowing assessment of the change of brightness with respect to polar
angle along the axis of the unimode.
As in Section~\ref{sec:1dconv}, the ejecta is expanded for 80 days, the
radial velocity spans 0 to 3.8$\times10^{9}$ cm/s, and the wavelength
($\lambda$) range for the luminosity and spectra is 10$^{3}$ to
3.2$\times10^{4}$ $\AA$.
The spatial cell volumes, $\Delta U^{3}$, and time step sizes,
$\Delta t$, are uniform, and the wavelength group intervals,
$\Delta\lambda$, are logarithmic.
We apply 1984 time steps and 500 wavelength groups.
The 3D result is physically different than the 1D results, since
in 3D holes and opaque blobs are not spread and smoothed out over
shells.
This has a significant impact on the radiative transfer, causing an
observable difference in the shape of the light curve between the
total (view-integrated) 3D and 1D results.

\begin{figure}
\includegraphics[height=70mm]{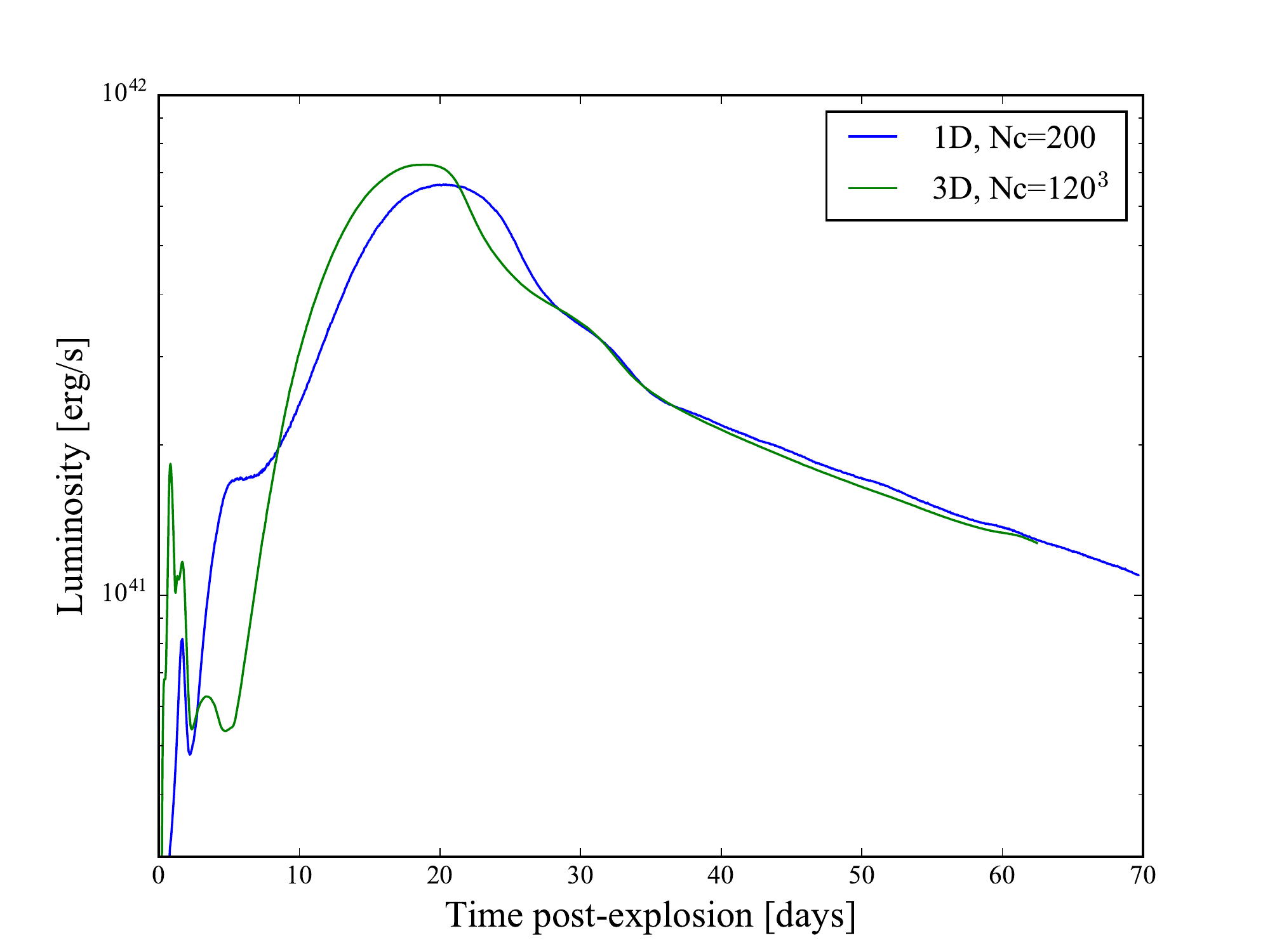}
\caption{
  Bolometric light curves from 1D spherical (blue) and
  3D Cartesian (green) simulations, where the 3D result is
  averaged over all viewing angles.
  The peak luminosity of the 3D result is $\sim$10\% higher
  than the 1D peak luminosity.
  Moreover, the light curve from the 3D simulation
  peaks $\sim2$ days earlier.
  The thermal light curves, from 0 to $\sim5$ days in 3D and
  from 0 to $\sim8$ days in 1D, are significantly different
  as well; these early differences are due in part to the
  low SPH particle count per cell in 3D near shock-breakout.
  The differences in the early and peak phases of the light
  curves are due ejecta morphology.
  In 3D, Monte Carlo particles can pass through lower density
  regions of the ejecta, avoiding the unimode.
  After $\sim30$ days, the Planck photosphere is near a radial velocity
  of $4\times10^{8}$ cm/s, and resolves the surface of the
  $^{56}$Ni core (see Fig.~\ref{fg4}).
  Consequently, the light curves appear roughly to converge.
}
\label{fg10}
\end{figure}

Figure~\ref{fg10} has bolometric luminosity of the 3D calculation
along with a 1D result with the same time and wavelength resolutions.
The viewing-angle averaged light curve from the 3D
simulation peaks earlier; the peak luminosity is $\sim$10\%
brighter than the 1D result.
This difference in the peak total luminosities are due to differences
in the 1D and 3D ejecta morphologies.
In 3D, the unimode has a high density, but does not comprise a large
portion of the ejecta volume; so photons traveling from the center
have access to a large range of directions that do not pass through
the unimode.
In 1D, all photons from the center have to pass through a region of
slightly higher density, since the unimode has effectively been spread
over all polar and azimuthal angles.
After $\sim$30 days, the light curves have a similar shape.
In the 1D simulation at $\sim$30 days, the Planck photosphere is roughly
at the surface of the $^{56}$Ni core (see Fig.~\ref{fg4}).

\begin{figure}
\subfloat[]{\includegraphics[height=70mm]{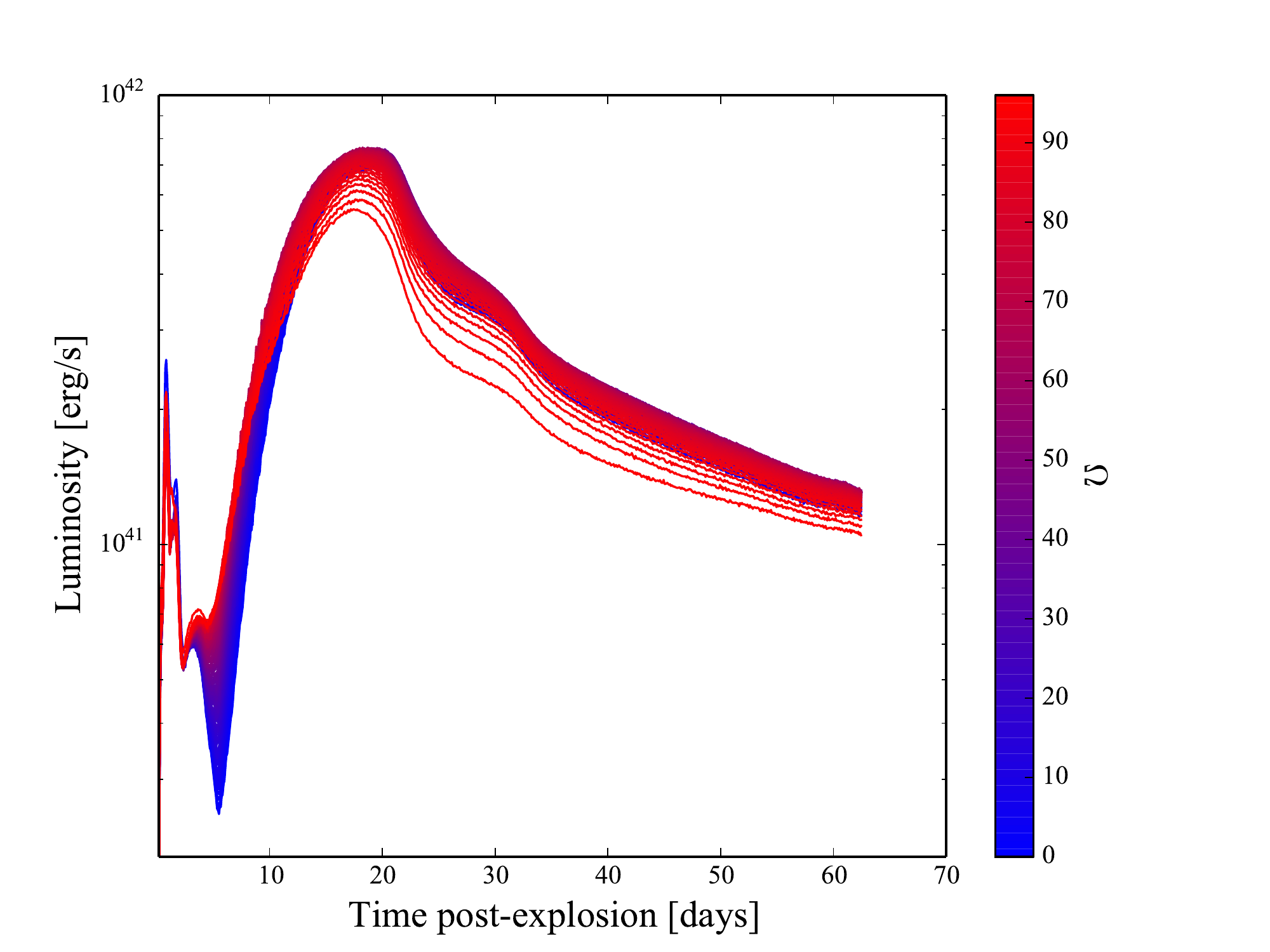}
  \label{fg11a}}\\
\subfloat[]{\includegraphics[height=70mm]{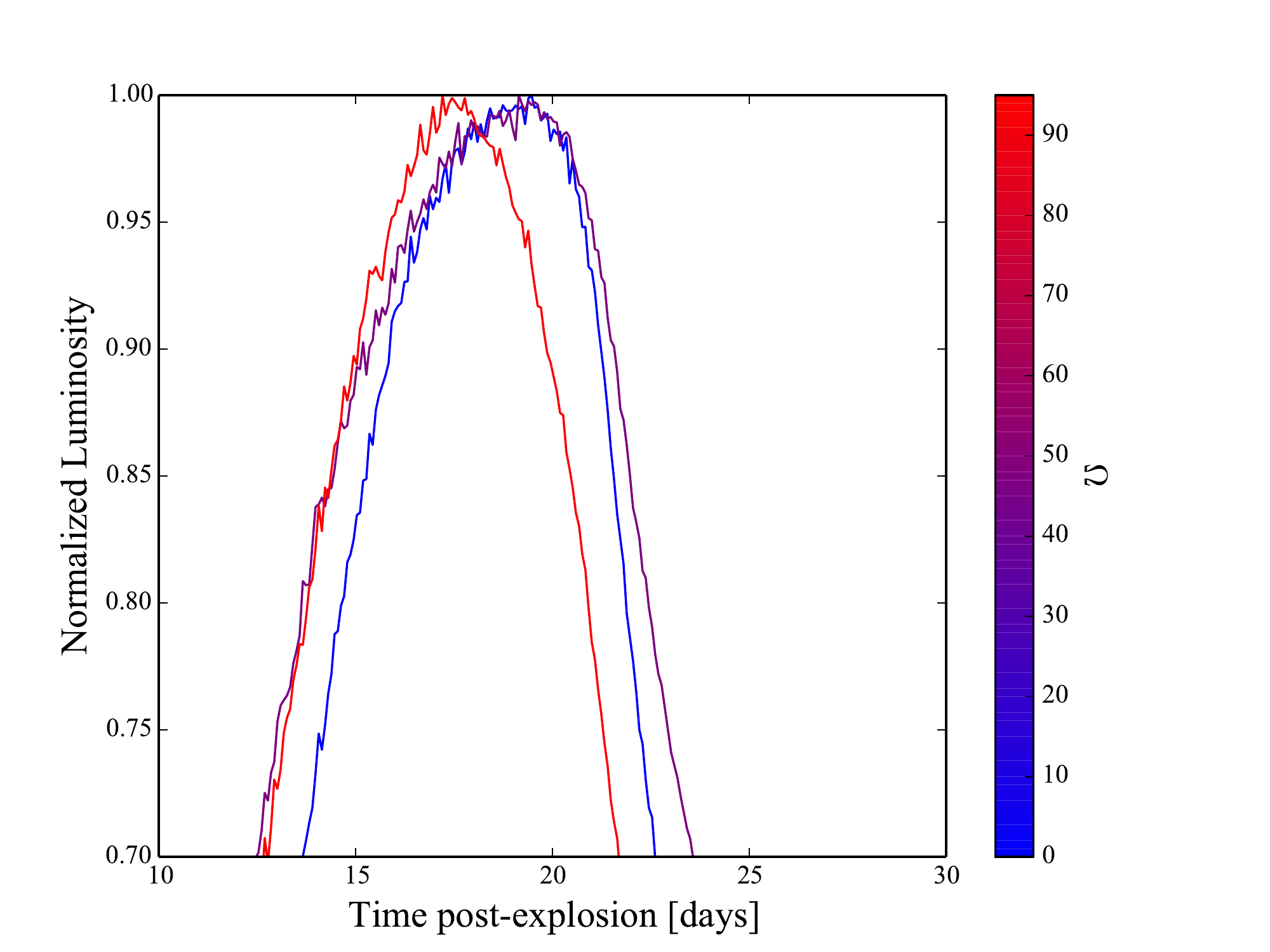}
  \label{fg11b}}
\caption{
  In Fig.~\ref{fg11a}, bolometric light
  curves for each of the 96 viewing-angle bins.
  Each viewing-angle bin corresponds to a range of polar views;
  each light curve is an average luminosity over a view.
  The highest peak luminosity (for a view orthogonal to
  the $z$-axis) is $\sim$36\% brighter than the lowest
  peak luminosity (a view down the positive $z$-axis,
  or aligned with the unimode in Fig.~\ref{fg5b}).
  In Fig.~\ref{fg11b}, light curves for bottom (blue),
  side (purple), and top (red) viewing-angle bins,
  each normalized by its peak luminosity.
  The shift in the peak luminosity time between the view
  aligned with the unimode and the others is only $\sim2$ days.
  The small scale fluctuations in the light curves are
  Monte Carlo noise.
}
\label{fg11}
\end{figure}

In Fig.~\ref{fg11a}, we plot all 96 light curves to see
the full dispersion in peak luminosities with respect to
viewing angle.
The ratio of the peak luminosity of the brightest viewing
angle (aligned with the $xy$-plane) to that of the dimmest
viewing angle (aligned with the north pole) is about 1.36.
Figure~\ref{fg11b} has light curves for viewing-angle bins
antiparallel, parallel, and orthogonal to the direction of the unimode.
The location of the peak luminosity in the angular view of
the unimode is only $\sim2$ days shifted behind the others.

\begin{figure}
\subfloat[]{\includegraphics[height=70mm]{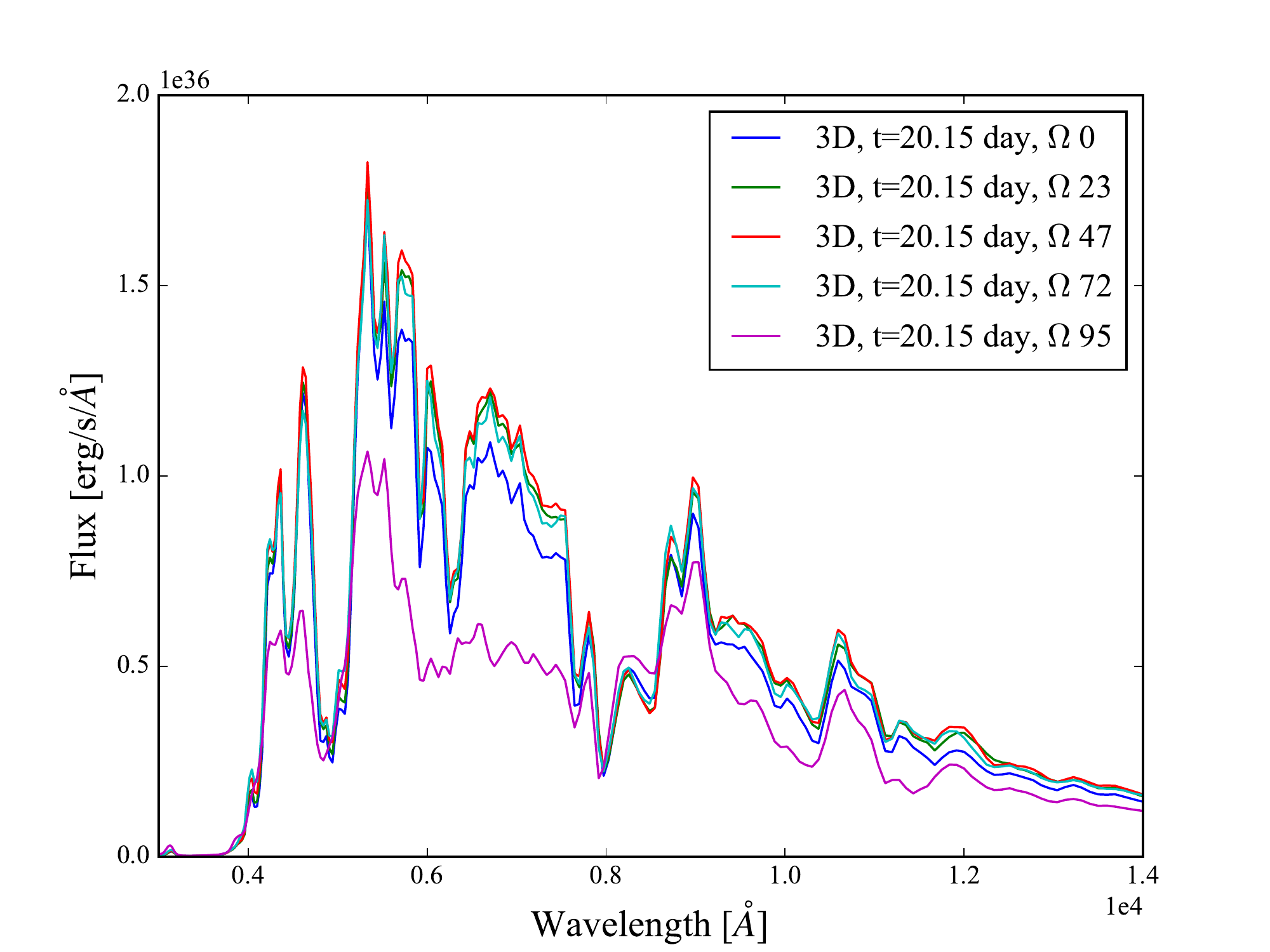}\label{fg12a}}\\
\subfloat[]{\includegraphics[height=70mm]{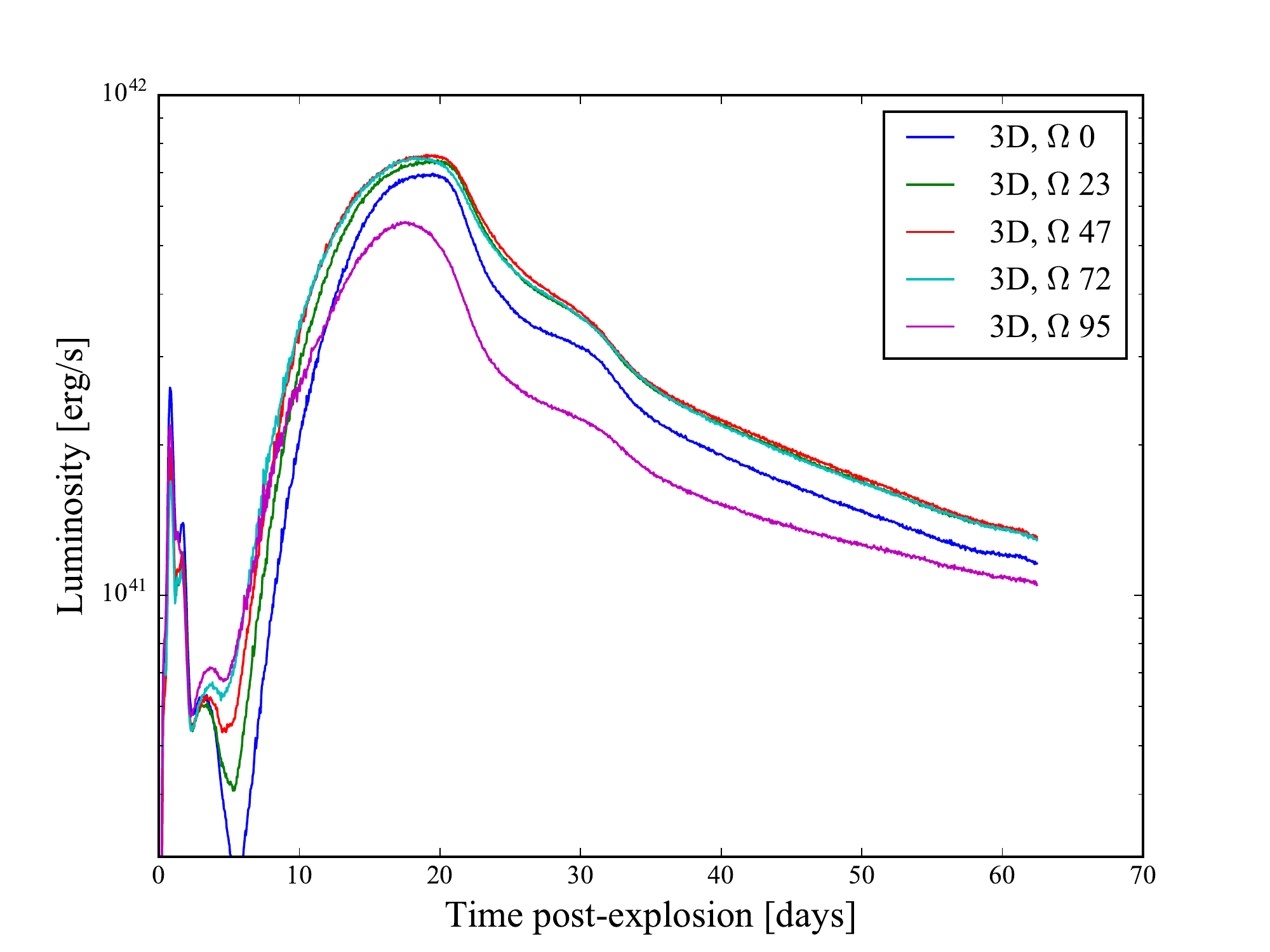}\label{fg12b}}
\caption{
  Light curves and spectra from several azimuthally integrated, polar flux
  bins selected from the 96 bin viewing-angle grid.
  The viewing-angle bins $\Omega\,95$ (purple) $\Omega\,72$ (light blue),
  $\Omega 47$ (red), $\Omega\,23$ (green), $\Omega\,0$ (blue) correspond
  to polar angle ranges $[0,11.7]^{\text{o}}$, $[60,61.4]^{\text{o}}$,
  $[91.2,92.4]^{\text{o}}$, $[121.4,122.8]^{\text{o}}$, $[168.3,180]^{\text{o}}$,
  respectively.
  The $\Omega\,95$ bin is aligned with the top of the ejecta, $z>0$ in
  Fig.~\ref{fg5}.
  The viewing-angle bins for these views have equal solid angle coverage.
  In Fig.~\ref{fg12a}, spectra for the selected viewing-angle bins
  time averaged over 8 days about day 20 post-explosion.
  In Fig.~\ref{fg12b}, bolometric light curves for the selected views.
  The actual luminosities from these views have been multiplied by 96.
  Compared to $\Omega\,0$, the emission in the $\Omega\,95$ is significantly
  lower.
  The dispersion in peak luminosities and spectra is a result of the
  asymmetric geometry of the ejecta.
}
\label{fg12}
\end{figure}

The total masses in the northern ($z>0$) and southern hemisphere are
both nearly $1.68$ M$_{\odot}$.
Moreover, the integrated ejecta compositions are very similar.
This similarity can be seen in the hemisphere-integrated light curves,
which have very close peak luminosities and times.
The dispersion in the light curves only becomes pronounced when
the hemispheres are divided into small viewing-angle bins.
Light curves and spectra for a subset of the 96 viewing-angle
bins are plotted in Fig.~\ref{fg12}.
The peak luminosity is largest from views near the equator and
systematically diminishes towards the poles, $\Omega\,0$ (looking
up the $z$-axis) and $\Omega\,95$ (looking down the $z$-axis).
The viewing-angle bins at the poles are darker since some of the
radioactive source is blocked in each of these views.
The view down the north pole, $\Omega\,95$, sees more emission
from the high-velocity unimode, producing a distinctly shaped light
curve.
The $\Omega\,95$ light curve is significantly dimmer than the
$\Omega\,0$ light curve.
The high density surface of the unimode both changes the diffusion time
scale and shades the central $^{56}$Ni source in the range of viewing
angles of $\Omega\,95$.

To remove MC noise, the spectra in Fig.~\ref{fg12b} have
been averaged over an 8 day interval about day 20 post-explosion.
The spectra are not varying strongly in this time frame, so the
averaged features should be representative of the spectra at day 20.
As in the 1D tests, the spectra from the 3D simulation has strong
Fe II line features in the 4-5$\times10^{3}$ $\AA$ range.
Additionally, the  $6.15\times10^{3}$ $\AA$ Si II line feature appears to
manifest for viewing angles out of alignment with the unimode.
The variation in the expression of an Si line with viewing angle is
consistent with the spatial abundance of Si depicted in Fig.~\ref{fg5},
where Si traces the unimode closely, and diminishes in mass fraction up the
$z$-axis.
Also of note is the variation in the $8.15\times10^{3}$ $\AA$ Ca II
feature, which shows higher emission in the unimode-aligned view at the
time given; this feature is examined in detail in Sec.~\ref{sec:caii}.
It should be noted that, except for the Ca II feature, we have not
performed a detailed ``knock-out'' spectrum study for each element,
where line contributions are removed from the opacity, as in the work
of~\cite{vanrossum2012b}.

\subsection{Gamma-Ray Light Curves}
\label{sec:gamma}

\begin{figure*}
\subfloat[]{\includegraphics[height=60mm]{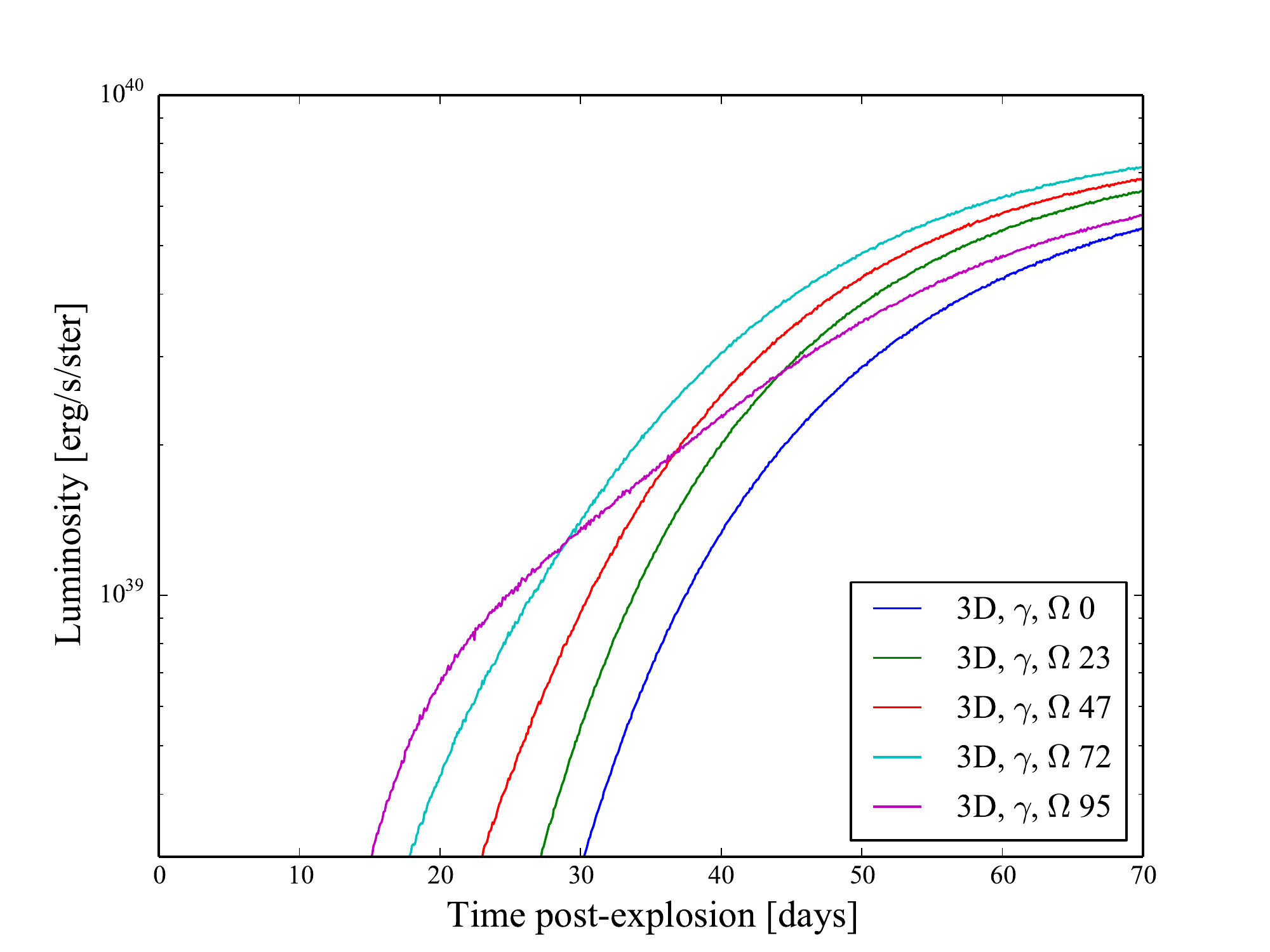}\label{fg13a}}
\subfloat[]{\includegraphics[height=60mm]{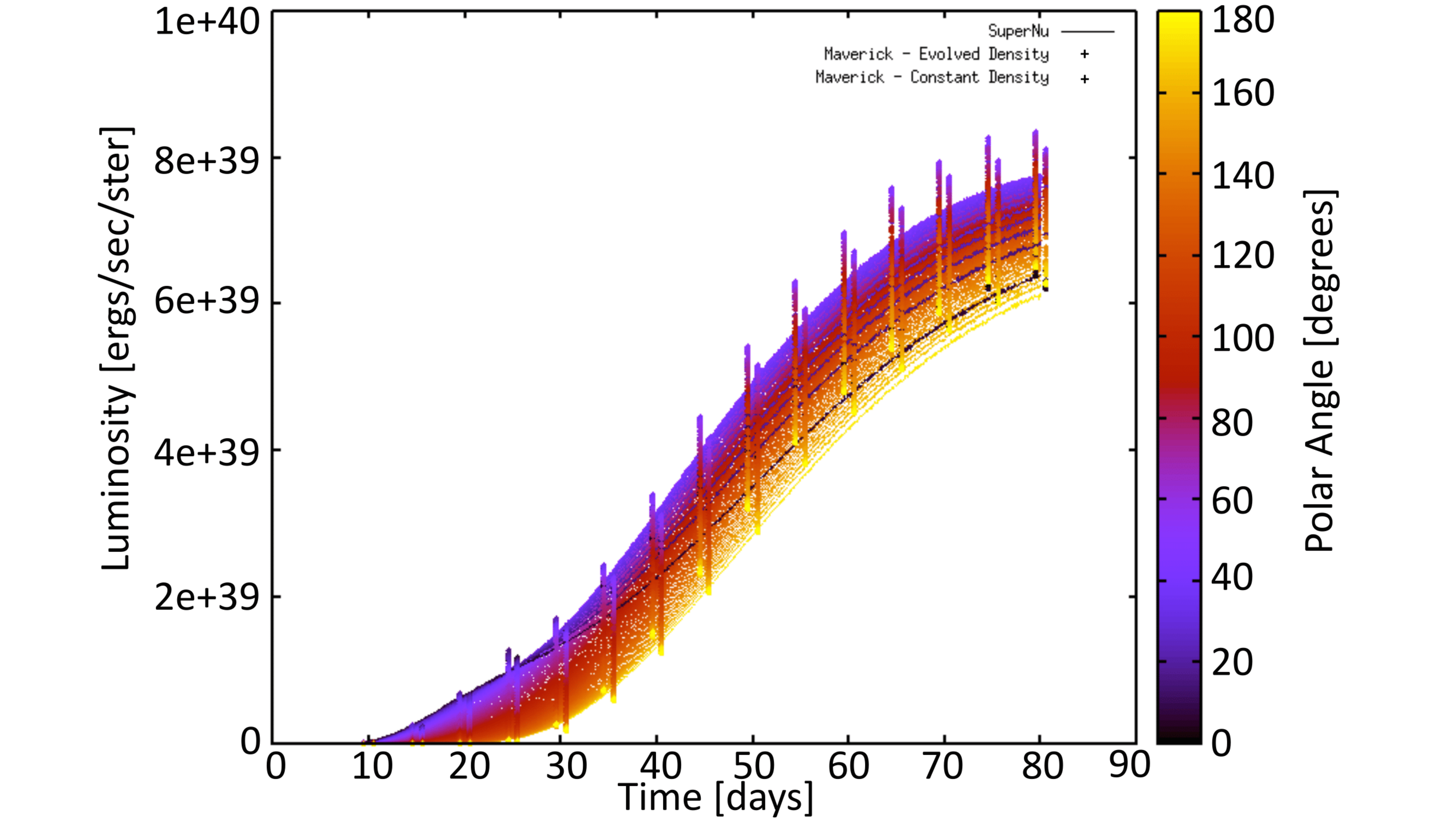}\label{fg13b}}
\caption{
  In Fig.~\ref{fg13a}, gamma-ray light curves for each of the
  viewing-angle bins depicted in Fig.~\ref{fg12}.
  These light curves exhibit monotonicity (in brightness) over
  a larger range of viewing angles than the UV/optical/IR curves.
  In Fig.~\ref{fg13b}, a comparison of the light curves
  between the grey treatment of \supernu\
  (solid lines) and two tests with the spectrally detailed
  treatment of \maverick\ (discrete points).
  The \maverick\ results for one test have been offset
  slightly backwards in time for clarity, but correspond to the same
  times as the other \maverick\ test.
  The set of \maverick\ points closer to the \supernu\
  curves have constant density per time step, while the offset
  higher luminosity \maverick\ points are from a test where
  the density was evolved during transport within the time step.
}
\label{fg13}
\end{figure*}

In Fig.~\ref{fg13a} are gamma-ray light curves for the same
viewing-angle bins as in Fig.~\ref{fg12}.
The gamma-ray light curves are about an order of magnitude dimmer
than the UV/optical/IR light curves over the time range simulated.
The \supernu\ gamma-ray transfer applies the grey, calibrated
prescription of~\cite{swartz1995} (see Section~\ref{sec:caveats} for details).
Consequently, we do not have gamma-ray spectra that were produced in-line
with the \supernu\ calculations, but we do have a reasonable estimate of
how the frequency-integrated gamma-ray luminosity changes with the polar
orientation of the observer view.
As with the UV/optical/IR results, the light curve for $\Omega\,95$ has
a distinct slope ($\Omega\,95$ is a polar range of $[0,11.7]^{\text{o}}$,
or a view of the top of the ejecta).
Over the time range simulated, the light curves monotonically increase
in brightness from bin $\Omega\,0$ to bin $\Omega\,72$.
Hence, the gamma-ray light curves appear to more closely correspond
to the inner unimode structure of the ejecta (see Fig.~\ref{fg5}).

To assess the magnitude of the errors in using a grey gamma-ray model,
we compare the gamma-ray light curves (in erg/s/steradian) of \supernu\
(solid lines) to those of the detailed gamma-ray transfer code \maverick\
(discrete points)~\citep{hungerford2003} in Fig.~\ref{fg13b}.
The \maverick\ data are for two tests; these points have been offset for
clarity, but correspond to the same times.
The dimming for low viewing angles in the simplified gamma-ray treatment in
\supernu\ is also observed in \maverick.
With \maverick, two test cases were simulated: one with constant density
in a time step and one with changing density within a time step.
The test of density change within a time step provides a measure of the importance
of time dependence in the gamma-ray transfer.
For the constant density test, there are $\sim30-40$\% differences in the
light curves between the codes.
For the test with changing density, \maverick\ produces a consistently
brighter light curve than the constant density test by up to $\sim30$\% in
the 20-60 day time range.
At 30 days, both codes find that the gamma-ray luminosities are $\sim3-4$\%
of the total source energy emitted in the form of gamma-rays, with
\maverick\ closer to 3\% and \supernu\ closer to 4\%.
Consequently, the gamma-ray energy content remaining in the ejecta available
for heating at times relevant to the optical light curves is similar between
the two codes.
The differences between the codes are attributable to the differing treatments
of opacity; the approximation in \supernu\ is discussed further in Section
\ref{sec:caveats}.
We can in principle calibrate the gamma-ray treatment in \supernu\ to
yield results closer to \maverick.
However, this would not necessarily furnish more insight to the conclusions,
since the code results have the same viewing-angle trend (moreover, at day 30,
a 30 to 40\% difference in the escaping gamma-ray energy, which is 3 to 4\% of
the total gamma-ray energy emitted at this time, corresponds to an
energy-exchange error on the order of 1\%).

We have found that the distinct trend of the low-angle light curves is
attributable to the gamma-rays escaping from the unimode early, making the
early low-angle light curves brighter, and the unimode acting to eclipse gamma-rays
from the core, making the later low-angle light curves dimmer.
Replacing the radioactive nickel with stable nickel in the unimode causes the
low-angle light curves (from views were the unimode partially eclipses the core)
to be dimmer than the mid-angle light curves for the entire time range simulated.

\begin{figure}
\subfloat[]{\includegraphics[height=70mm]{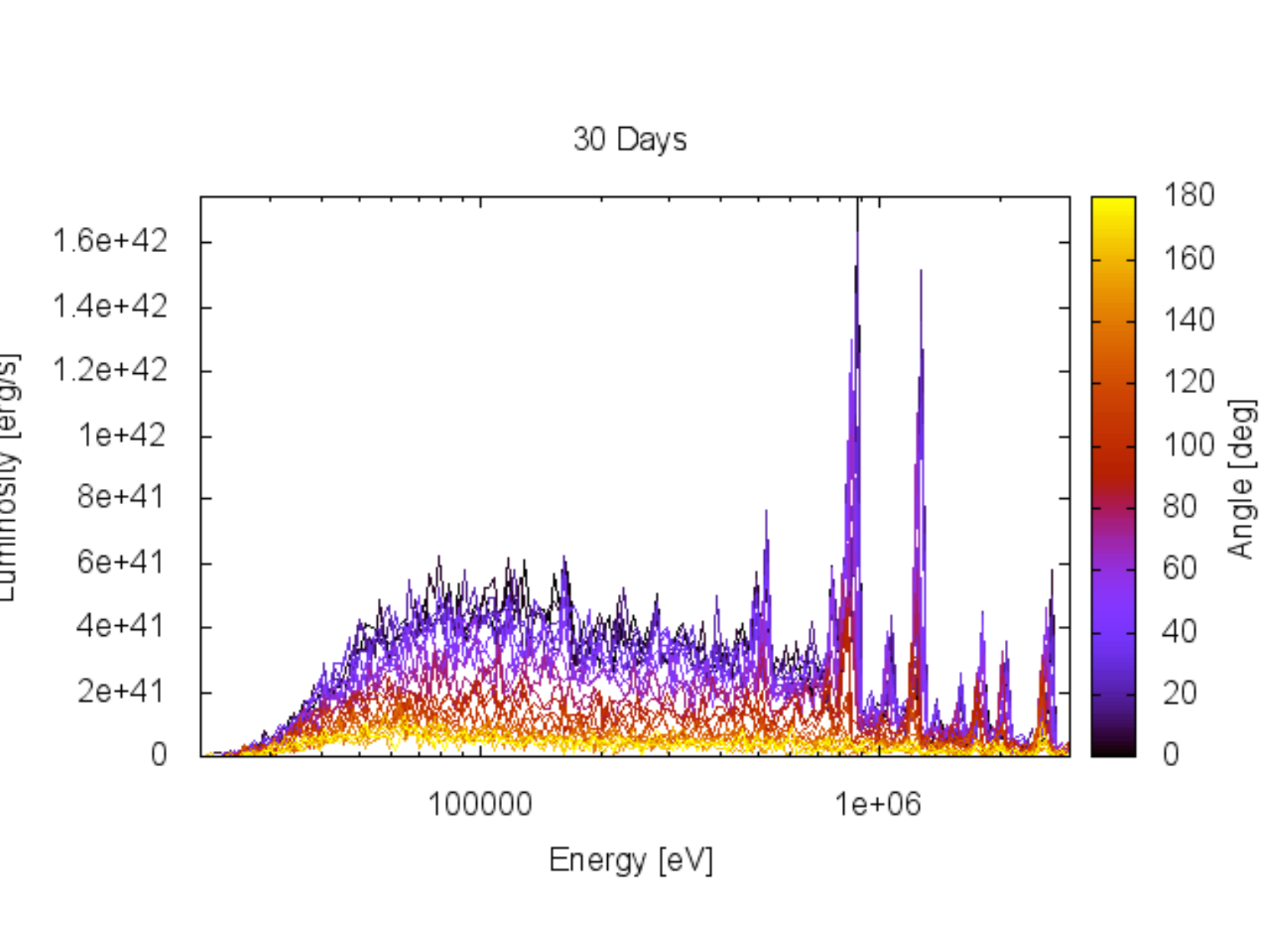}\label{fg14a}}\\
\subfloat[]{\includegraphics[height=70mm]{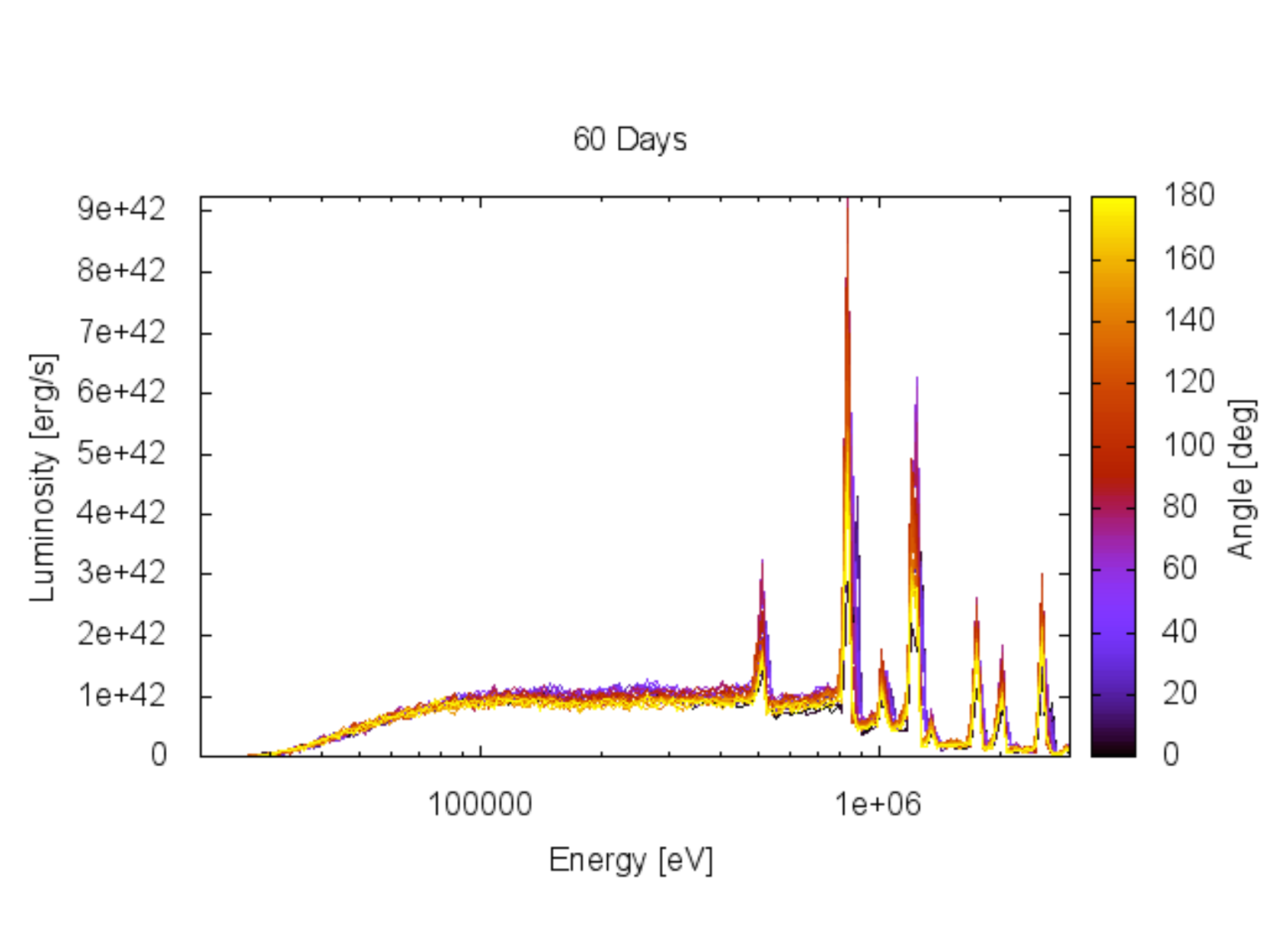}\label{fg14b}}
\caption{
  \maverick\ spectral snapshots at day 30 and day 60
  post-explosion for all polar views.
  In Fig.~\ref{fg14a}, at day 30, the low-angle views (that are
  aligned with the unimode) generally have higher luminosity
  across the span of gamma-ray energies.
  In Fig.~\ref{fg14b}, at day 60, the lowest-angle views are
  dimmer across the spectral features.
}
\label{fg14}
\end{figure}

Gamma-ray spectra from \maverick\ for each polar angle are plotted in
Fig.~\ref{fg14} at 30 and 60 days.
These spectra snapshots at the two times, and are not
integrated over time ranges.
The change with polar viewing angle shows that the entire spectrum is
affected by the ejecta geometry, similar to the optical spectra depicted
in Fig.~\ref{fg12a}.
Moreover, the shift in line features is consistent with the $^{56}$Ni mass
distribution in velocity space.

The gamma-ray study is meant to show that
the heating model in \supernu\ is reasonable compared
to \maverick.
However, it is also possible to estimate detection prospects,
similar to those given by~\cite{hungerford2005}, for the model
examined here.
We take the time of detection to be day 30 post-explosion, about
where the unimode-aligned gamma-ray light curve is overtaken in
brightness by the other viewing angles.
At this time, the total gamma-ray luminosity in photons/s
(ph/s) is $1.28\times10^{46}$.
Assuming continuum sensitivity for the SPI telescope of the
INTEGRAL mission~\citep{winkler2003}, $\sim10^{-6}$ ph/s/cm$^{2}$,
and further assuming this sensitivity is applicable to the entire
energy range of the simulated spectrum, we estimate the distance
limit for the detection at day 30 is O(10) Mpc.
Further, assuming a rate of 1 CCSN per Milky Way size galaxy
per century~\citep{adams2013}, a 10 Mpc observable volume would furnish
$\sim 1$ detectable occurrence per decade.
However, a detectable unimode-aligned orientation would then only
occur at a rate lower than 1 per century.

\subsection{Ca-II Line Emission}
\label{sec:caii}

\begin{figure}
  \subfloat[]{\includegraphics[height=70mm]{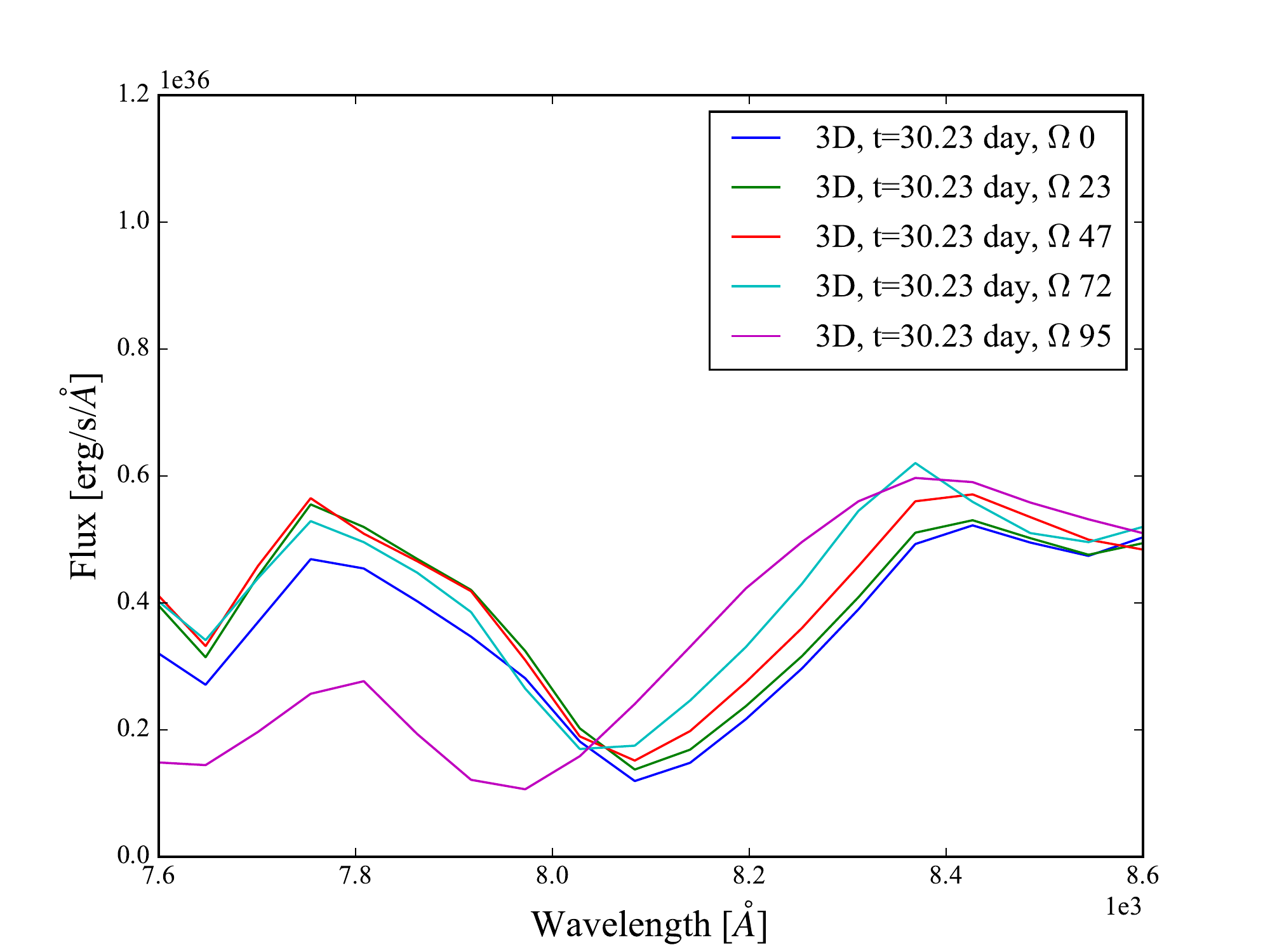}
    \label{fg15a}}\\
  \subfloat[]{\includegraphics[height=70mm]{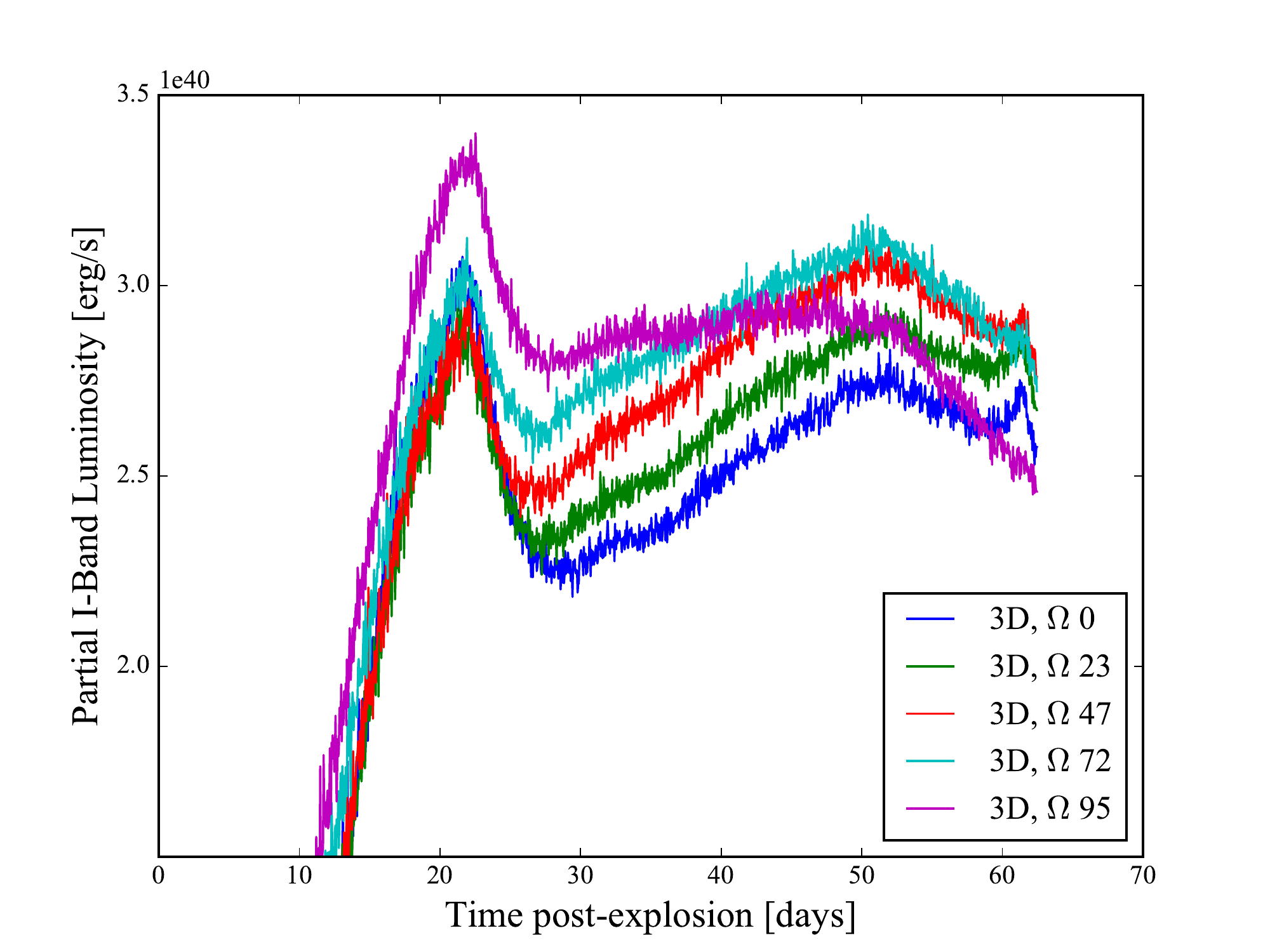}
    \label{fg15b}}
  \caption{
    In Fig.~\ref{fg15a}, close-ups of spectra near a Ca II IR line
    feature for several viewing-angle bins (see Fig.~\ref{fg12}
    for angular ranges).
    The spectra have been time-averaged over 8 days around day 30.
    Since calcium is only in the interior of the unimode, the Ca II
    emission feature is bluer for viewing-angle bins closer
    in alignment with the unimode.
    In Fig.~\ref{fg15b}, partial I-band light curves, integrated
    from $8\times10^{3}$ to $8.6\times10^{3}$ $\AA$ in wavelength,
    for each of the for the selected viewing-angle bins.
    These light curves exhibit monotonicity (in brightness) over
    a larger range of viewing angles than the total UV/optical/IR
    curves, similar to the grey gamma-ray result.
    Statistical noise is apparent over this wavelength range.
  }
  \label{fg15}
\end{figure}

Considering the UV/optical/IR, near peak luminosity, the line features
of the spectra align closely for all views.
Shortly after peak luminosity, emission features corresponding to
elements in the unimode show a monotonic shift with respect to viewing
angle, where views along the unimode are blue-shifted for elements that
the unimode contains, in particular in the near IR Ca II line feature
near $8.15\times10^{3}$ $\AA$.
Integrating from $8\times10^{3}$ to $\sim8.6\times10^{3}$ $\AA$, a
part of the I-band wavelength range, we also find a light
curve viewing angle dependence similar to the gamma-ray result, despite
the significant differences in time-dependence.
The integral of the spectrum from $8\times10^{3}$ to
$\sim8.6\times10^{3}$ $\AA$ is equivalent to applying a rectangular
response function over this range, and we refer to the result as a
``partial I-band'' light curve.
This near-IR region corresponds to a Ca II emission feature.
Calcium is only found in the unimode (Fig.~\ref{fg6}), near $^{56}$Ni.
The Ca II emission feature exhibits a higher degree of monotonicity
in brightness with respect to viewing angle.
Specifically, like the gamma-ray light curves, the Ca II emission
feature in the $\Omega 95$ (top) view is brighter than in the other
views past day 20, and also makes a departure in slope from the
other curves.
Figure~\ref{fg15} has close-up spectra and light curves for the
partial I-band.
Evidently, in order to obtain the trend in the light curves with
viewing angle in Fig.~\ref{fg15b}, the near IR Ca II lines are
necessary contributions.
Following~\cite{vanrossum2012b}, we remove the oscillator strengths
for Ca II lines and recover the trend in the total light curves,
depicted in Fig.~\ref{fg12b}, for the partial I-band light curves.

\subsection{UBVRI Light Curves}
\label{sec:ubvri}

\begin{figure}
\centering
\includegraphics[height=70mm]{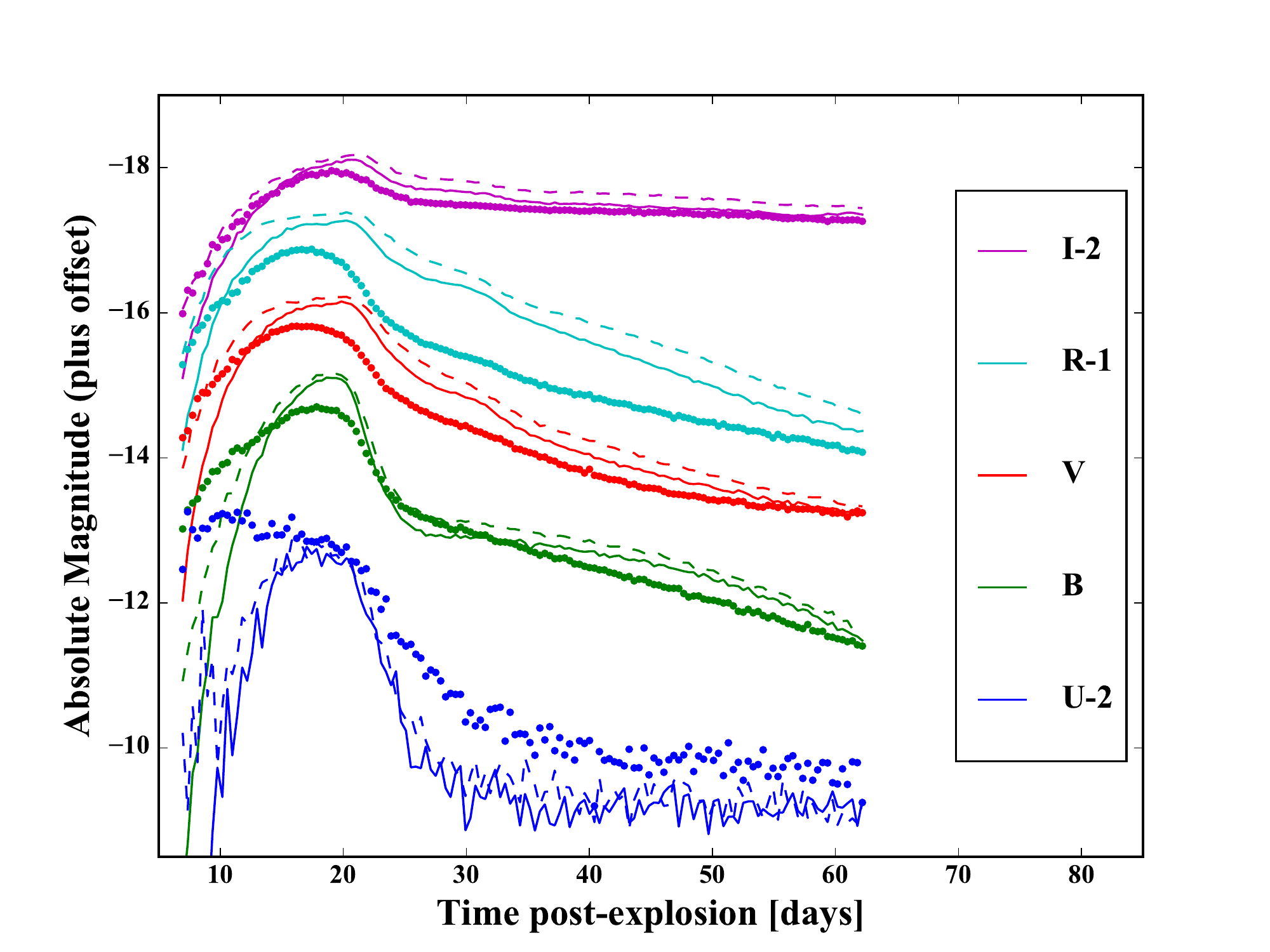}
\caption{
  AB absolute magnitude UBVRI broad band light curves for
  bottom (solid, $\Omega 0$), side (dashed, $\Omega 47$), and top
  (dotted, $\Omega 95$) views of ejecta.
  Response bands and broad band light curve formulae are
  from~\cite{bessell2012}.
}
\label{fg16}
\end{figure}

We briefly discuss the detection prospects for the radioactive unimode
signature in the UBVRI broadband light curves of our model
(following the outline of~\cite{grossman2014}).
Figure~\ref{fg16} has UBVRI absolute AB magnitude light curves for
bottom ($\Omega 0$), side ($\Omega 47$), and top ($\Omega 95$)
views of the ejecta.
Since thermal emission is low in the UV range for this problem,
in the time range simulated, \supernu\ samples relatively few source
particles in the U-band.
Consequently, the U-band suffers more Monte Carlo noise than
the other bands.
The AB magnitudes are computed with the formulae and response
functions of~\cite{bessell2012}.
In Fig.~\ref{fg16}, the contribution of the $^{56}$Ni in the unimode can
be seen in the top view for the U and B bands around day 10
post-explosion.
At day 10, the absolute magnitudes in the U-band and B-band are
$\sim -11$ and $\sim -14$, respectively.
We assume a telescope detection horizon of $\sim 24$ magnitudes in the
UV/optical range~\citep{roming2005}.
Assuming no reddening from dust, the distance moduli for our model at
the detection horizon for the early radioactive excess are
\begin{subequations}
  \label{eq4}
  \begin{align}
    & \mu_{U} = 35 = 5\log_{10}(d)-5
    \rightarrow d_{U} \approx 100 \,\text{Mpc}\;\;,\\
    & \mu_{B} = 38 = 5\log_{10}(d)-5
    \rightarrow d_{B} \approx 398 \,\text{Mpc} \;\;,
  \end{align}
\end{subequations}
where $\mu_{U}$, $\mu_{B}$, $d_{U}$, and $d_{B}$ are the U and B
band distance moduli and distances.
We suspect that obtaining light curves for the low-magnitude U and B
bands at early time, along with follow-up observations, would be
difficult for this CCSN model.

Apart from the U-band, and to a lesser extent the B-band, there are
no other strong signs of asymmetry in Fig.~\ref{fg16}.
The Ca II feature in the near-IR, discussed in Section~\ref{sec:gamma},
does not manifest significantly in the I-band light curve, where it is
subsumed by other emission contributions.
The change in the I-band light curves with viewing angle is small
(Figs~\ref{fg16}).
Hence, the I-band light curve would not be a good measure for the
asymmetry of our ejecta either.

\subsection{Comparison to SN 2002ap}
\label{sec:sn2002ap}

\begin{figure}
\subfloat[]{\includegraphics[height=70mm]{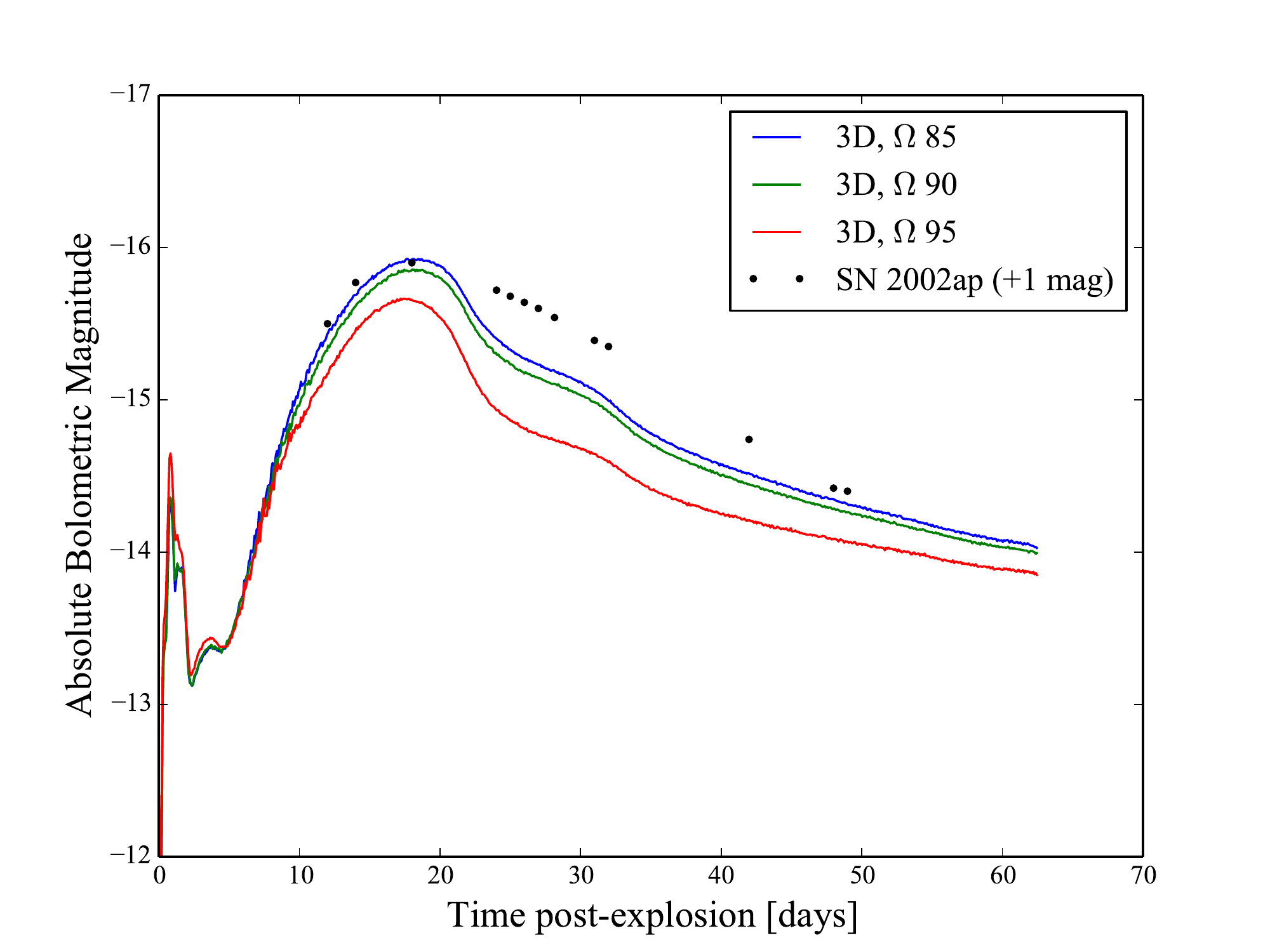}
  \label{fg17a}}\\
\subfloat[]{\includegraphics[height=70mm]{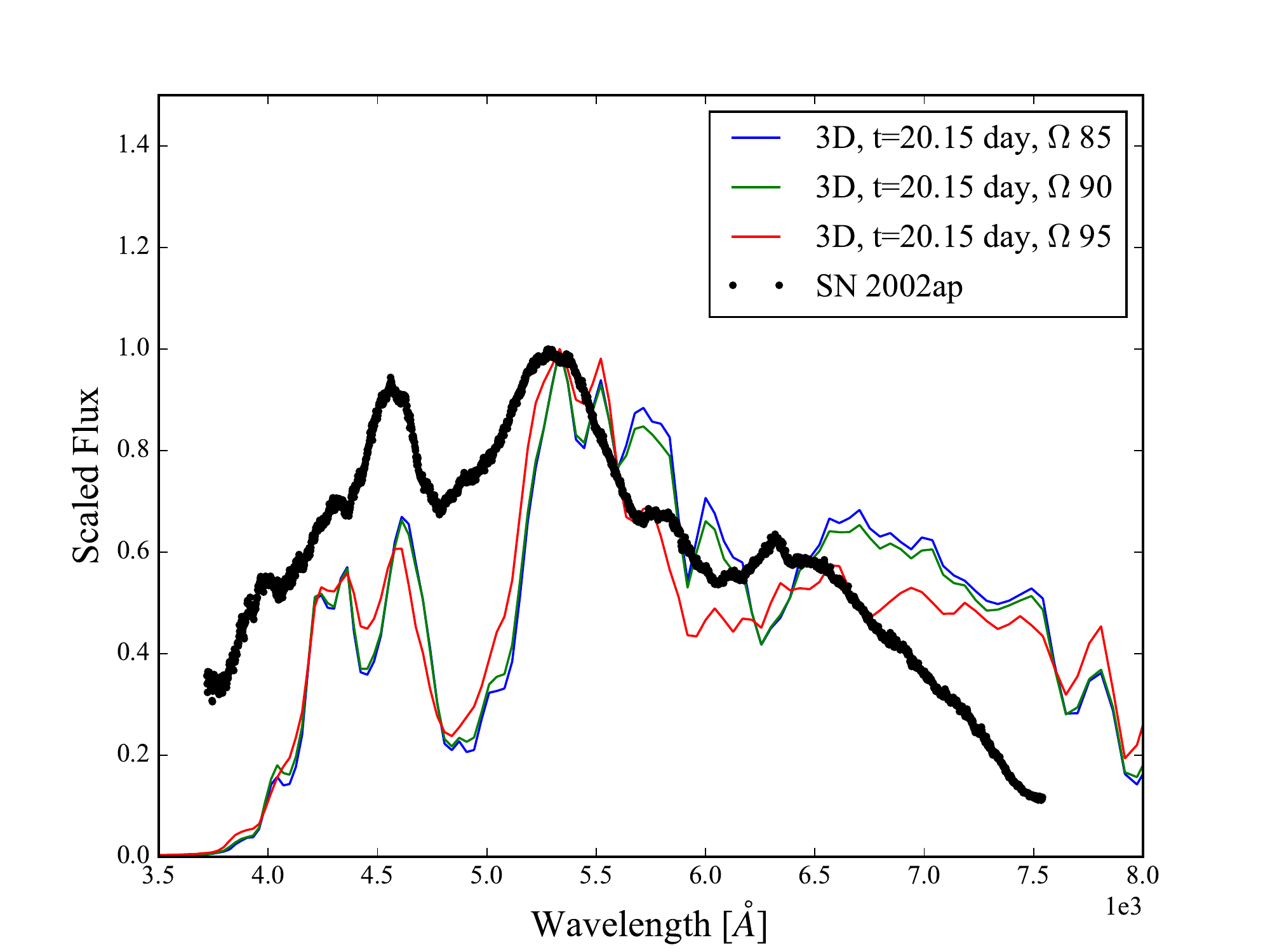}
  \label{fg17b}}\\
\subfloat[]{\includegraphics[height=70mm]{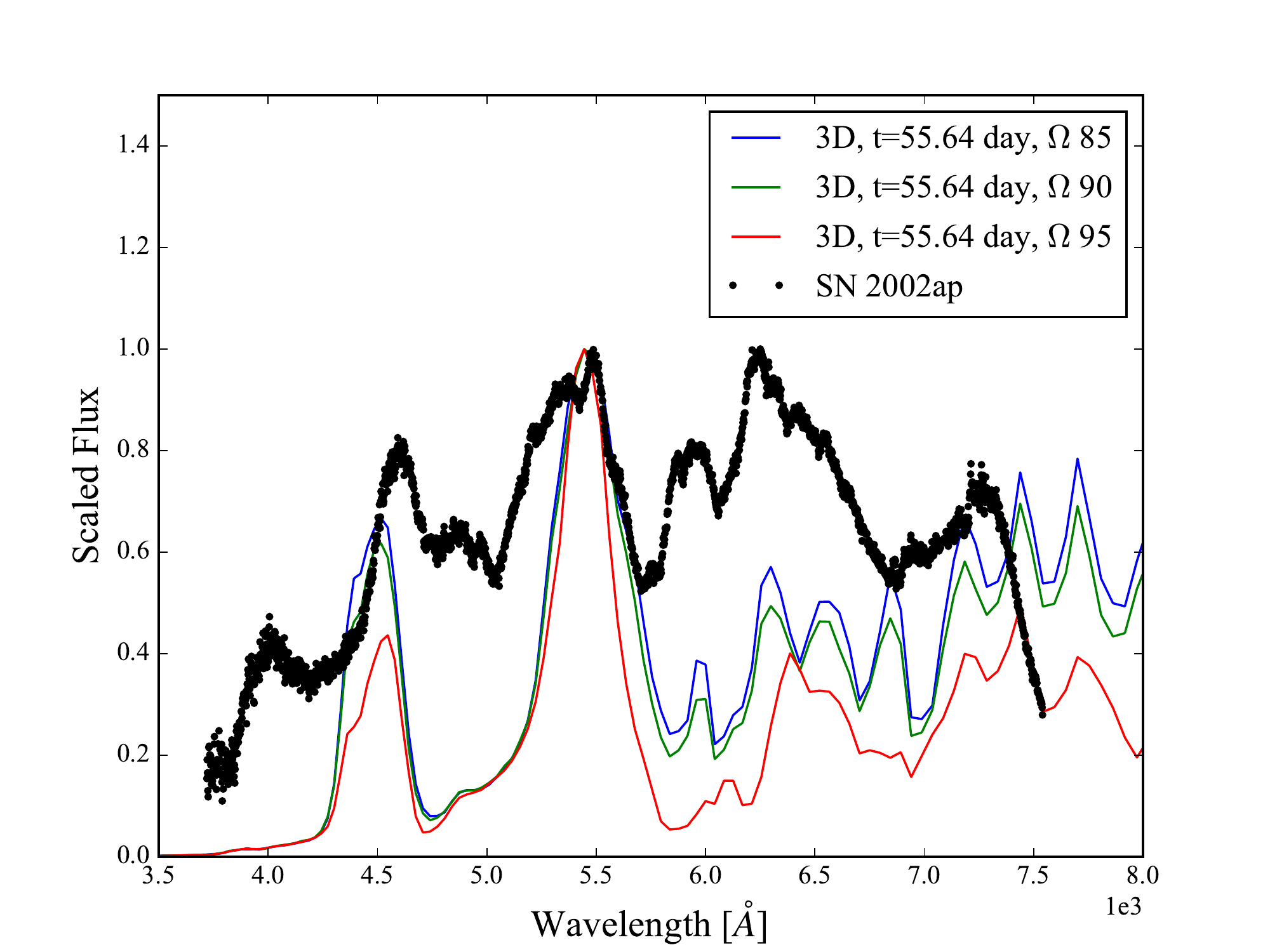}
  \label{fg17c}}
\caption{
  Bolometric light curve and spectra at peak and $\sim$26
  days past peak luminosity of SN 2002ap, an low-energy, low-mass
  SN Ic hypernova, along with the light curve and spectra at similar
  times for the unimodal CCSN model.
}
\label{fg17}
\end{figure}

We compare our result to SN 2002ap, a low-energy, low-mass SN Ic
hypernova possibly having a kinetic energy of $\sim 4-10\times10^{51}$ erg,
an ejected mass of 2.5-5 M$_{\odot}$, and $\sim0.07$ M$_{\odot}$ of
$^{56}$Ni~\citep{mazzali2002,yoshii2003}.
\cite{kawabata2002} find that the polarization profiles of a
O I multiplet and a Ca II IR triplet indicate SN 2002ap has a
high-velocity unimode containing $^{56}$Ni.
Our unimodal CCSN model has a kinetic energy of $6\times10^{51}$ erg,
an ejected mass of 3.36 M$_{\odot}$, and $0.024$ M$_{\odot}$ of $^{56}$Ni.
Consequently, we expect some differences in the light curves and spectra.
To compare, we select angular views that are close to aligning with the unimode.
Figure~\ref{fg17} has light curves and spectra for SN 2002ap, along with
high-latitude light curves and spectra produced by \supernu.
The spectral data for SN 2002ap is from a set of low-redshift
($z\lesssim 0.05$) stripped-envelope CCSNe presented by
\cite{modjaz2014}\footnote{https://www.cfa.harvard.edu/supernova/SNarchive.html}
and~\cite{bianco2014}.
The bolometric light curve data is obtained from~\cite{yoshii2003}
(see their Table 3).
No redshift or dust corrections were applied to the spectral data
\citep{bianco2014}; SN 2002ap has a redshift of $z\approx 0.002$ (galaxy M74)
\citep{modjaz2014}.
As expected, we obtain a dimmer light curve by an amount consistent with
the difference in $^{56}$Ni abundances~\citep{arnett1982}.
In the spectra, we clearly see the Fe II lines, indicative of Type Ic,
align between the unimodal CCSN model and SN 2002ap.
However, since SN 2002ap has a higher luminosity, the ejecta temperature
is likely higher than that of our model.
The emission of SN 2002ap is bluer than the model,
which is apparent in the shallower Fe II feature and steeper decline
from $\sim6500-7500$ $\AA$ in Fig.~\ref{fg17b}.

\subsection{Caveats}
\label{sec:caveats}

Here we discuss some limitations of the current work in order to
illuminate potential areas for improvement.
Possibly the largest limitation in the utility of our work is the
lack of shock heating due to the ejecta passing through circumstellar
media (CSM).
A CSM may be produced by wind from the SN progenitor, and
the amount of wind depends on the metallicity of the progenitor
\citep{heger2003}.
Shock heating from the CSM can contribute significantly to
the light curve~\citep{chatzopoulos2012}.
For models with explosion properties similar to ours but with
a sufficiently dense CSM~\citep{chugai1990}, the luminosity
contribution of H$\alpha$ recombination from the pre-shocked
region of the CSM can be $\sim10^{40}-10^{41}$ erg/s for $\lesssim100$
days post-explosion~\citep{chugai1997}.
Such H$\alpha$ luminosity ranges are within an order of magnitude
of the peak luminosities for our CCSN model.

There are several numerical limitations as well.
For the explosion model, asymmetry is introduced artificially
in the kinetic energy to produce the unimodal structure.
A full 3D treatment of the progenitor mass loss and collapse may
significantly impact the ejecta and in turn the observables.
For the radiative transfer phase, we have performed
resolution tests in space, time, and wavelength for the 1D spherical
ejecta, and consequently do not assess the impact of resolution
on small scale ejecta features in 3D.
Additionally, our multigroup wavelength grid does not resolve
inertial (or thermal) line shapes.
However, to obtain more detailed spectral synthesis over
narrower wavelength groups, we need to implement thermal line
broadening.
For the gamma-ray transfer, \supernu\ only has a grey
absorption treatment to get energy deposition.
Implementations of continuous line opacity sampling and detailed
gamma-ray transfer already exist (see, for instance,
\citet{hauschildt1999,kasen2006,baron2007,kromer2009,vanrossum2012}
for different methods of directly treating optical lines, and
\citet{hungerford2003} for a multi-frequency gamma-ray treatment).

%-- multigroup treatment
In \supernu, for the UV/optical/IR radiative transfer, the
opacity has a multigroup treatment, where
the wavelength-dependent opacity is integrated over wavelength
ranges; this is favorable for the DDMC implementation.
The opacity array is computed in the comoving frame, meaning
the opacity used in the radiative transfer does not encode
broadening from differential velocities due to the ballistic
expansion of the ejecta.
Instead, opacity is computed for each cell from the temperature,
density, and composition of the cell as though the cell is static.
Following~\cite{abdikamalov2012}, the IMC-DDMC transport scheme
keeps track of frame transformations, which effectively broaden the
expanse of each group in an inertial frame.
The opacity is transformed to the inertial frame of the center
of the ejecta for IMC particles.
To treat redshift over a multigroup structure,
\cite{wollaeger2013} introduce a transport particle distance
in velocity space~\citep{kasen2006} to redshift across comoving
group edges,
\begin{equation}
  \label{eq5}
  u_{\text{Dop}}=
  c\left(1-\frac{\lambda_{p}}{\lambda_{g+1/2}}\right)-
  \vec{U}_{p}\cdot\hat{\Omega}_{p} \;\;,
\end{equation}
where $u_{\text{Dop}}$ is the distance an IMC particle has to
stream in order to redshift into a longer wavelength group,
and $\lambda_{g+1/2}$ is the target wavelength group bound.
For IMC particle $p$, $\lambda_{p}$, $\text{U}_{p}$, and
$\hat{\Omega}_{p}$ are the wavelength, the gas velocity at the
particle location in the ejecta, and the particle direction,
respectively.
Equation~\eqref{eq5}, along with the frame transformation of
the opacity (where the inertial opacity is used to calculate
a collision distance), replaces the Sobolev approximation
implemented in other MC codes (for a detailed discussion of the
Sobolev approximation, see~\citet{castor2004}).

Notwithstanding the ability to treat expansion effects, our
multigroup treatment does not take into account the effect
of lines spanning multiple wavelength groups due to the thermal
motion of atoms and ions.
Since our light curves are in the UV/optical/IR range, this motion
is an important contributor to line broadening relative to
the ``observational'' resolution in wavelength.
From~\cite{wollaeger2014}, for a group with index $g$, the total
bound-bound contribution to the multigroup opacity array is
\begin{multline}
  \label{eq6}
  \sigma_{a,g,bb} = \\
  \frac{1}{\Delta\lambda_{g}}\sum_{s}\sum_{i}\sum_{i'>i}
  \left(\frac{\pi(e^{-})^{2}}{m_{e^{-}}c}\right)f_{i,i',s}
  \frac{\lambda_{i,i',s}^{2}}{c}\times\\
  (\Theta(\lambda_{i,i',s}-\lambda_{g-1/2})-
  \Theta(\lambda_{i,i',s}-\lambda_{g+1/2}))\times\\n_{i,s}
  (1-e^{hc/kT\lambda_{i,i',s}}) \;\;,
\end{multline}
where $\sigma_{a,g,bb}$ is the bound-bound absorption opacity,
$e^{-}$ and $m_{e^{-}}$ are the electron charge and mass,
$\Delta\lambda_{g}=\lambda_{g-1/2}-\lambda_{g+1/2}$ is the wavelength
band of the group, $f_{i,i',s}$ is the non-dimensional oscillator strength
for transitions from atomic state $i$ to $i'$ of species $s$,
$\lambda_{i,i',s}$ is the wavelength at the center of the line, $n_{i,s}$ is
the total density of species $s$ occupying state $i$.
The $\Theta$ are again Heaviside step functions constraining the sum
to opacity profiles centered in the group.
Since the standard LTE opacity formula for a single line is
\citep{mihalas1984}
\begin{equation}
  \label{eq7}
  \sigma_{a,\lambda(i,i',s),bb}=\frac{\pi(e^{-})^{2}}{m_{e^{-}}c}f_{i,i',s}
  \frac{\lambda_{i,i',s}^{2}}{c}(1-e^{hc/kT\lambda_{i,i',s}})\varphi(\lambda)
  \;\;,
\end{equation}
where $\varphi(\lambda)$ is a Gaussian function for thermal broadening,
we are effectively compressing line wings into the group containing
the line center, $\lambda_{i,i',s}$ (or treating $\varphi(\lambda)$ as a
Dirac delta distribution), in Eq.~\eqref{eq6}.
In order to estimate an upper bound for the group resolution, we can
compare the ratio of the full width at half maximum of a line to the
width of a group,
\begin{equation}
  \label{eq8}
  \frac{\Delta\lambda_{g}}{\lambda_{g}}\gtrsim
  \sqrt{8\ln(2)\frac{kT}{m_{A}c^{2}}} \;\;,
\end{equation}
where $m_{A}$ is the mass of a relevant element $A$; we have
required in Eq.~\eqref{eq8} that the ratio of the size of the
group to the wavelength center, $\lambda_{g}$, must be greater than
or comparable to the characteristic ratio of a thermal line width
to the line wavelength center (for the right side of Eq.~\eqref{eq8}
see~\citet{griem2005}).
For our logarithmic group structures, the left side of Eq.~\eqref{eq8}
has the property of being constant with respect to group index, $g$.
Conservatively, with $A=$Hydrogen and $T=10^{5}$ K, for the right
side of Eq.~\eqref{eq8} we obtain $\sim2.3\times10^{-4}$.
This conservative estimation would yield an upper bound of 2000 groups.
More realistically, using $m_{A}\sim10$ and $T=10^{4}$ K, a proper
upper bound would be 20,000 groups.
This is more groups than are typically applied with multigroup
SN codes, since the broadening due to the high ejecta velocities
has a larger wavelength scale.
Consequently, the absence of a thermal broadening treatment does
not impact the spectra presented for our CCSN.
However, this is another obstacle to treating narrow line emission
from a CSM, which might move at much lower speeds.

%-- gamma-ray treatment
For the gamma-ray deposition source, we apply the prescription of
\cite{swartz1995}.
In this approach, gamma-ray transfer is treated efficiently as a
pure-absorption Monte Carlo process.
The absorption opacity is calibrated from exact transfer of gamma-rays
emitted from the Eq.~\eqref{eq3} decay chain in SN conditions, where
Compton scattering dominates~\citep{swartz1995}.
\cite{swartz1995} find the calibrated absorption opacity to be
\begin{equation}
  \label{eq9}
  \kappa_{\gamma}\approx 0.06Y_{e} \;\;,
\end{equation}
where $\kappa_{\gamma}$ is in cm$^{2}$/g and $Y_{e}$ is the electron
fraction (computed per cell based on the cell's composition).
The physical motivation for Eq.~\eqref{eq9} is the small (large)
loss of energy from a gamma-ray when it undergoes a small (large)
angle Compton scattering~\citep{swartz1995}.
Equation~\eqref{eq9} was shown to produce accurate energy
deposition profiles for 1D models that resemble SNe Ia (W7) and
CCSNe (10H)~\citep{swartz1995} and hence should be reasonable
for our ejecta.
However, since this is a calibrated result, we do not dismiss the
possibility of a difference in the optimal value of $\kappa_{\gamma}$
for our ejecta or even between our 1D and 3D simulations.
The best way to test Eq.~\eqref{eq9} is to perform
multi-frequency gamma-ray transfer for our 3D structure and
perform the best-fit procedure of~\cite{swartz1995} to
$\kappa_{\gamma}$, which we do not provide in this work.
Also, despite being able to obtain light curves from the
pure absorption gamma-ray transfer model, we cannot obtain
spectra.
Considering the results of~\cite{hungerford2003} and
\cite{hungerford2005}, the gamma-ray spectra provide a promising
avenue to examining the asymmetry in SNe.

%-- space-time resolution test
Lastly, we briefly remark on the slight increase in percentage
change in peak luminosity for the spatial resolution test in
section~\ref{sec:1dconv}.
The operator-split particle advection prescription of \supernu,
described by~\cite{wollaeger2013}, reverse-advects IMC particles in
velocity space before and after the MC transport step.
It can be shown that this prescription produces the correct $1/t^{4}$
dependence of comoving radiation energy density for an optically thick
expanding atmosphere~\citep{mihalas1984,abdikamalov2012}.
Generally, the operator split requires the time step to be small
or comparable to the fluid time scale for each cell, or
\begin{equation}
  \label{eq10}
  U\Delta t\lesssim\Delta U t \rightarrow
  \frac{1}{n}\lesssim\frac{1}{j} \;\;,
\end{equation}
where $n$ and $j$ are the current time step and radial cell at time
$t$ and velocity $U=|\vec{U}|$, respectively.
The implied inequality between $n$ and $j$ in Eq.~\eqref{eq10}
assumes uniform time steps and spatial cells, as applied in Section
\ref{sec:1dconv}.
In the $N_{r}=400$ cell 1D calculation, the peak luminosity is
averaged from $n=400$ to $n=600$ (which is the case for all
resolution tests where $N_{t}=1984$).
Slightly before peak luminosity, the Planck photosphere is located
at $j\gtrsim 200$.
Thus, without making changes to the time step size, the
400 cell calculation of the $N_{r}$ resolution test approaches
a regime where the operator split fluid advection is becoming
insufficient near peak luminosity.
Hence, we see the percent change in peak luminosity go from
-.04\% for $100\rightarrow 200$ cells to .4\% for
$200\rightarrow400$ cells.
For the spatial and temporal resolutions tested, the small changes
in the peak luminosity demonstrate the operator split fluid
advection scheme performs well for SN domains.

\section{Conclusions and Future Work}
\label{sec:conc}

We have simulated time-dependent radiative transfer in a 3D
core-collapse explosion structure generated by the \snsph\
software.
The ejecta produced by \snsph\ has a unimodal
geometry with unimode along the positive $z$-axis.
From the radiative transfer simulation, performed by the
implicit Monte Carlo code \supernu\ and the detailed gamma-ray
Monte Carlo code \maverick, we have obtained light curves
and spectra.
We have performed several 1D simulations, which lend evidence
that the 3D results are insensitive to changes in the
temporal and wavelength resolutions.
We have also lent some justification for the point-approximation
used to map SPH data to our spatial (velocity) grids.

For the light curves and spectra from the 3D simulation,
the total dispersion in light curves with respect to viewing
angle is significant; the ratio between the maximum and minimum
peak luminosities is $\sim1.36$.
From either pole, the UV/optical/IR light curves increase in
brightness towards the views aligned with the equator (the
$xy$-plane in Fig.~\ref{fg5}) of the ejecta.
The dimmer light curves at the poles are a result of
a smaller visible projected area of the $^{56}$Ni region, where
radioactive decay heats the plasma to produce thermal UV/optical/IR
photons.
Our grey, multidimensional gamma-ray treatment also provides
light curves; these exhibit monotonicity in brightness versus
viewing angle over a larger range of viewing angles.
We have compared our gamma-ray light curves to those of the
detailed gamma-ray transfer code \maverick, and find
agreement in the trend in brightness with viewing angle.
Consequently, the gamma-ray light curves are potentially a
deeper probe of the ejecta than the UV/optical/IR light curves.

However, we find that a Ca II emission feature in the near-IR,
from $8\times10^{3}$ to $8.6\times10^{3}$ $\AA$,
follows a similar trend as the gamma-ray light curves.
For our CCSN, calcium is exclusively in the unimode.
Moreover, if we remove Ca II line contributions from the opacity
calculation~\citep{vanrossum2012b}, we see that this partial
I-band no longer follows the gamma-ray trend with viewing angle,
but instead matches the trend of the total light curves.
This finding appears to agree with the spectropolarimetric
observations of~\cite{kawabata2002} for the Type Ic SN 2002ap,
which show strong polarization the Ca IR triplet emission feature
for about one month after peak luminosity.
Unfortunately, we did not simulate polarized transport, forbidding
direct comparison.

The CCSN simulation from \snsph\ produced an extreme level
of asymmetry in the ejecta.
Other CCSN models do not produce isolated unimodes.
Furthermore, we did not include shock heating or hydrodynamical
effects from a CSM.
Despite these features, our radiative transfer simulation
only produced a modest dispersion in the brightness and shape
of the light curves.
For our CCSN structure, the most promising band to observe is part
of the near-IR, from $8\times10^{3}$ to $8.6\times10^{3}$ $\AA$.
From Fig.~\ref{fg15}, based on the shape of the $\Omega\,95$ light
curve, it may be possible to infer if the axis of the unimode is in
alignment with Earth's line of sight.
This could be accomplished by incorporating a partial I-band filter
into a telescope or by reconstructing the partial I-band
light curve from a time series of the spectrum.
Considering the results of~\cite{hungerford2005}, which
demonstrate a close connection between redshifted iron group
gamma-ray line emission and modal mixing/morphology, we expect
gamma-ray spectra are more useful probes of asymmetry in the
innermost ejecta, assuming the supernova is Galactic
\citep{hungerford2005}.

%-- future work
Future work involves investigating the topics described in
Section~\ref{sec:caveats}.
Additionally, and most critically, we must upgrade the point
particle approximation used in the SPH particle-to-grid mapping,
described in Section~\ref{sec:sphgrid}, and perform a more
focused study on the shock-breakout contribution to the light
curve.
Additionally, it would be interesting to perform radiative
transfer on other CCSN SPH simulations to obtain a more
comprehensive picture of how asymmetries affect UV/optical/IR
light curves.
We might then submit these light curves to SN analysis
tools, which compare simulated data to observations
and perform model verification~\citep{bayless2016}.
A more detailed treatment of gamma-ray transfer, that
furnishes spectra, would help to supplement our gamma-ray
light curves and lend further evidence to the utility
of gamma-rays as a probe for asymmetry.

\nocite{*}
\bibliography{Bibliography}

\end{document}